\documentclass[longauth]{aa}  

\usepackage[colorlinks=true,linkcolor=blue,citecolor=blue,urlcolor=blue]{hyperref}
\usepackage{url}
\usepackage{graphicx}
\usepackage{txfonts}
\usepackage{epstopdf}
\usepackage{lipsum}
\usepackage{subcaption}         
\usepackage{lscape}             
\usepackage{placeins}           

\usepackage{xcolor}

\newcommand\ujy{{$\mu$Jy}}

\newcommand{\thisSN}[0]{SN\,2025ulz}

\newcommand{\revone}[1]{#1}

\begin{document}

   \title{ENGRAVE follow-up of a type IIb supernova spatially coincident with the sub-threshold gravitational wave trigger S250818k}
   \titlerunning{ENGRAVE follow-up of \thisSN}


\author{
{Kendall} {Ackley}\inst{1},
{Maria Teresa} {Botticella}\inst{2},
{Andreas} {Boye}\inst{3},
{Marica} {Branchesi}\inst{4},
{Gabriele} {Bruni}\inst{5},
{Enrico} {Cappellaro}\inst{6},
{Sylvain} {Chaty}\inst{7},
{Ting-Wan} {Chen}\inst{8},
{Filippo} {D'Ammando}\inst{9},
{Valerio} {D'Elia}\inst{10},
{Massimiliano F.} {De Pasquale}\inst{11},
{} {Dimple}\inst{12},
{Rob A.~J.} {Eyles-Ferris}\inst{13},
{Morgan} {Fraser}\inst{14},
{Giulia} {Gianfagna}\inst{5},
{James H.} {Gillanders}\inst{15},
{Giuseppe} {Greco}\inst{16},
{Mariusz} {Gromadzki}\inst{17},
{Claudia P.} {Guti\'errez}\inst{18,19},
{Aprajita} {Hajela}\inst{20},
{Luca} {Izzo}\inst{2},
{Peter G.} {Jonker}\inst{21},
{Shiho} {Kobayashi}\inst{22},
{Rubina} {Kotak}\inst{23},
{Gavin  P.} {Lamb }\inst{22},
{Giorgos } {Leloudas}\inst{3},
{Andrew J.} {Levan}\inst{21,1},
{Joe D.} {Lyman}\inst{1},
{Kate} {Maguire}\inst{14},
{Antonio} {Martin-Carrillo}\inst{24},
{Andrea} {Melandri}\inst{25},
{Michał J.} {Michałowski}\inst{26},
{Samantha R.} {Oates}\inst{27},
{Francesca} {Onori}\inst{25},
{Barbara} {Patricelli}\inst{28,29,30},
{Elena} {Pian}\inst{31},
{Giuliano} {Pignata}\inst{32},
{Silvia} {Piranomonte}\inst{25},
{Luigi} {Piro}\inst{5},
{Quentin} {Pognan}\inst{33},
{Maria L.} {Pumo}\inst{34,35},
{Samuele} {Ronchini}\inst{4},
{Andrea} {Rossi}\inst{31},
{Rupak} {Roy}\inst{36,37},
{Andrea} {Saccardi}\inst{38,39},
{Om Sharan} {Salafia}\inst{40,41},
{Ruben} {Salvaterra}\inst{42},
{Nikhil} {Sarin}\inst{43,44},
{Steve} {Schulze}\inst{45},
{Stephen J.} {Smartt}\inst{15},
{Rhaana L.~C.} {Starling}\inst{13},
{Danny} {Steeghs}\inst{1},
{Nial R.} {Tanvir}\inst{13},
{Aishwarya Linesh} {Thakur}\inst{5},
{Susanna D.} {Vergani}\inst{46},
{Shuxu} {Yi}\inst{47,48},
{David R.} {Young}\inst{49}
}
\institute{
University of Warwick, Gibbet Hill Road, Coventry, CV4 7AL UK \and
INAF - Osservatorio Astronomico di Capodimonte , Salita Moiariello, 16, 80131 Napoli NA Italy \and
DTU Space, Department of Space Research and Space Technology, Technical University of Denmark, Elektrovej 327, 2800 Kgs. Lyngby, Denmark \and
Gran Sasso Science Institute, I-67100 L'Aquila, Italy \and
INAF - Istituto di Astrofisica e Planetologia Spaziali, Via del Fosso del Cavaliere, 100, I-00133 Rome (RM), Italy \and
INAF - Osservatorio Astronomico di Padova, vicolo dell'Osservatorio 5, I-35122 Padova (PD), Italy \and
Université Paris Cité, CNRS, Astroparticule et Cosmologie, F-75013 Paris, France, 85 boulevard Saint-Germain - 75270 PARIS Cedex 06 — France \and
Graduate Institute of Astronomy, National Central University, 300 Jhongda Road, 32001 Jhongli, Taiwan \and
INAF - Istituto di Radioastronomia, Via P. Gobetti 101, I-40129, Bologna, Italy \and
ASI Space Science Data Centre, Via del Politecnico, snc, I-00100 Rome (RM), Italy \and
University of Messina, Via F. S. D’Alcontres 31, Messina, 98166 \and
School of Physics and Astronomy, University of Birmingham, Edgbaston, Birmingham, B15 2TT, UK. \and
School of Physics and Astronomy, University of Leicester, University Road, Leicester, LE1 7RH, UK \and
School of Physics, O’Brien Centre for Science North, University College Dublin, Belfield, Dublin 4, Ireland \and
Astrophysics, Department of Physics, University of Oxford, Keble Road, Oxford, OX1 3RH, UK \and
INFN - Sezione di Perugia, Via A. Pascoli, I-06123 Perugia (PG), Italy \and
Astronomical Observatory, University of Warsaw, Al. Ujazdowskie, 00-478 Warszawa, Poland \and
Institute of Space Sciences (ICE, CSIC), Campus UAB, Carrer de Can Magrans, s/n, E-08193 Barcelona, Spain \and
Institut d'Estudis Espacials de Catalunya (IEEC), Edifici RDIT, Campus UPC, 08860 Castelldefels (Barcelona), Spain \and
DARK, Niels Bohr Institute, University of Copenhagen, Jagtvej 155, 2200 Copenhagen, Denmark \and
Radboud University, Department of Astrophysics/IMAPP, Radboud University, P.O. Box 9010, 6500 GL, Nijmegen, The Netherlands \and
Astrophysics Research Institute, Liverpool John Moores University , IC2 Liverpool Science Park, 146 Brownlow Hill, Liverpool L3 5RF UK  \and
University of Turku, Dept. of Physics \& Astronomy, Vesilinnantie 5, 20014, Turku, FI \and
School of Physics and Centre for Space Research, University College Dublin, Belfield, Dublin 4, Ireland \and
INAF - Osservatorio Astronomico di Roma, Via di Frascati 33, I-00078 Monteporzio Catone, Italy \and
Astronomical Observatory Institute, Faculty of Physics and Astronomy, Adam Mickiewicz University, ul. Słoneczna 36, 60-286, Poznań, Poland \and
Lancaster University, Department of Physics, Lancaster University, Lancaster, LA1 4YB, UK \and
University of Pisa, Largo B. Pontecorvo 3, I-56127 Pisa (PI), Italy \and
INFN - Sezione di Pisa, Largo B. Pontecorvo 3, I-56127 Pisa (PI), Italy \and
INAF - Osservatorio Astronomico di Roma, Via Frascati 33, I-00078 Monte Porzio Catone (RM), Italy \and
INAF - Osservatorio di Astrofisica e Scienza dello Spazio, via Piero Gobetti 93/3, I-40129 Bologna, Italy \and
Instituto de Alta Investigación, Universidad de Tarapacá, Casilla 7D, Arica, Chile \and
Max Planck Institute for Gravitational Physics (Albert Einstein Instiute), Am Mühlenberg 1, Potsdam-Golm,14476, Germany \and
Dipartimento di Fisica e Astronomia “Ettore Majorana”, Università degli Studi di Catania, Via Santa Sofia 64, I-95123 Catania (CT), Italy \and
INAF - Osservatorio Astrofisico di Catania, Via Santa Sofia 78, I-95123 Catania (CT), Italy \and
Institute of Astronomy Space and Earth Science (IASES), P 177, CIT Road, Scheme 7m, Kolkata-700054, West Bengal, India \and
Inter-University Centre for Astronomy \& Astrophysics, Post Bag 4, Ganeshkhind, Pune 411 007, India \and
Université Paris-Saclay, Université Paris Cité, CEA, CNRS, AIM, 91191, Gif-sur-Yvette, France \and
Centre national d’études spatiales (CNES), Paris, France \and
INAF - Osservatorio Astronomico di Brera, via Brera 28, I-20121 Milan (MI), Italy \and
INFN - Sezione di Milano-Bicocca, piazza della Scienza 3, I-20126 Milan (MI), Italy \and
INAF - Istituto di Astrofisica Spaziale e Fisica Cosmica di Milano, Via A. Corti 12, 20133 Milano, Italy \and
Kavli Institute for Cosmology, University of Cambridge, Madingley Road, Cambridge CB3 0HA UK \and
Institute of Astronomy, University of Cambridge, Madingley Road, Cambridge CB3 0HA UK \and
Department of Particle Physics and Astrophysics, Weizmann Institute of Science, 234 Herzl St, 76100 Rehovot, Israel \and
LUX, Observatoire de Paris, Université PSL, CNRS, Sorbonne Université, 5 place Jules Janssen, Meudon 92190, France \and
Key Laboratory of Particle Astrophysics, Institute of High Energy Physics, Chinese Academy of Sciences, Beijing 100049, China \and
University of Chinese Academy of Sciences, No.1 Yanqihu East Rd, Huairou District, Beijing, PR China, 101408 \and
Astrophysics Research Centre, School of Mathematics and Physics, Queen's University Belfast, Belfast BT7 1NN, UK
}

\authorrunning{K.\ Ackley et al.}

   \date{Received September 30, 20XX}

 
  \abstract{
  The candidate gravitational wave (GW) event S250818k was one of only three non-retracted LIGO-Virgo-KAGRA public alerts issued during the fourth observing run of the network (O4) with a binary neutron star (BNS) merger classification probability exceeding one percent. This triggered a prompt search for a potential electromagnetic (EM) counterpart in the large localisation error region (949 deg$^2$ projected in the sky at 90\% credible level). The transient \thisSN, discovered by the Zwicky Transient Facility (ZTF) during the search, attracted a great deal of attention due to a potential spatial and temporal coincidence, and due to its initial fast decay and featureless spectrum. Here, we report on the follow up of this transient by the Electromagnetic counterparts of gravitational wave sources at the Very Large Telescope (ENGRAVE) Collaboration. We conducted an extensive multi-wavelength observational campaign, which led to the spectral classification of the transient as a type IIb supernova (SN), indicating that it is unrelated to the candidate GW event. In this article, we describe our observing strategies, data reduction, and interpretation. All of our results confirm and strengthen our classification of the source, and also show that shock cooling tails associated with type IIb SNe are one of the most prominent contaminants in kilonova searches. 
  }
  

   \maketitle
   \nolinenumbers

\section{Introduction}


The current international ground-based gravitational wave (GW) detector network comprises the two advanced Laser Interferometer Gravitational wave Observatories (aLIGO, \citealt{Aasi2015}), the advanced Virgo \citep{Acernese2015} and KAGRA \citep{Somiya2012}, operated by the LIGO-Virgo-KAGRA Collaboration (LVK hereafter). During the fourth observing run (O4) of the network, which concluded on 18 November 2025, the LVK issued public alerts\footnote{\url{https://emfollow.docs.ligo.org/userguide/index.html}} for GW candidates found during real-time processing of detector data through a number of search pipelines. The pipelines produced triggers with an associated false alarm rate (FAR): a compact binary coalescence (CBC) GW candidate was considered `significant' if the associated FAR was less than \revone{a threshold that depends on the specific search}, and `low-significance' if the FAR was higher, but still lower than two per day. \revone{For searches that targeted binary neutron star (BNS) mergers, the threshold was one event per seven months; for searches that targeted sub-solar mass (SSM) binary mergers, it was one per two years.} Over the 2.5 years of O4,  only two significant public alerts (excluding retracted ones) for GW candidates with a BNS merger classification probability larger than one percent were issued: S230529ay \citep{2023GCN33889} and S250206dm \citep{2025GCN39175}.

On 18 August 2025 at 01:20:06 UT, a low-significance public alert \revone{(with a FAR of about one every six months)} was issued for a CBC GW candidate, S250818k, found in a search on aLIGO and Virgo data. Two LVK Preliminary GCN notices\footnote{\url{https://gcn.gsfc.nasa.gov/notices_l/S250818k.lvc}} were sent out within 5 minutes. The online analysis attributed the event to a terrestrial source (i.e.\ noise) with 71\% probability, or to a BNS merger with 29\% probability. A neutron star - black hole (NSBH) or a binary black hole (BBH) binary merger origin both resulted as highly unlikely ($<1\%$ probability). 

For low-significance alerts in O4, no GCN circular was normally issued by the LVK Collaboration. However, after rapidly covering the GW localization region (skymap hereafter), the Zwicky Transient Facility (ZTF) announced the discovery of an optical transient plausibly associated with S250818k \citep{2025GCN41414}. This led the LVK to release a GCN Circular to confirm the properties of the GW candidate event \citep{2025GCN41437}. Assuming an astrophysical origin, the quoted probability that at least one of the compact objects was a neutron star (NS) with mass above one solar mass (\texttt{HasNS}) is $>99\%$, and the probability that matter remained outside the final compact object (\texttt{HasRemnant}) is $>99\%$. According to the circular, the source chirp mass\footnote{The chirp mass of a binary is defined as $m_\mathrm{c}=(m_1m_2)^{3/5}/(m_1+m_2)^{1/5}$, where $m_1$ and $m_2$ are the gravitational masses of the binary components.} falls with highest probability in the range $(0.1, 0.87)\,\mathrm{M_\odot}$, indicating that the lighter component would be of sub-solar mass. 

An additional GCN circular was sent on 20 August 2025 with an updated sky localization and other inference based on parameter estimation \citep{2025GCN41440}. The updated skymap has a $90\%$ credible region of $949$\,deg$^{2}$ (somewhat larger than the 786 deg$^2$ of the previous skymap), and the posterior mean and standard deviation of the luminosity distance\footnote{We note that the sky-position-conditional distance posterior in LVK skymaps is represented as a skewed Gaussian, $p(d)\propto d^2 \exp(-(d-\mu)^2/2\sigma^2)$, where $d$ is the distance and $\mu$ and $\sigma$ are sky-position-dependent location and scale parameters \citep{Singer2016_3D}. The average over the whole sky has a more general shape, but is also most often best described by the same skewed Gaussian.} are $237\pm 66$\,Mpc (a more detailed account of the S250818k localisation is reported in Appendix \ref{sec:localisation}). 

The ZTF follow-up observations began approximately 2.7\,hours after the GW trigger, using the Palomar 48-inch telescope (P48), and covered 25.2\% of the localisation probability \citep{2025GCN41414}. 
A total of 58 transient candidates were identified and vetted by the ZTF team. A single transient passed all vetting criteria: \thisSN\,(initially identified as AT\,2025ulz), located in what initially appeared to be a passive galaxy with only a photometric redshift of $z=0.091\pm0.016$ (corresponding to a luminosity distance $D_{\rm L}=435^{+82}_{-80}$\,Mpc, adopting the  cosmological parameters from \citealt{Planck2020}),  

From a Keck spectrum, \citet{2025GCN41436} determined the presumed host galaxy  redshift to be $z\simeq0.0848$ based on host emission lines ($D_\mathrm{L}\sim 399$ Mpc). This made it consistent within 2.5$\sigma$ of the distance estimate for S250818k at that sky position, which improved to 2$\sigma$ in the updated skymap from the offline parameter estimation \citep{2025GCN41440}. Further imaging by Gemini GMOS-N and Pan-STARRS confirmed the fading of the source \citep{2025GCN41452,2025GCN41453,2025GCN41454}, strengthening the case for a rapidly fading, reddening source spatially and temporally coincident with S250818k. Our ENGRAVE Collaboration observed the source with the Very Large Telescope (VLT; see section \ref{sec:obs}), confirming the host redshift of $z=0.0848$ (\citealt{2025GCN41476}; in Sect.\ \ref{sec:xshooter} we provide a refined estimate with error bars), but any assessment of the source at that stage was made challenging by the host galaxy contamination, which proved difficult to estimate in absence of deep pre-explosion templates. During these first 2-4 days, the source was observed intensively from the X-ray to the radio, but none of the GCN announcements reported definitive multi-wavelength detections or kilonova (KN) like signatures. 
Finally, a photometric rebrightening was initially detected by two separate instruments and announced on 23 and 24 August \citep{2025GCN41507,2025GCN41518} and confirmed with consistent daily Pan-STARRS coverage \citep{2025GCN41540}.  On the latter day, we also announced the results of our observations with the Multi-Unit Spectroscopic Explorer (MUSE) integral-field spectrograph \citep{2025GCN41532}, which allowed for a better handling of the host galaxy contamination (see Section \ref{sec:muse}). The spectrum at the source position showed a prominent broad feature that we identified as a P-Cygni profile of H$\alpha$, similar to that observed in young type II or IIb supernovae (see Section \ref{sec:spectral_classification}).

Possible transient radio emission at the position of \thisSN\,was reported based on MeerKAT observations \citep{2025GCN41500,2025GCN41594},  but later reprocessing of the datasets suggested instead a constant flux density, with no clear evidence for variability. This pointed to the more likely scenario where the radio emission is due to star formation in the host galaxy \citep[][see also sect.\ \ref{sec:radio_from_SFR}]{2025GCN41666}. Together with the lack of significant
X-ray emission \citep{2025GCN41453,2025GCN41528,OConnor2025}, the supernova (SN) nature of \thisSN\,emerged as the most likely explanation. After an intense phase of global observations and GCN reporting over the first 10 days following S250818k, preliminary results were reported in a number of preprints: \citet{Gillanders2025} reported the Pan-STARRS and Asteroid Terrestrial-impact Last Alert System (ATLAS; \citealt{Tonry2018,Smith2020}) coverage of the skymap and a well-sampled \thisSN\ lightcurve which clearly showed a behaviour similar to known Type IIb SNe; \cite{Franz2025}, \citet{Yang2025} and \citet{Hall2025a}
also favoured the SN interpretation over a KN-like signature, while \citet{Hall2025b} analysed the host galaxy spectra in comparison to other DESI galaxy spectra and potential hosts of the source of S250818k; \citet{OConnor2025} presented radio and X-ray limits, ruling out an afterglow interpretation of the re-brightening and excluding a GW170817-like afterglow within viewing angles of $\theta\lesssim12.5^{\circ}$ (see also our analysis in sect.\ \ref{sec:radio_limits_on_jet}). 
Finally, \citet{Kasliwal2025}, while also favouring the type IIb SN interpretation, drew attention to the low chirp mass of S250818k and qualitatively discussed their multi-wavelength dataset in the context of a `superkilonova' scenario, where sub-solar mass NSs form and merge via disk fragmentation within a core-collapse SN  \citep{Piro2007,Metzger2024,Chen2025}. 

In this work, we present \thisSN\,observations primarily obtained through the ENGRAVE VLT Large Program and supplemented with other associated datasets. Our data ultimately favour a type IIb SN interpretation for the source, in chance temporal coincidence with the GW candidate event. 

Throughout the paper we adopt the best-fit cosmological parameters from \citet{Planck2020}.

\section{Observations and data reduction}
\label{sec:obs}

\subsection{Optical and NIR imaging}

We imaged \thisSN\,with the FORS2 and HAWK-I instruments at the ESO Very Large Telescope (VLT) at Cerro Paranal, Chile, and with the Alhambra Faint Object Spectrograph and Camera (ALFOSC) on the Nordic Optical Telescope (NOT) in La Palma, Spain. Additionally, we made use of publicly available \textit{Hubble Space Telescope} (HST) observations of the source. We describe below our observing campaign and data reduction.  The results of these observations are summarized in Table \ref{tab:photometry}.

\subsubsection{VLT/FORS2}

We observed \thisSN\,on 21 and 22 August 2025, approximately 3 and 4 days after the GW event, using VLT/FORS2. Observations were obtained in the $R$, $I$, and $z$ bands, with exposure times of 180, 180 and 120 seconds, respectively. 
All imaging data were reduced with \texttt{ESOReflex} \citep{Freudling2013}, employing bias and flat-field frames acquired on the same night and with the same instrumental configuration.

\begin{figure}
\centering
\includegraphics[width=\columnwidth]{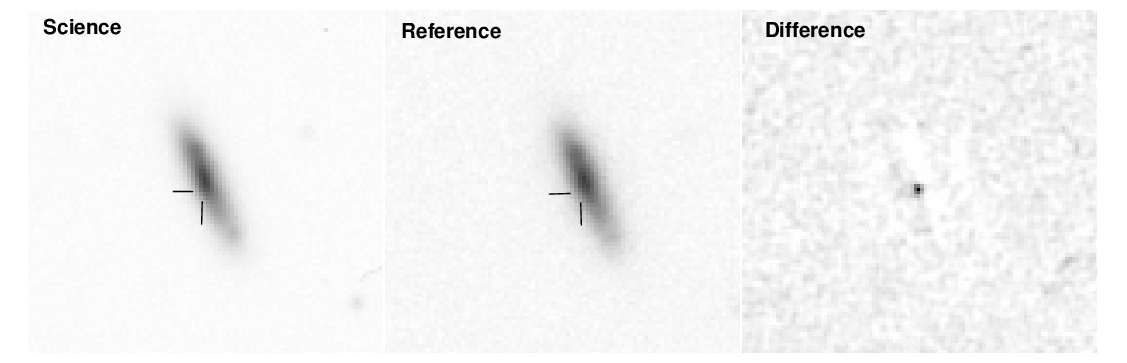}
\caption{VLT/FORS2 difference imaging. The left-hand panel shows a $25''\times 25''$ cutout of our FORS2 $I$-band image of \thisSN\,and its host galaxy, taken on 21 August. The middle panel shows the deep template constructed from Pan-STARRS data (see text). The residual from difference imaging is shown in the right-hand panel.}
\label{fig:FORS2_diff_imaging}
\end{figure}

Astrometric and photometric calibration, as well as template subtraction and PSF-forced photometry, were performed using \texttt{EC-SuperNOva PhotometrY} (ECSNOoPY\footnote{ECSNOoPY is a Python package developed by E.~Cappellaro for transient photometry, PSF fitting and reference image subtraction. It extensively uses the \texttt{astropy} ecosystem \citep{astropy2013}, particularly \texttt{photutils} \citep{Bradley2024}, and employs \texttt{HOTPANTS} \citep{Becker2015} for PSF-matched image differencing, SExtractor \citep{Bertin1996} for source extraction, SCAMP \citep{2006ASPC..351..112B} for astrometric calibration and SWARP for image resampling and coaddition. 
A detailed description of the package is available at \url{http://sngroup.oapd.inaf.it/ecsnoopy.html}.}).
For template subtraction, we used the Pan-STARRS deep reference images presented by \cite{Gillanders2025}.\footnote{The templates are publicly available at \url{https://ora.ox.ac.uk/objects/uuid:624c1bc5-b841-4da0-9c56-c8683454da7f}.}
These stacked reference images have total exposure times of 960~s, 4445~s, and 6720~s in the $r$, $i$ and $z$ bands, respectively. 
Using these deep templates enabled robust host-galaxy subtraction and accurate photometric measurements of the residual transient emission in the difference images (see Fig.\ \ref{fig:FORS2_diff_imaging}).

PSF fitting on the difference images was performed with ECSNOoPY, using its \texttt{photutils}-based implementation. 
The PSF model was constructed from isolated field stars automatically selected in each frame, and the fitting was iteratively refined at the position of \thisSN. 
The residuals were visually inspected to validate the fits. A faint but clear residual was detected at the position of \thisSN\,in both epochs. The corresponding photometric measurements are reported in Table \ref{tab:photometry}.

Photometric uncertainties were estimated through artificial star experiments: a synthetic star with brightness similar to that of the source was injected into the PSF-subtracted image at positions near, but not coincident with, the source. 
The simulated frames were processed through the same PSF fitting pipeline, and the dispersion of the recovered magnitudes from multiple realizations was adopted as the instrumental magnitude error. 
Finally, this value was combined in quadrature with the uncertainty from PSF fitting. 

\subsubsection{VLT/HAWK-I}
\label{sec:hawki}

\begin{figure}
\centering
\includegraphics[scale=0.45]{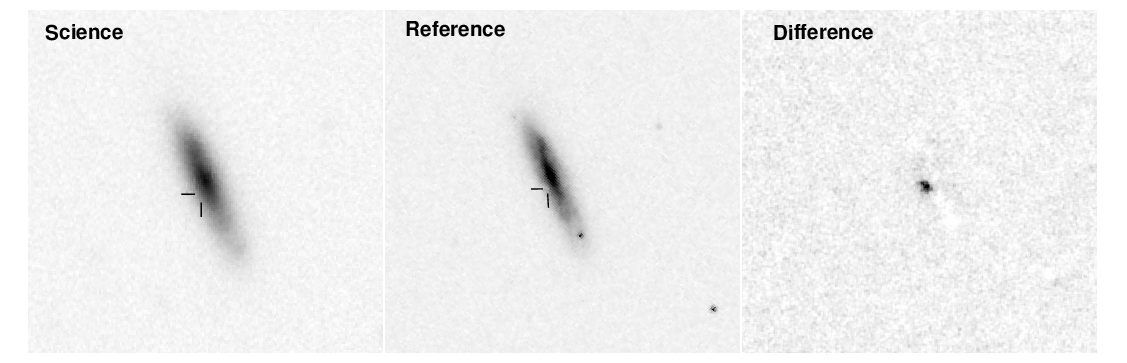}
\caption{VLT/HAWK-I difference imaging. Similar to Figure \ref{fig:FORS2_diff_imaging}, but showing a $24''\times 24''$ cutout of our HAWK-I $K$-band image taken on 22 August (left-hand panel), the template based on the F160W HST image (middle panel, see text) and the resulting difference imaging residual (right-hand panel).}
\label{fig:HAWKI_diff_imaging}
\end{figure}

We also observed \thisSN\,on the 20 and 22 August 2025, approximately 2 and 4 days after the GW event, using VLT/HAWK-I. 
Observations were conducted in the $K$ band, with a total exposure of 690 and 2160 s in the first and second epoch, respectively. 
All imaging data were reduced using \texttt{ESOReflex}. 
As no pre-explosion $K$-band images were available for host-galaxy subtraction, we used the available near-infrared observations of the transient to aid this process: 
the field of \thisSN\ was observed at multiple epochs with the HST using the wide-field camera (WFC3), as part of the GO-17450 and GO-17805 programs (PI: Troja). 
The transient and its host galaxy were imaged in the F336W, F606W, F110W, and F160W bands between 22 and 28 August 2025.
All HST images were retrieved from the Mikulski Archive for Space Telescopes (MAST) and processed with \texttt{hst123} \citep{Kilpatrick2022}, including frame alignment with \texttt{TweakReg}, image combination with \texttt{AstroDrizzle} \citep{STSCI2012}, and photometry on calibrated frames using \texttt{DOLPHOT} \citep{Dolphin2016}.

We used the F160W image obtained on 28 August 2025 to model and subtract the source at the source position. 
The host offset is approximately 0.9~arcsec, allowing an effective removal of the SN contribution from the host galaxy. 
We used ECSNOoPY to perform PSF fitting on the transient position in the F160W image, and the resulting residual was adopted as the reference frame for host-galaxy subtraction in the HAWK-I $K$-band data. 
Despite the mismatch in bandpass, the subtraction successfully revealed the transient in the HAWK-I image (see Fig.\ \ref{fig:HAWKI_diff_imaging}), and the PSF fitting was used to measure its magnitude in the difference images.

The photometric calibration was based on the 2MASS catalog \citep{Skrutskie2006}, without applying any color-term corrections. 
Uncertainties arising from local background variations in the difference images (obtained by subtracting the F160W frame from the $K$-band data) were propagated into the total photometric error budget. 
The final photometric measurements are presented in Table \ref{tab:photometry}.

\subsubsection{NOT/ALFOSC}

We obtained 6 epochs of $r$-band photometry using the ALFOSC on the NOT, starting from 20 August 2025 \citep[][see Table \ref{tab:photometry} for the full list of observation times]{2025GCN41492}. The images were reduced, aligned and stacked using standard procedures. We performed image subtraction using \texttt{HOTPANTS} with an $r$-band template from the Legacy Survey DR10 \citep{Dey2019}. Even if the Legacy image is the deepest available for the field in the $r$-band, the NOT images in many epochs had superior seeing. This required us to convolve the NOT images with a Gaussian filter to match the Legacy full width at half maximum (FWHM), rather than the other way around. We visually inspected the subtraction residuals for correlated noise and other artifacts, which confirmed clean subtractions. In a few cases, the host galaxy subtraction was not perfect, but the residuals left were outside the radius where we performed photometry, and we found they were not correlated with the transient brightness. In particular, the cleanest subtractions were obtained for the first epochs, when the transient was faintest. 
We performed PSF photometry using \texttt{photutils}, calibrated against PS1 photometry of field stars, using a small radius (1 FWHM). 
Our NOT photometry is included in Table \ref{tab:photometry}. 

\subsubsection{HST}
\label{sect:HST}

To further aid our interpretation of the source, we used DOLPHOT \citep{Dolphin2016} to perform photometry on the HST images described in section \ref{sec:hawki}, obtaining magnitudes of F336W = 23.49$\pm$0.21~mag (22 August) and F606W=21.91$\pm$0.02~mag (26 August). These measurements are consistent with those reported by \citet{2025GCN41506} and \citet{Franz2025} on the same data to within 0.1 mag, with the difference likely due to a different choice of background parameters and aperture radii.

\begin{figure}
    \centering
    \includegraphics[width=\columnwidth]{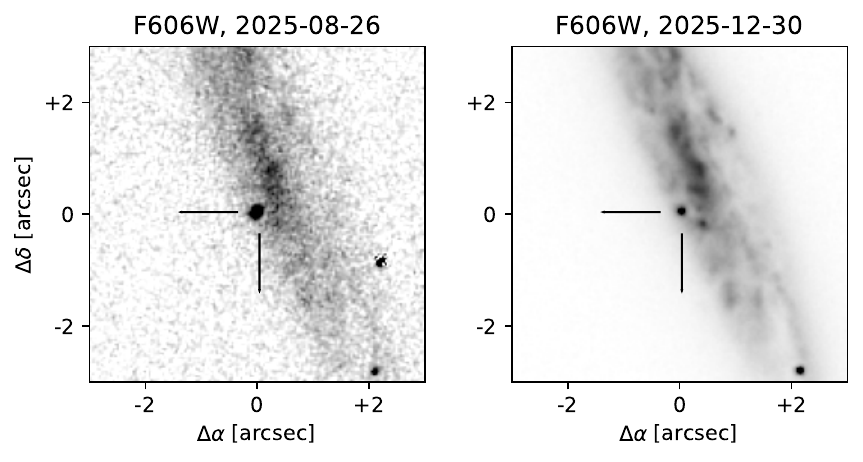}
    \caption{HST/WFC3 F606W imaging. Both panels show images of the same sky position and with the same scale, with the \thisSN\,position indicated with tick marks. The image in the left-hand panel was taken on 26 August, while that in the right-hand panel is from 30 December. The disparity in quality between the two images is due to the different exposure times.}  
    \label{fig:hst}
\end{figure}

We also examined a HST/WFC3/UVIS image in F606W taken on 2025 Dec 30 (PI: O'Connor), approximately 4.5 months after the discovery of \thisSN. A source is still clearly visible at the position of the source (see Fig. \ref{fig:hst}), with a magnitude F606W=23.82$\pm$0.01. This corresponds to an absolute magnitude of -14.1, in good agreement with that of the Type IIb SN 2011dh at a similar phase \citep{Ergon2014}. 

\subsection{Spectroscopy}

Spectroscopic follow up of \thisSN\ was obtained with VLT/FORS2, X-shooter and MUSE. A log of all spectroscopic observations is reported in Tab. \ref{tab:spectroscopy}, while details of the reduction and calibration of these data are given below.

\begin{table}[t]
\centering
\caption{Log of spectroscopic observations of \thisSN.}
\label{tab:spectroscopy}
\begin{tabular}{cccccc}
\hline\hline
Night  & Mid-time MJD & Instrument & Grism  \\
\hline
Aug 20 & 60908.02 & X-shooter  & UVB/VIS/NIR \\
Aug 20 & 60908.03 & FORS2      & 300I        \\
Aug 24 & 60911.02 & MUSE       & WFM         \\
Aug 28 & 60915.00 & MUSE       & WFM         \\
\hline
\end{tabular}
\end{table}

\subsubsection{VLT/X-Shooter}
\label{sec:xshooter}

\begin{figure*}
    \centering
    \includegraphics[width=1\linewidth]{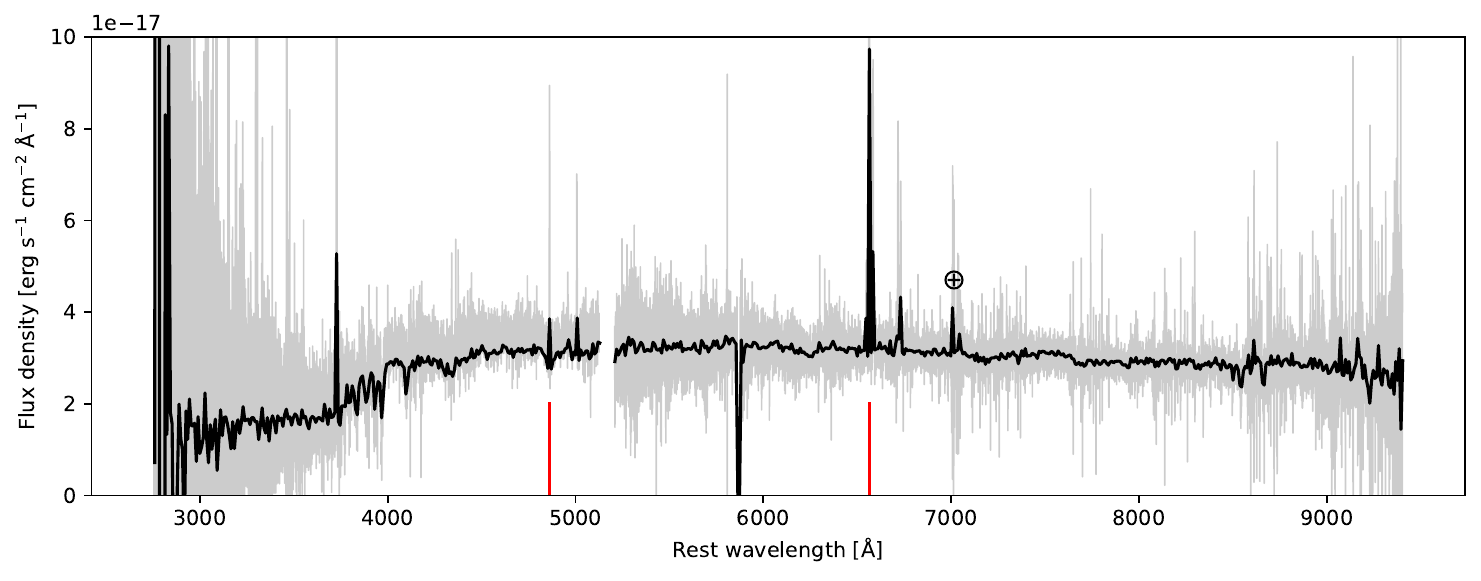}
    \includegraphics[width=1\linewidth]{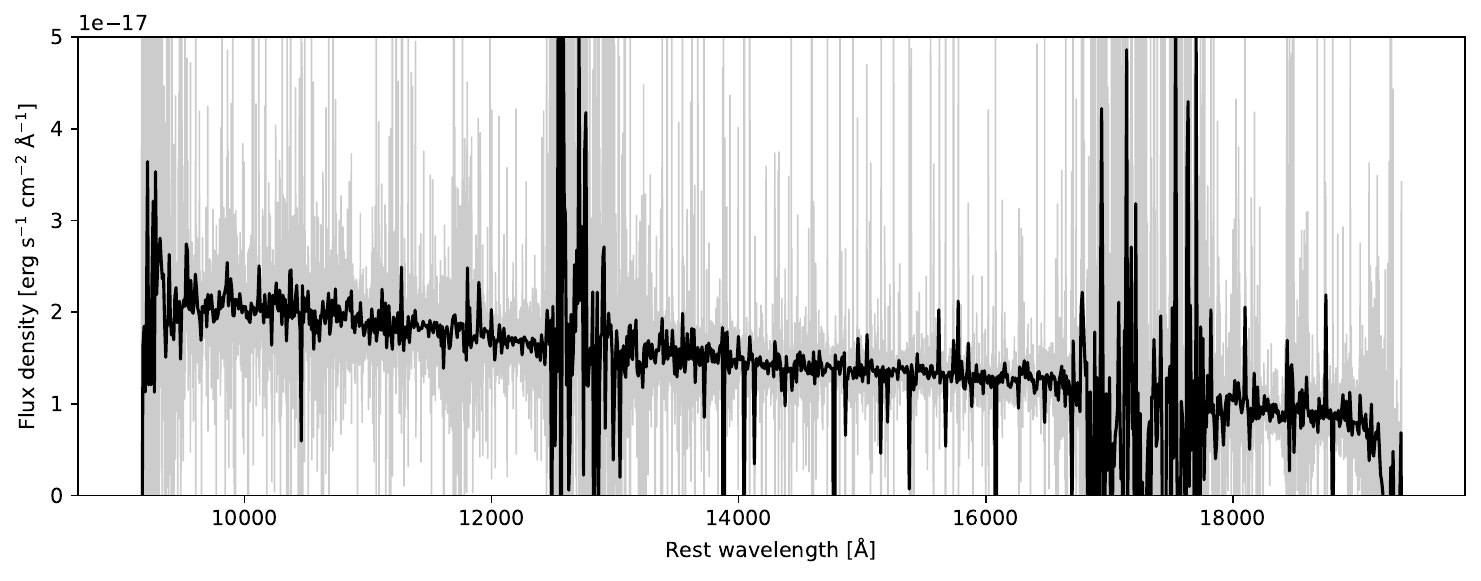}
    \caption{The VLT/X-shooter spectrum of \thisSN\,taken on the night of Aug 20, plotted in the rest frame. The upper panel shows the UVB and VIS arms, while the lower panel shows the NIR arm. In both cases, the reduced spectrum at the native pixel sampling is shown in light grey, the black line is after the spectrum is averaged and rebinned to 10 $\AA$ sampling. Red lines mark the position of H$\alpha$ and H$\beta$, while the residual left after correction of the telluric A band is marked with an $\oplus$ symbol.
    }
    \label{fig:xsh}
\end{figure*}

X-Shooter observations of \thisSN\,were obtained on the night of 20 August. 6$\times$580~s and 6$\times$600~s exposures were taken in the UVB and VIS arms, respectively, while 12$\times$300~s exposures using on-slit nodding were used in the NIR, giving an approximate total exposure time of 1 hr. A one arcsec slit was used for the UVB arm, and 0.9 arcsec slits in the VIS and NIR arms, which were oriented at a position angle of -160.7 deg (i.e. parallactic angle). The X-Shooter data were reduced following standard techniques, using the ESO X-Shooter pipeline running under \texttt{esorex} \citep{ESO2015}. Processed calibrations appropriate to the science data were downloaded from the ESO archive using the \texttt{calselector} service and used during the reduction. To correct for telluric absorption, we used the \texttt{molecfit} package \citep{Smette2015} to model the atmospheric transmission. The reduced spectrum is shown in Fig.\ \ref{fig:xsh}. Using H$\alpha$, we measured the redshift of the \thisSN\ host galaxy to be $z=0.0849\pm 0.0004$.

\subsubsection{VLT/FORS2}


Two exposures of 1500~s each were also obtained, at a relatively high airmass ($\sim$ 2.0), with FORS2 using the 300I grism, which covers the wavelength range from 6,000 $\AA$ to 10,000 $\AA$. 2D spectra were  obtained using standard {\tt esorex} recipes, while 1D single spectra were  derived using {\tt IRAF} after mitigating contamination from bright sky background features. We finally stacked both exposures, producing the final spectrum visible in Fig.\ \ref{fig:FORS_spec}. 

\subsubsection{VLT/MUSE}
\label{sec:muse}

\begin{figure*}
    \centering
    \includegraphics[width=1\linewidth]{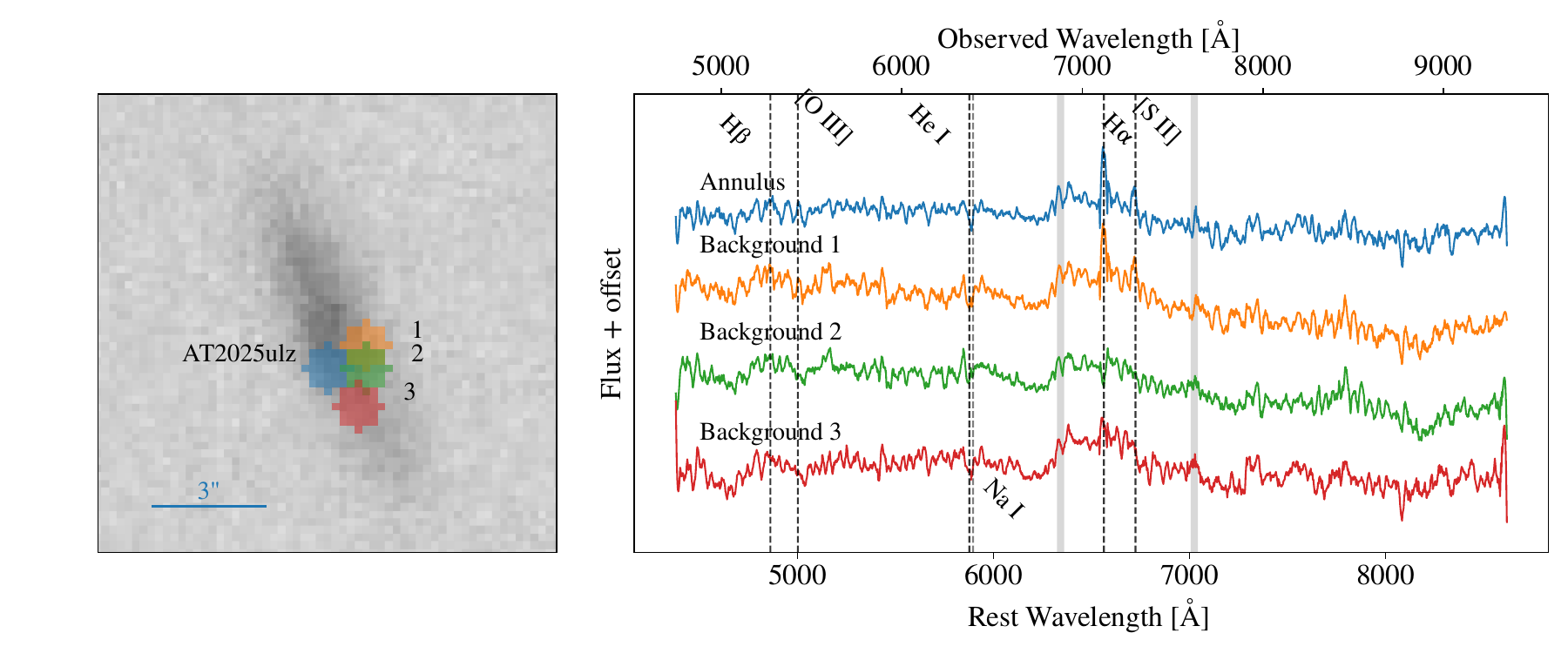}
    \caption{\textbf{Left}: The host of \thisSN\,observed with MUSE on the night of Aug 24. The blue aperture marks the spaxels for extracting the transient spectrum at the location of \thisSN. The other regions were used to estimate and remove the host galaxy background.  
    \textbf{Right}: Spectra of \thisSN\,for different extractions and host galaxy removal, including a circular annulus and the background apertures shown on the left panel (colour-coded similarly). All extraction methods result in broadly similar spectra with the same features, including a prominent broad H$\alpha$ P-cygni profile. Gray shaded bands indicate regions of telluric absorption.}
    \label{fig:musespec}
\end{figure*}

We  also used the Multi-Unit Spectroscopic Explorer \citep[MUSE;][]{Bacon2010} to observe \thisSN\,on the nights of 24 and 28 August. We reduced the data using a custom version of the ESO pipeline and standard {\tt esorex} recipes, and used the {\tt IFUanal} package \citep{Lyman2018} to analyse and post-process the data.
The MUSE cube was astrometrically aligned with images containing \thisSN\,by using stars that were common to the MUSE white image and reference images. 
We extracted the spectrum in a circular aperture centered at the astrometrically transformed location of \thisSN. The aperture diameter was chosen approximately equal to the FWHM as measured on point sources in the field ($1.2''=6\,\mathrm{pixels}$ for 24 August; $1.8''=10\,\mathrm{pixels}$ for 28 August). We experimented with different ways of removing the galaxy background: a circular annulus centred on \thisSN\ with inner radius of $1.2''$ and outer radius of $1.8''$; three different apertures at nearby galaxy locations (marked with different colours in Fig.\ \ref{fig:musespec}), after correcting for the rotation of the galaxy by applying velocity offsets. Without applying these offsets, host galaxy lines such as H$\alpha$ would subtract badly, contributing noise and residuals to the final spectra. 
Figure~\ref{fig:musespec} shows the resulting spectra for \thisSN\,for all methods, demonstrating that all choices produce similar results. We conclude that the transient spectrum is not dependent on the details of the background removal.

\subsection{Radio observations}

We observed the position of \thisSN\ at radio wavelengths with the Karl G. Jansky Very Large Array (VLA), the MeerKAT, the upgraded Giant Metrewave Radio Telescope (uGMRT) and the enhanced Multi Element Remotely Lindked Interferometer Network (e-MERLIN). A summary of our observation epochs, setup, and results is given in Table \ref{tab:radio}. Below we detail the observing strategies and data reduction.

\subsubsection{VLA}
\label{subsec:vla}
We acquired a broadband observation of \thisSN\,with the VLA on 29 August 2025 at 01:49:37.4 UT ($\delta t = 10.9$\,days) as part of our program 22A-414 (PI: Schulze). The observation was obtained in the B configuration, in a standard phase referencing mode, for a total on-source time of $\sim$50 minutes at the mean frequency of 3\,GHz (S-band), and for 18-20 minutes at the mean frequencies of 6\,GHz (C-band), 10\,GHz (X-band) and 15\,GHz (Ku-band). We used 8-bit samplers for S-band and 3-bit samplers for C, X and Ku-bands; 3C\,286 = J0137$+$3309 was selected as the bandpass and flux density calibrator, and J1602$+$3326 as the complex gain calibrator. We reduced the data using the VLA calibration pipeline packaged with \texttt{CASA} v.6.6.1.17 \citep{casa}. After manually inspecting the data, we further flagged antennae with bad solutions as well as additional weak radio-frequency interference and then re-ran the pipeline. We imaged each band individually using \texttt{wsclean} \citep{wsclean1,wsclean2} with Briggs weighting of 0.0. We further performed phase-only self-calibration on S- and C-band data using a sky model dominated by the nearby bright source ($\sim 2.5\arcmin$ away, visible in Fig.\ \ref{fig:VLA_Sband_image_wide}) to suppress its residual sidelobe artifacts at the location of \thisSN. We found no evidence of a significant detection at the location of the source (i.e.\ within one synthesized beam) in any of the images, down to 3$\sigma$ upper-limits of $\sim$ 21, 24, 27, and 21\,\ujy\ in S, C, X, and Ku-band, respectively (also reported in Table \ref{tab:radio}), consistent with the limits reported by \citealt{Franz2025} and \citealt{OConnor2025}. 

\begin{figure}
\centering
\includegraphics[width=\columnwidth]{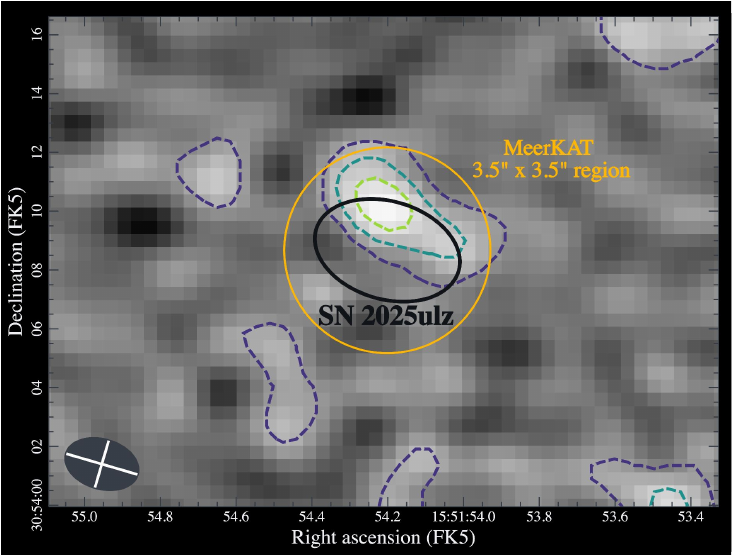}
\caption{Cleaned VLA S-band image using Briggs weighting of 0.5. Self-calibration was used to reduce contamination from the bright source visible in the lower-right corner of Fig.\ \ref{fig:VLA_Sband_image_wide}. At the location of \thisSN, no significant emission is present within one synthesized beam (black ellipse). The 1, 2 and 3$\sigma$ contours are respectively shown as purple, blue and green dashed lines, respectively. The emission enclosed within the $2\sigma$ (blue-dashed) contour has an integrated flux density of 37 $\pm$ 8 uJy. For comparison, MeerKAT synthesized beam for the observation of 28 August is shown as a yellow circle, and the VLA synthesized beam is displayed in the bottom-left corner of the inset.}
\vspace{-10pt}
\label{fig:VLA_Sband_image}
\end{figure}

Statistically significant radio emission from the host galaxy, localized in a region slightly north of \thisSN, is evident in the S-band image. To better recover this diffuse emission and enhance surface brightness sensitivity, we re-imaged each band using a Briggs weighting of 0.5 (the resulting image is shown in Figure \ref{fig:VLA_Sband_image}). Even with this weighting, no radio emission above $2\sigma$ is detected from the galaxy in C, X, and Ku bands. In the new S-band image, we used the \texttt{Contours} generator in \texttt{CARTA} to visualize the emission morphology, and defined an elliptical region enclosing all emission within the $2\sigma$ contour level (shown with dark green dashed lines in Fig.~\ref{fig:VLA_Sband_image}). This region, centred at 15:51:54.19 +30:54:10.02 (at a projected distance of $\sim 2.5$\,kpc from \thisSN), has dimensions of approximately $2.7\arcsec \times 1.1\arcsec$ in our image. We used the \texttt{imstat} task in \textsc{CASA} to measure an integrated flux density of $37 \pm 8$\,uJy within this region, where the uncertainty includes statistical and 5\% systematic contributions added in quadrature. For a quick comparison with MeerKAT results in Section \ref{subsec:meerkat}, we also measured the integrated flux density within a $3.5\arcsec \times 3.5\arcsec$ region (which corresponds to the MeerKAT synthesized beam for the observation of 28 August, see Table~\ref{tab:radio}), obtaining 60 $\pm$ 20\,uJy from the VLA data, consistent with MeerKAT measurements.  

\subsubsection{MeerKAT}
\label{subsec:meerkat}


We acquired observations of \thisSN\,with MeerKAT at three epochs from 21 August 2025 to 13 September 2025 (3 to 26 days after the GW) under program ID SCI-20241101-GB-01 (PI: Bruni) at 3 GHz (S4 band). Observations were also performed at 1.28 GHz (L-band) in the last epoch. The first epoch in the S-band had a total duration of $\sim$2 hours, while the second and third lasted $\sim$4 hours. The L-band observation lasted $\sim$2.5 hours. We used J1939-6342 as flux and bandpass calibrator, while J1609+2641 as phase reference at both frequencies. Data were processed with the {\tt{oxkat}} pipeline \citep{2020ascl.soft09003H}. We obtained a detection in all epochs and frequencies. The emission position is consistent with the host galaxy of \thisSN, and is marginally resolved only at S-band. However, the angular resolution was not sufficient to resolve a possible transient radio counterpart. Since S-band flux densities for the different epochs are consistent within errors, we also produced a deeper image combining the calibrated visibilities for the second and third epochs, which had the best UV coverage. In this way, we improved the RMS to about 3 $\mu$Jy/beam, and derived a flux density of 71$\pm$8 $\mu$Jy. The flux density in the L-band is $116\pm 14\,\mathrm{\mu Jy}$.

\subsubsection{uGMRT}
Single-epoch observations with the uGMRT were carried out 8 days post GW event, at 1.4 GHz (L-band), under project 48\_097 (PI: Bruni). The total duration of the observation was $\sim$2 hours. We used 3C\,286 as flux and bandpass calibrator. Data were reduced with the {\tt{SPAM}} pipeline \citep{Intema2014}. Image was restored with a Briggs 1.0 weighting scheme, allowing us to reach an RMS of 23 $\mu$Jy/beam. No detection of the transient or the host was achieved, resulting in a 5-$\sigma$ upper limit of 115 $\mu$Jy.

\subsubsection{e-MERLIN}
Observations were also performed with the e-MERLIN under program CY20210 (PI: Bruni). They were carried out at C-band (5.1 GHz) for one epoch on 19 September 2025 (31 days post-GW), including the following antennas: Lo, Mk2, Pi, Da, Kn, De, Cm. The phase calibrator was 1605+3001, while 1331+3030 was adopted for amplitude calibration. The total duration of the observing run was $\sim11$ hours. Data were processed with the e-MERLIN pipeline \citep{Moldon2021}, and imaging was performed with CASA \citep{casa} at the central frequency of 5.1 GHz, adopting natural weighting. The RMS was 26 $\mu$Jy/beam. The source was not detected, and the corresponding 5-$\sigma$ flux density upper limit is 130 $\mu$Jy.

\section{Interpretation}
\label{sec:interpretation}

\subsection{Light curve and colour evolution}
\label{sec:lc_and_colour}

\begin{figure*}
    \centering
    \includegraphics[width=17cm]{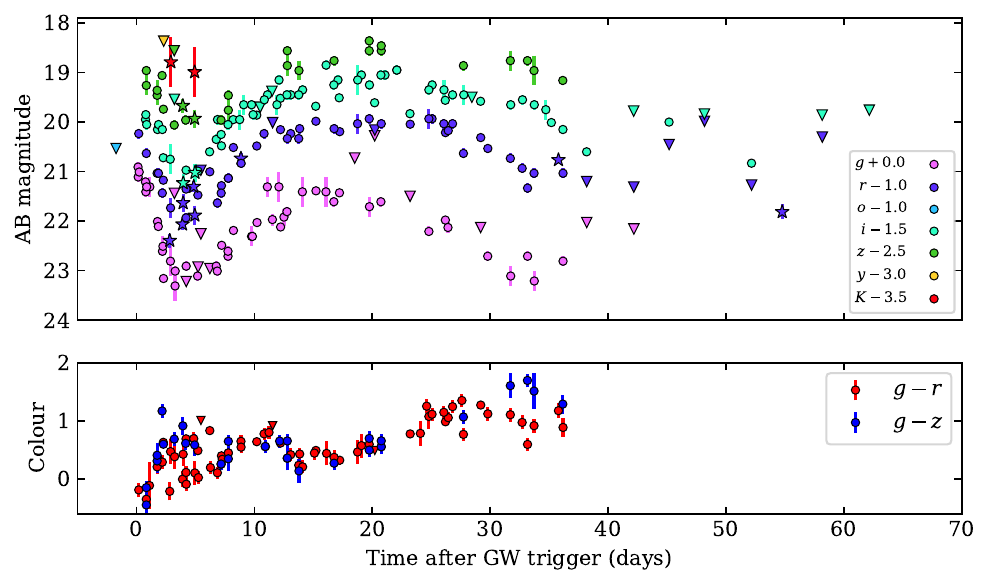}
    \caption{Multi-band light curves and colour evolution of \thisSN. \textit{Top panel:} estimated AB magnitudes of \thisSN\ in selected bands (color-coded and offset, for presentation purposes, as indicated in the legend -- $o$ is the ATLAS `orange' band \citep{Tonry2018}, where the most constraining pre-explosion limit is available) are shown by star symbols (data from this work) and circles (data compiled from table 2 of \citealt{Gillanders2025} and table 2 of \citealt{Kasliwal2025}), with error bars showing the one-sigma errors. Triangles represent three-sigma limiting magnitudes from observations that did not yield a detection. The data are corrected for dust extinction in the Milky Way, assuming $E(B-V)=0.028$ \citep{Schlegel1998,Schlafly2011}. \textit{Bottom panel:} $g-r$ (red) and $g-z$ (blue) colour evolution of \thisSN\ constructed subtracting the $r$ and $z$ band magnitudes from a piecewise-polynomial fit of the $g$ band data (see text). Error bars represent 68\% credible ranges, while triangles are 95\% credible upper limits.} 
    \label{fig:compiled_light_curve}
\end{figure*}

The top panel of Fig.\ \ref{fig:compiled_light_curve} shows the light curves of \thisSN\ in multiple bands, constructed by combining our data with those from \citet[][their table 2]{Gillanders2025} and \citet[][their table 2]{Kasliwal2025}. To visualise the colour evolution of the source, we divided the $g$-band light curve in three time segments (0 to 3.5 days; 3.5 to 17 days; 17 to 35 days) and fitted a third-degree polynomial to each segment, requiring continuity at the matching ends. We then subtracted the $r$ and $z$ band magnitude measurements from the $g$-band model evaluated at the corresponding time to construct the $g-r$ and $g-z$ colour estimates shown in the bottom panel of Fig.\ \ref{fig:compiled_light_curve}. We note that the colour evolution constructed in this way looks somewhat noisy: this is likely the consequence of the presence of data obtained with different host galaxy subtraction approaches in our diverse dataset.

The early evolution of \thisSN\,was very rapid: the initial detections by ZTF in the \textit{g} and \textit{r} bands, taken 0.13 and 0.19\,d after the GW trigger respectively, were the brightest in the first few days, at $g\sim 21$ and $r\sim 21.3$ magnitudes, respectively \citep{Kasliwal2025}; subsequent photometry in $g$, $r$, $i$ and $z$ bands all showed rapid fading in the first 3 days. The colour evolution in the first 2 days showed rapid reddening, with both $g-r$ and $g-z$ going from around $-0.5$ at 1\,d to $g-r \approx 0.2$ and $g-z \approx 0.5$ at 2\,d, respectively. From 4\,d onwards, a rebrightening was observed in all optical bands until $\sim 20$\,d, with the colour evolution turning slightly towards bluer colours in the 10 - 20 day range, before continuing to redden more slowly until the last detections around $\sim$37\,d, as shown in the bottom panel of Fig. \ref{fig:compiled_light_curve}. 

The initial rapid fading and colour evolution towards the red of \thisSN\,are reminiscent of AT\,2017gfo \citep{Gillanders2025}, though some details are critically different \citep[e.g.][]{Tanvir2017}. Notably, the $g$, $r$ and $i$-bands show a much shallower decline in \thisSN\,(see e.g.\ figure 3 of \citealt{Hall2025b}), while only the $z$-band has a similar slope. In other words, though the initial colour evolution revealed significant reddening, it was not as extreme as that of AT\,2017gfo, particularly past the first 2 days. However, theoretical modelling of KN emission \citep[e.g.][]{Kawaguchi2020} suggests a relatively diverse range of observable evolutions, depending on merger scenario, viewing angle, ejecta mass, composition and the nature of the remnant. Given that the only confirmed KN with data at these early times is AT\,2017gfo, such differences in the initially qualitatively similar evolution do not immediately invalidate the possibility that \thisSN\,was a KN. 

Conversely, the subsequent rebrightening across all bands is not predicted by any current KN model, particularly on a timescale of $\sim$30 days \citep[e.g.][]{Pognan2026}. Both AT\,2017gfo and the KN AT\,2023vfi \citep{Gillanders2023, Gillanders2025_AT2023vfi, Levan2024, Yang2024} show rapid fading on these timescales, and consistently redder colours in time. Considering \thisSN\,as an isolated event with no other emitting components, the optical light curves are therefore not compatible with KN models (see section \ref{sec:lc_modeling}). The presence of a gamma-ray burst (GRB) afterglow may be invoked in order to explain the optical evolution past 4 days (e.g.\ figure 5 of \citealt{Hall2025b}). However, such an afterglow component is ruled out by non-detections in radio and X-ray \citep{OConnor2025}.

The initial fast fading in all bands, followed by a rebrightening to a second peak at $\sim$15 - 20 days after the first detection, assumed to be at or close to the moment of explosion, is similar to previously observed type IIb SNe. Comparison of broadband light curve evolution reveals that \thisSN\,evolved in a qualitatively similar fashion to several other IIb SNe: 1993J, 2008ax, 2011dh, and 2016gkg \citep{Gillanders2025}. While some details differ between these objects, such as first and second peak absolute magnitudes, and exact timescales on which they evolve, this is to be expected. The first light curve peak and rapid decline is expected to arise from shock breakout and cooling, while the second peak arises from $^{56}$Ni decay heating of the ejecta, the evolution of which is significantly different from the expected r-process heating of KN ejecta.

\subsection{Spectra}
\label{sec:spectral_classification}

\subsubsection{Transient classification}
While the first spectra obtained on the night of Aug 20 (2 days after the GW trigger) are mostly featureless (Figs.~\ref{fig:xsh} and \ref{fig:FORS_spec}), the spectra obtained with MUSE on the nights of Aug 24 and 28 (6 and 10 days after the GW trigger) present sufficient structure to allow a spectroscopic classification. \thisSN\,shows a broad H$\alpha$ emission feature, accompanied by clear P-Cygni absorption, as well as H$\beta$ at the same velocity. These are features characteristic of Type~II SNe, including Type IIb at early phases.
We have cross-correlated the \thisSN\,spectra with those of other transients using the SuperNova IDentification - Spectral Analysis and Guided Exploration (SNID-SAGE; \citealt{2026arXiv260328741S}; \citealt{SNID}). We found good agreement with early spectra of SNe IIb, including those of SN~2016gkg and SN~2011dh, which have shown prominent shock cooling phases before their main radioactive peak \citep{Arcavi2011,Bersten2012,Ergon2014,Arcavi2017,Tartaglia2017}. 

\begin{figure}
    \centering
    \includegraphics[width=1\linewidth]{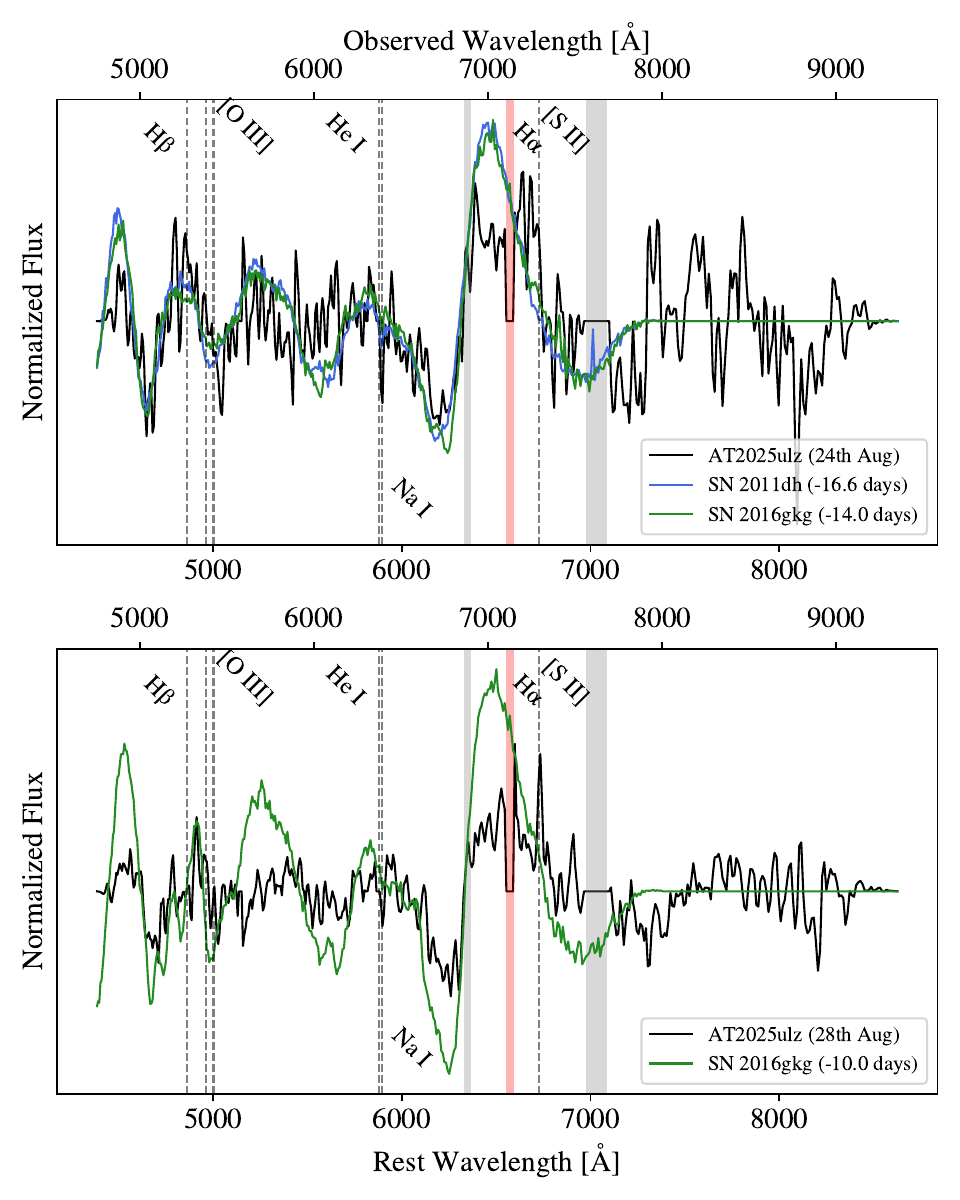}
    \caption{Comparisons of the MUSE spectra obtained on Aug 24 (top) and Aug 28 (bottom) with early spectra of type IIb SNe, produced by SNID-SAGE. The shaded regions mark host galaxy lines (red) and telluric features (gray) that have been masked out. Dashed lines mark the locations of typical SN features. The overall match is very good and the classification of \thisSN\,as a Type II SN, with IIb as the favored subtype, is unambiguous.}
    \label{fig:specsnidsage}
\end{figure}

For the spectrum of Aug 24, SNID-SAGE
provides a Type~II classification with high confidence, and a preferred subtype of IIb. Eight out of twelve top-matches are SNe IIb and their phases cluster between $-17$ and $-14$ days, with one outlier at $+5$ days. The confidence provided by SNID-SAGE for a Type II classification versus the next-best alternative (a tidal disruption event, TDE; \citealt{Gezari2021}) is $+99\%$. 
For the spectrum of Aug 28, SNID-SAGE formally provides a medium confidence for a Type II classification, with only $+42\%$ relative to the next-best alternative (again a TDE). We note, of course, that a TDE classification is inconsistent with every other observable of this transient and we do not consider it further. In fact, this just highlights that there is no other viable SN classification beyond a Type II for \thisSN. This time, 18 of the 31 top matches are of Type IIb, while the remaining 13 are of Type IIP or IIL. The phases for SN IIb top-matches are distributed within $-8.0 \pm 4.7$ days. This is perfectly in line with the date of observation, which is $-8$ days before the main peak of \thisSN.
Figure \ref{fig:specsnidsage} shows the spectral matches of \thisSN\,with those of well-studied SNe IIb at early phases. The spectroscopic classification of \thisSN\,is therefore secure and unambiguous.

\subsubsection{Host galaxy metallicity and star formation rate}
\label{sec:MUSE_Z_and_SFR}

From the integral-field MUSE spectra, we constructed two oxygen abundance maps of the galaxy based on the empirical calibration of the O3N2 and N2 indicators provided by \citet{Marino2013}. Both maps are nearly uniform, with $8.46 \lesssim 12 +\log(O/H)\lesssim 8.54$, indicating a moderately sub-solar metallicity of $0.6-0.7\,\mathrm{Z_\odot}$ (assuming $12 +\log(O/H)= 8.69\pm 0.04$ for the Sun, see \citealt{Asplund2021}). At the position of \thisSN, the metallicity is $12 +\log(O/H)= 8.51\pm 0.01$.

From the H$\alpha$ emission line luminosity in the spatially integrated spectrum, we estimated the star formation rate (SFR) of the galaxy based on equation 2 of \citet{Kennicutt1998}, obtaining
$\mathrm{SFR}=1.6\pm 0.5\,\mathrm{M_\odot\,yr^{-1}}$. The error is conservatively taken as 30\% of the central value, reflecting the dispersion in the calibration that follows from using different initial mass functions and models of stellar evolution  and atmosphere.

\subsubsection{Radio limits on SN/KN emission}
\label{sec:radio_limits_on_jet}

\begin{figure}
\centering
\includegraphics[scale=0.3]{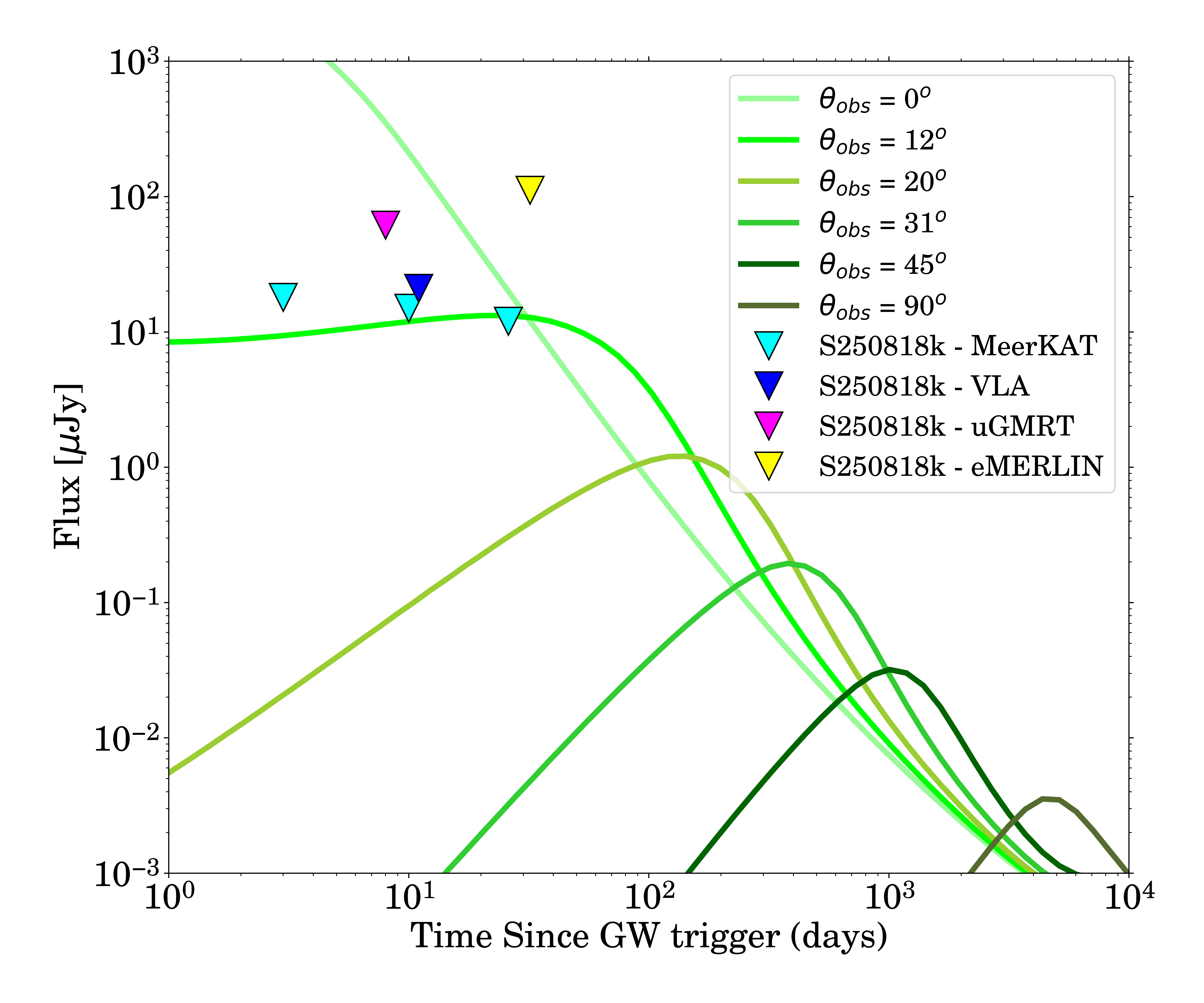}
\caption{Gaussian jet afterglow model light curves at 3 GHz. Different viewing angles (given in the legend) are represented with different shades of green. 3-$\sigma$ upper limits from VLA and MeerKAT at 3 GHz are represented in blue and cyan inverted triangles respectively.
We note that the e-MERLIN (in yellow) and uGMRT (in magenta) observations are rescaled to 3 GHz assuming an afterglow spectral slope of -0.6 \citep{Troja2019, Hajela2022}.}
\label{fig:ag_models_gw17}
\end{figure}

If \thisSN\,was associated with a BNS merger that produced a GRB jet, early radio observations would allow constraints on the jet observing angle. Assuming a Gaussian jet expanding into a uniform interstellar medium \citep{Lamb2017}, with the same global and microphysical parameters as those found for GW170817 in previous works \citep{Gianfagna2024, Ryan2024}, the uGMRT, e-MERLIN, MeerKAT\footnote{MeerKAT formally yields a detection, but this is consistent with constant emission from star formation in the host galaxy (see section \ref{sec:radio_from_SFR}). Hence we use 3$\times$RMS as an upper limit in this section, assuming that variable radio emission at this level would have been distinguishable from the constant component.}, and VLA observations constrain, under our simplifying assumptions, the inclination of the jet to be larger than about $12^{\rm o}$ (see Fig.\ \ref{fig:ag_models_gw17}). Hence, radio observations alone cannot rule out the possibility that an off-axis GRB was produced. This conclusion agrees with \citet{OConnor2025}. 

KN ejecta are also expected to produce a radio afterglow. However, the fastest tail of the KN ejecta is expected to decelerate on timescales of $\sim$months to years \citep[e.g.][]{NakarPiran2011}, making it irrelevant to test KN afterglow models with the available observations.

\begin{figure}
\centering
\includegraphics[scale=0.3]{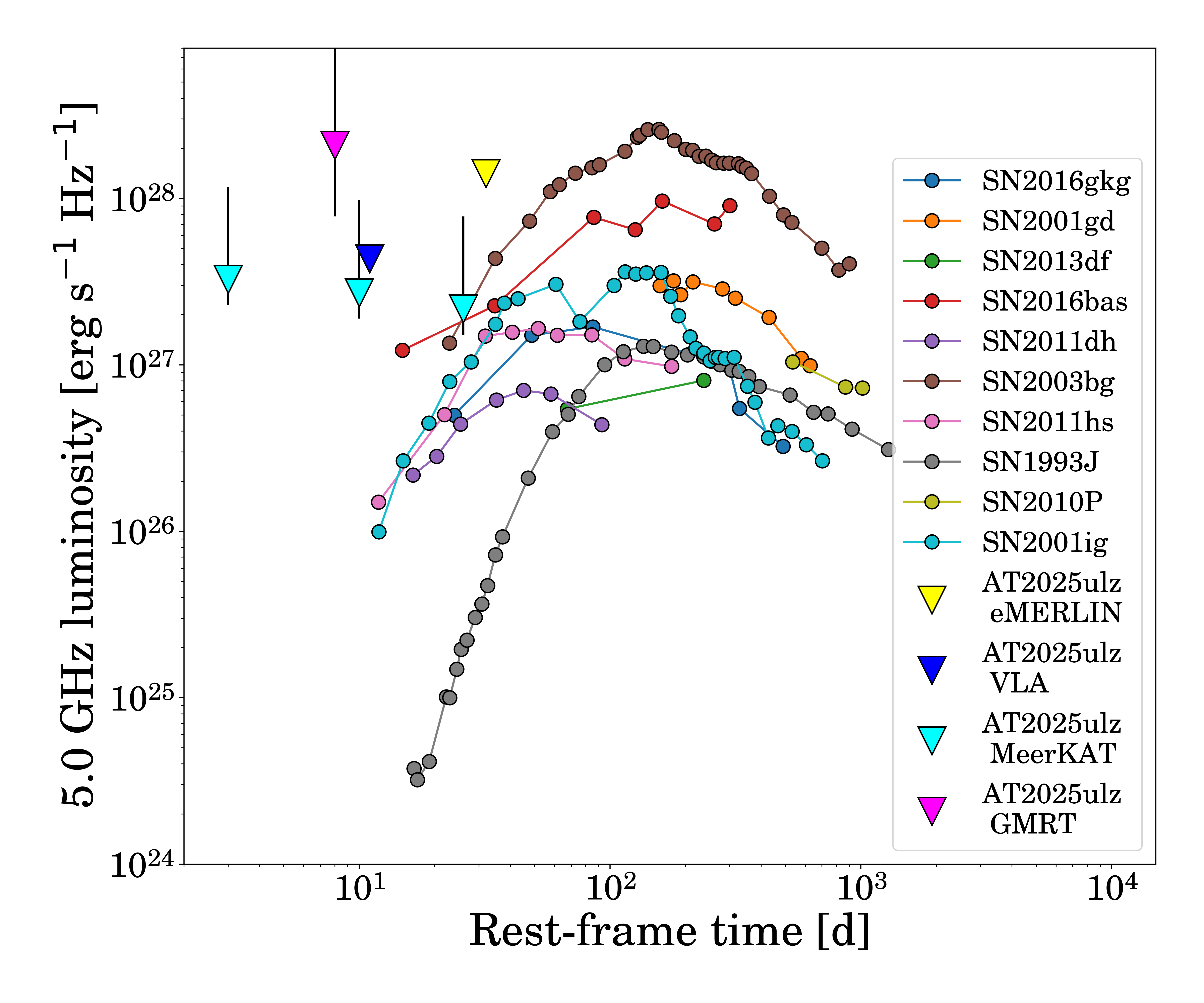}
\caption{5 GHz luminosity density of the radio emission from type IIb SNe. Upper limits for \thisSN\,are represented with inverted triangles. The e-MERLIN upper limit at 5.1 GHz is represented in yellow, while the VLA upper limit at 6 GHz is represented in blue. The uGMRT (1.2 GHz) and MeerKAT (3 GHz; estimated as 3$\times$RMS) upper limits are in magenta and cyan respectively. The vertical error bars show the flux variation assuming two different spectral slopes of -0.7 and +2.5, typical of radio SN.
Data for the SNe are from \citet{Nayana2022, Weiler2007, Soderberg2006, Ryder2004, Bietenholz2021, Kamble2016, Krauss2012, Bufano2011, Stockdale2007, Romero-Canizales2014}. }
\label{fig:SN_radio_lightcurves}
\end{figure}

Supernovae IIb are sometimes detected in radio, although this is highly dependent on the density of the circumstellar medium into which they explode \citep{Chevalier1982, Chevalier10}. Radio emission of a sample of SNe IIb is represented in Fig.\ \ref{fig:SN_radio_lightcurves}. Given the early time of the radio observations and the relatively large distance, the upper limits obtained with the VLA, MeerKAT, e-MERLIN, and uGMRT do not conflict with the interpretation of \thisSN\,as a Type IIb SN

\revone{In a recent preprint, \citet{ODwyer2026} report later detections in radio at 6 GHz and 10 GHz, around 100 days after the GW. As shown in their Figure 2, the implied luminosity is in good agreement with that of other type IIb SNe, where the emission most likely arises from the interaction of the ejecta with a circum-stellar medium. Hence, these observations support our interpretation of the source. The authors also claim that the emission can be alternatively explained by a relativistic jet with a viewing angle $\theta_\mathrm{obs}\sim 32^\circ$. This is consistent with our conclusion that a jet, if present, should have $\theta_\mathrm{obs}\gtrsim 12^\circ$, although we note that they assume different jet structure and microphysical parameters than those adopted here.}


\subsection{Radio emission as due to star formation}
\label{sec:radio_from_SFR}
While some of our radio observations did provide evidence of emission close to the location of \thisSN, overall we did not find clear signs of variability. It is instructive to estimate the SFR of the galaxy that would be needed to produce the emission observed. 
Following \citet{Murphy2011}, the host galaxy SFR can be estimated as 
\begin{equation}
    \mathrm{SFR} \sim 0.64 \left(\frac{L_{1.4\ \mathrm{GHz}}}{10^{28}\,\mathrm{erg\,s^{-1}\,Hz^{-1}}}\right)\ \mathrm{M_{\odot}\,yr^{-1}}.
\end{equation}
Taking the MeerKAT flux measurement of $F_{1.3\ \mathrm{GHz}} = 116 \pm 14$ \ujy\, and assuming a luminosity distance $D_\mathrm{L} = 399$ Mpc, we estimate a total SFR = $1.3\pm0.2\ \mathrm{M_{\odot}\,yr^{-1}}$. This estimate is in excellent agreement with that obtained from MUSE spectroscopy in Sect.\ \ref{sec:MUSE_Z_and_SFR}, strengthening the interpretation of the radio emission detected in our observations as being entirely produced by SFR.

\subsection{Volumetric rate of \thisSN-like events}
\label{sec:rate}


We can roughly estimate the volumetric rate of \thisSN-like events by assuming this to be the only such transient within the ZTF S250818k search volume. The ZTF search covered a sky area $\Omega_\mathrm{ZTF}\sim 168$ deg$^2$ looking for fast-evolving transients associated to galaxies at a distance compatible with the GW event. Taking the 2-$\sigma$ distance uncertainty range of S250818k as being comprised between the $2.5^\mathrm{th}$ and $97.5^\mathrm{th}$ percentiles of the sky-marginalised distance posterior, $[d_\mathrm{2.5\%},d_\mathrm{97.5\%}]\approx[120,410]\,\mathrm{Mpc}$ (from the \texttt{Bayestar} skymap available at that time, see also Appendix \ref{sec:localisation}), the search volume is $V_\mathrm{ZTF}\sim (\Omega_\mathrm{ZTF}/4\pi\,\mathrm{sr})\times \frac{4\pi}{3}(d_\mathrm{97.5\%}^3-d_\mathrm{2.5\%}^3)\approx 1.1\times 10^6\,\mathrm{Mpc^3}$. Available pre-discovery upper limits constrain the age of \thisSN\,at discovery to $\lesssim 1$ day \citep{Gillanders2025}. Given the observed light curve, the source would have been detectable at the ZTF search typical limit \citep[22 mag, ][]{2025GCN41414} for around two more days. The likelihood of finding one event with a volumetric rate density $R$ within a volume $V_\mathrm{ZTF}$ and a time $\Delta t \sim 3\,\mathrm{days}$ or less after its explosion is given by the Poisson probability $p(1\,|\,\lambda)=\lambda\exp(-\lambda)$ with expected number $\lambda = V_\mathrm{ZTF} \Delta t R$. From Bayes' theorem, adopting the Jeffreys prior for the Poisson probability $\pi(\lambda)\propto \lambda^{-1/2}$, the posterior probability on the volumetric rate is then
\begin{equation}
    p(R) \propto \pi(\lambda)\lambda \exp(-\lambda) \propto R^{1/2}\exp\left(-V_\mathrm{ZTF}\Delta t R\right).
\end{equation}
At the 90\% credible level, the volumetric rate of \thisSN-like events is therefore in the range $2\times 10^{-5}\lesssim R/\mathrm{Mpc^{-3}\,yr^{-1}}\lesssim 4\times 10^{-4}$.  This is in good agreement with the $z\lesssim 0.1$ core-collapse supernova (CCSN) volumetric rate $R_\mathrm{CCSNe}\sim 10^{-4}\,\mathrm{Mpc^{-3}\,yr^{-1}}$ \citep[e.g.][]{Frohmaier21,Ma25,Pessi25}, with the lower end of the estimate being roughly compatible with the rate of type IIb SNe, which is around ten percent of the total CCSNe.

\subsection{Light curve modelling}
\label{sec:lc_modeling}

As reconstructed in the introduction, early observations of \thisSN\ initially attracted the attention of the community due to the rapid brightness and colour evolution, which was considered by some as consistent with a KN \citep{2025GCN41436,2025GCN41480,2025GCN41538}. While the rebrightening between 3.5 and 20 days is not expected in a KN scenario, one could hypothesize that a long-lived NS merger remnant with a long spin-down timescale could provide the necessary energy to power the second peak \citep[e.g.][]{Sarin2022}, or that the afterglow of an off-axis GRB jet could be responsible for that emission \citep[but see][]{OConnor2025}. To test such a scenario and the general compatibility with a KN hypothesis, we started by fitting a KN model to data up to 4 days post GW trigger. We employed a two-component ejecta model following prescriptions in \citet{Villar2017} with updated heating rates \citep{Sarin2024_caution}. We included an additional contribution from cooling emission of putative material that was shocked and cast into a cocoon following the propagation of a relativistic jet through the KN ejecta \citep{Piro2018,Nicholl2021}. Each ejecta component is described by an ejecta mass, a characteristic velocity, and a gray opacity. The mass of the shocked material in the cocoon was assumed to be a fraction, $f_{\rm shock}$ (between 0 and 1) of the mass of the first ejecta component, with an opening angle $\theta_{\rm cocoon}$ (between 15 to 30$\deg$ and a shock breakout timescale (effectively the timescale for the jet to reach the breakout radius) $t_{\rm shock}$ (between 0.1 to 20 s).  
To test our preferred scenario, that of a type IIb SN, we used a one-zone radioactive decay model combined with shock cooling following prescriptions in \citet{Arnett1982, Pinto2000, Piro2021}. Both models are implemented in {\sc Redback} \citep{Sarin2024} and predict a luminosity and temperature evolution, but not a detailed spectrum. For simplicity, we adopted a blackbody description of the spectrum at all times. We left the explosion time as a free parameter in the SN scenario, while we fixed it to the GW trigger time in the KN model fit. In all fits we included the host-galaxy extinction, parameterised by $A_{\rm V, host}$ as a free parameter. 

The models were tested against the combined photometry from this work, \citet{Gillanders2025} and \citet{Kasliwal2025} (a subset of which is shown in Fig~\ref{fig:compiled_light_curve}), with a systematic error contribution of $0.05$ mag added in quadrature to the statistical errors of all data points to capture differences due to filter transmission curves. For the KN model, we only considered data up to 4 days after the GW trigger. We defined the posterior probability density of the model parameters assuming a standard Gaussian likelihood, and used {\sc Redback} to evaluate the models and perform a nested sampling of the posterior through the {\sc pymultinest} \citep{Feroz2009, Buchner2016} sampler, using the {\sc Bilby} interface \citep{Ashton2019}. 

\begin{figure*}
\centering
  \includegraphics[width=0.49\textwidth]{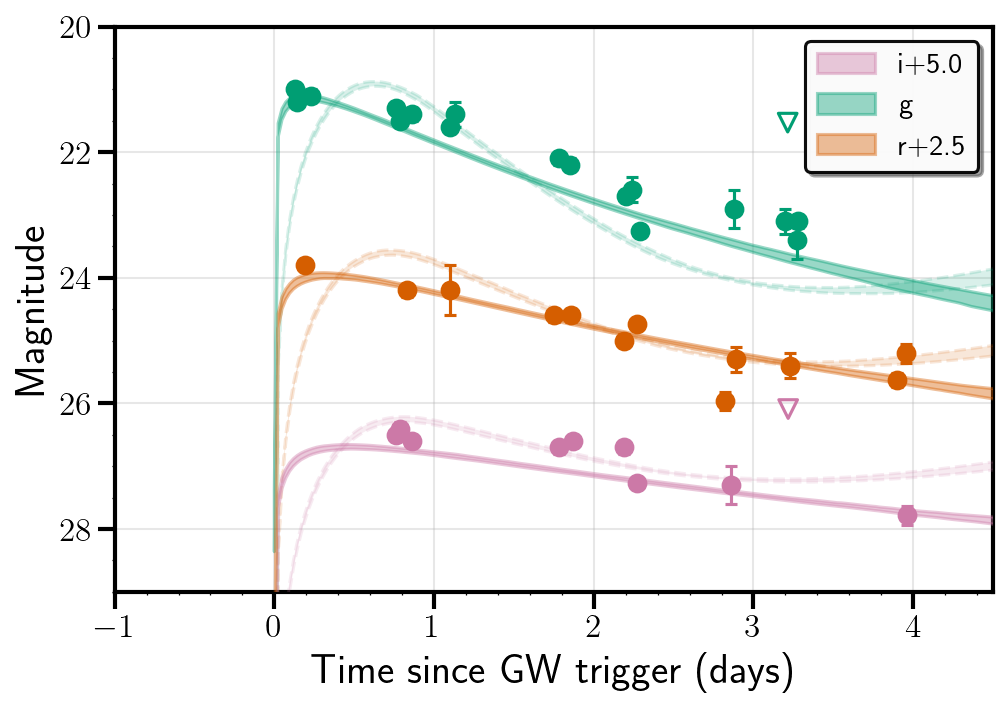}\includegraphics[width=0.49\textwidth]{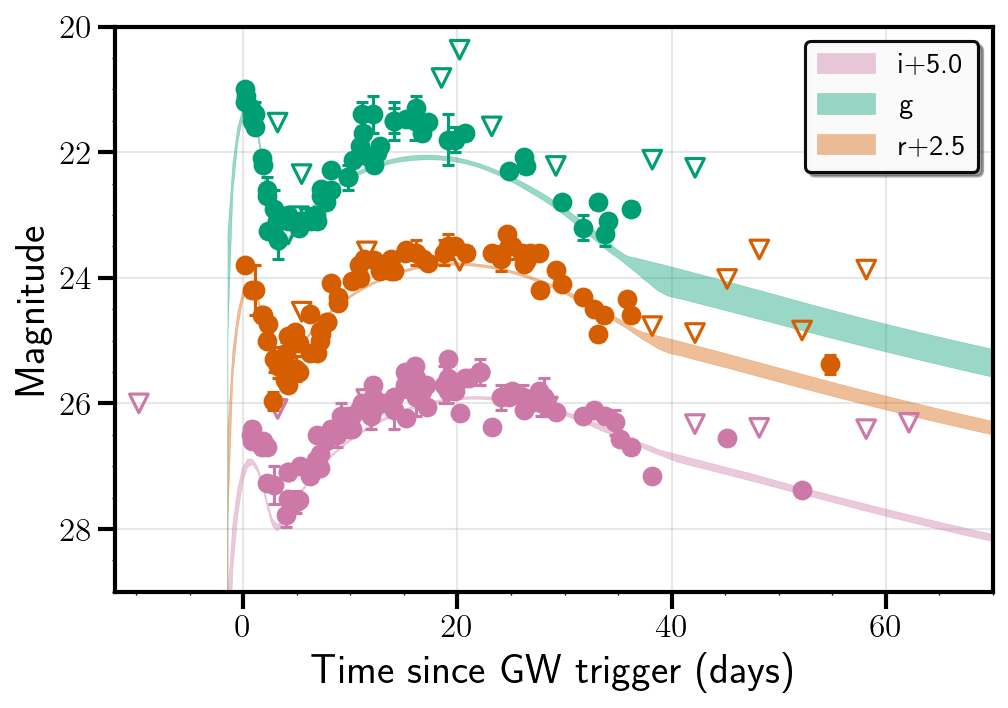}
  \caption{\textit{Left-hand panel}: fit to the first four days of photometry post GW trigger of \thisSN\ with two-component KN models. Solid lines refer to the KN plus cocoon cooling model, while the dashed lines are for the KN-only model (see text). The shaded bands contain, at each fixed time, the $68\%$ credible interval of the posterior samples in the $g$ band (green), $r$ band (orange, offset by +2.5 mag for presentation purposes) and $i$ band (offset by +5.0 mag). Circles with error bars show detections from our compiled dataset; triangles show 3-$\sigma$ upper limits. \textit{Right-hand panel:} similar to the left-hand panel, but in this case the model is shock cooling plus radioactive decay, and the fit is performed including all data up to 60 days after the GW trigger (see text).}
  \label{fig:fittedlc}
\end{figure*}

The left-hand panel of Fig~\ref{fig:fittedlc} shows the results of fitting the KN plus cocoon cooling emission model (solid lines) and the KN-only model (dashed lines) to the data of the first 4 days. While the KN plus cocoon cooling provides a relatively good fit by eye, the inferred parameters are not consistent with expectations for a BNS or NS-BH merger (see the corner plot in Fig.\ \ref{fig:kn_model_corner}), where we would expect masses on $\lesssim 0.05$ and opacities above $1\,\mathrm{cm^2g^{-1}}$ for both components \citep{Metzger2020}. 
In particular, the posterior median and 68\% credible ejecta mass, velocity and grey opacity of the first component are $M_\mathrm{ej,1}=0.10_{-0.03}^{+0.04}M_{\odot}$, $v_\mathrm{ej,1}=(0.05 \pm 0.003)$c and $\kappa_1 = 0.78_{-0.12}^{+0.14}\,\mathrm{cm^{2}g^{-1}}$. 
The corresponding parameters of the second component are $M_\mathrm{ej,2}=0.18^{+0.14}_{-0.11}M_{\odot}$, $v_\mathrm{ej,2}=(0.21^{+0.09}_{-0.08})$c, with an opacity of $\kappa_{2} = 27\pm 11\,\mathrm{cm^2g^{-1}}$. 
Finally, the fraction of material in the shocked cocoon is $f_\mathrm{shock}=0.28_{-0.16}^{+0.10}$, with an opening angle of $24_{-5}^{+4}\deg$ and a shock time, $t_{\rm shock} = 15.8_{-3.2}^{+2.6}$s. 
Turning the cocoon shock cooling emission off results in a KN prediction that does not fit the first $g$ and $r$ band data points post GW trigger (see the dashed lines in Fig.\ \ref{fig:fittedlc}) but otherwise has consistent parameters. 
The combined ejecta properties of the two components are not consistent with expectations for dynamical ejecta or winds produced by an accretion torus in the aftermath of a BNS merger, even in the presence of a long-lived NS~\citep[e.g.,][]{Metzger2020}. In particular, the inferred opacity for the first component is lower than expected for r-process elements and far more consistent with a supernova, meanwhile the second component is also extremely massive. When stricter priors that reflect realistic expectations for a KN are imposed, such as ejecta masses below $0.1\,\mathrm{M_\odot}$ and opacities above $1$ cm$^2$/g, then the KN model struggles to explain the first peak and subsequent decline without a significant additional emission component, such as shock-heated cocoon from jet-ejecta interaction~\citep{Nicholl2021, Gillanders2025}.  These results disfavour a KN interpretation for the data, at least with models with grey opacity and diffusion-based one-zone approximations applied here.

Fig.~\ref{fig:fittedlc} shows a comparison of our Type IIb SN shock cooling and radioactive decay model with the $g$, $r$ and $i$ band light curves of \thisSN. Only these bands are shown in the figure for simplicity, but the fits were performed to the entire dataset. The figure demonstrates that a SN model describes the data well. We provide a corner plot of the posterior probability of the model parameters in Figure \ref{fig:model_corner}. 
We derived a total SN ejecta mass of $M_{\rm ej} = 2.2_{-0.20}^{+0.21}~\mathrm{M_{\odot}}$, with a $^{56}{\rm Ni}$ fraction of $f_\mathrm{Ni}=0.07^{+0.004}_{-0.007}$ (corresponding to a mass $M_\mathrm{Ni}=0.14\pm0.01\,\mathrm{M_\odot}$). 
The characteristic ejecta velocity is $v_{\rm ej} = 11{,}270^{+410}_{-434}~{\rm km~s^{-1}}$. The gray opacity is $\kappa = 0.32^{+0.02}_{-0.03}~{\rm cm^2~g^{-1}}$, and the kinetic energy is $E_\mathrm{K}= 1.7_{-0.24}^{+0.27} \times 10^{51}~{\rm erg}$. The shock cooling model requires an envelope radius of $106_{-85}^{+457} R_{\odot}$. We also constrained the explosion time of the supernova to be $-1.6 \pm 0.15$ days before the GW trigger. 

The inferred ejecta mass, $^{56}{\rm Ni}$ mass, kinetic energy, and the envelope radius are broadly consistent with those inferred by \citet{Kasliwal2025} and with previous estimates for Type IIb supernovae \citep[e.g.][]{Taddia2018}. We find that the posterior rails against the prior limit on the opacity of $0.35~{\rm cm^2~g^{-1}}$, but this is likely due to the simplifications in the modelling, such as the assumption of a blackbody SED.

\section{Discussion and conclusions}

\subsection{Viability as a GW counterpart}

As shown in Sec.~\ref{sec:spectral_classification}, despite the challenges associated with the background removal, the spectroscopic classification of \thisSN\,as a type II SN, with SN IIb as the preferred subtype, is solid.
This classification is further strengthened by our rate calculations (sect.\ \ref{sec:rate}) and by the results of light curve modelling (sect.\ \ref{sec:lc_modeling}). Therefore, we can conclude that the appearance of \thisSN\,is entirely consistent with a type IIb SN. 
\revone{The probability of a chance association between such a SN and the candidate GW event can be estimated as follows. Assuming 10\% of the CCSN population to be of type IIb (see sect.\ \ref{sec:rate}) yields a volumetric rate density $R_\mathrm{SNIIb}\sim 10^{-5}$~yr$^{-1}$~Mpc$^{-3}$. The probability of having at least one such event in the ZTF search volume within $\Delta t= 3$ days of any random point in time by chance is then equal to $p(N_\mathrm{IIb}>0)= 1-\exp(-V_\mathrm{ZTF}\Delta t R_\mathrm{SNIIb})\approx 9\%$.}

\revone{More generally, the probability that \thisSN\ and S250818k are unrelated, factoring in the fact that the GW event is most likely a noise artifact, can be computed as $p_\mathrm{unrel}=p_\mathrm{astro}p_\mathrm{chance}+(1-p_\mathrm{astro})\approx 73.5\%$, where $p_\mathrm{astro}=29\%$ is the probability that S250818k is of astrophysical origin. These calculations show that the simplest and most conservative conclusion is that \thisSN\ is not related to S250818k.}

\revone{Given that another association between a candidate SSM BNS merger and a type IIb SN has been claimed recently (S251112cm and SN2025adtq, with a similar $p_\mathrm{chance}$; \citealt{Hall2026arXiv}), it is instructive to estimate how many coincidences are needed in order to claim a certain statistical significance of the association between the two classes of events. If we assume, for simplicity, that all coincidences have the same probability $p_\mathrm{unrel}$ of being unrelated, then the probability of having $N$ random coincidences is simply\footnote{\revone{Of course, this does not consider a trials factor that would arise in the case where a search for an association had been carried out in $M>N$ cases. Then, the post-trial probability of $N$ random coincidences would be increased by an $M$-choose-$N$ factor.}} $p_\mathrm{unrel}^N$. The minimum number of such coincidences needed to claim an association between the two classes with a significance $s$ (such that $p_s$ is the corresponding p-value) is $N=\log(\mathrm{p_s})/\log(p_\mathrm{unrel})$. Assuming optimistically that $p_\mathrm{astro}\sim 100\%$ for all events, and that $p_\mathrm{chance}=5\%$, then $N=2$ associations are needed to claim a 3-sigma significance ($p_s=2.7\times 10^{-3}$), or 5 associations for a 5-sigma significance ($p_s=5.7\times 10^{-7}$). This agrees with the calculations of \citet{Hall2026arXiv}. If $p_\mathrm{astro}=50\%$ for all events\footnote{\revone{The S251112cm event was found by the SSM search of the MBTA pipeline \citep{Allene2025}, with a FAR of about $1/6$ yr$^{-1}$. No $p_\mathrm{astro}$ was reported for this event by the LVK collaboration. If the relation between $p_\mathrm{astro}$ and FAR for SSM events is similar to that of BNS events, then the FAR of S251112cm corresponds to a $p_\mathrm{astro}\sim 50\%$, indeed (see Figure 11 of \citealt{Allene2025}).}}, on the other hand, a 3-sigma significance would require 10 associations. Hence we conclude that, from the statistical point of view, the available observations do not yet support an association between SSM BNS mergers and type IIb SNe.}

\subsection{Superkilonova scenario}

\revone{Our conclusions on the \thisSN --S250818k association do not provide support for the `superkilonova' scenario\footnote{\revone{This should not be confused with the processes described in \citet{Siegel2022}, for which the term `super-kilonova' was initially coined.}} proposed by \citet{Kasliwal2025}. Since this scenario has recently gained increasing attention \citep{Metzger2024,Chen2025,Kasliwal2025,Hall2025b,Hall2025a,Hall2026arXiv,Baibhav2026arXiv}, we further discuss here its viability as an explanation of our event and of the other GW events S251112cm and GW190814, which have been proposed to be associated with SN2025adtq (which exploded $\sim 2$ days prior to S251112cm, in a time-volume that corresponds to a similar chance association probability $p_\mathrm{chance}\sim 2-9\%$; \citealt{Hall2026arXiv}) and SN2019npv ($\sim 60$ days before the GW, $p_\mathrm{chance}\sim 4-5\%$; \citealt{Baibhav2026arXiv}), respectively. First, we summarize the main ingredients of the scenario, for clarity. A stellar mass black hole (BH) forms from the collapse of the iron core of a rotating massive star -- a collapsar \citep{MacFadyen1999}. Due to the fast rotation of the star, the collapsing envelope forms an accretion disk around the BH. The outer rims of the disk (at around one hundred gravitational radii and beyond) are prone to fragmentation. Due to the partial neutronization of the disk, fragments somewhat lighter than a solar mass can overcome the electron degeneracy pressure and form SSM NSs \citep{Piro2007}. Some of these light NSs can end up in a binary due to migration or three body interactions, and merge before spiralling into the central BH \citep{Metzger2024}. For what concerns GW emission, the basic predictions of this scenario are one or more SSM BNS mergers, followed by the eventual inspiral of their remnants (which can be NSs or light BHs) into the central BH. The time separation between a SSM BNS merger and the final high-mass-ratio merger in this scenario can range from hours \citep{Metzger2024} to days \citep{Hall2026arXiv} or even months \citep{Baibhav2026arXiv}.}

\revone{For what concerns electromagnetic emission, the collapsar scenario is generally expected to lead to a stellar explosion driven by a jet launched by the central BH or high-velocity outflows from the accretion disk \citep{MacFadyen1999,Metzger2024}. SSM BNS merger remnants themselves could take the form of highly magnetized, highly spinning proto-NSs, which could release a large fraction of their rotational energy in the form of a relativistic wind \citep{Metzger2024}, further powering the explosion. Such an engine-driven explosion is usually expected to be more energetic than a standard neutrino-driven SN, and could present peculiar features such as non-thermal emission (in the radio and/or the X-rays) arising from relativistic external shocks or long-lasting central engine activity, in addition to the usual nickel-decay-powered emission, but no definite predictions have been made yet in this respect, to our knowledge.}

\revone{In the case of S250818k and also S251112cm, the predicted GW emission from the high-mass-ratio merger between the SSM BNS remnant and the central BH has not been detected. Proponents of the scenario attributed this to the duty cycle limitations of the GW detectors, which presumably caused these signals to be missed \citep{Hall2026arXiv}. Similarly, the lack of SSM BNS merger signals prior to the high-mass-ratio GW190814 event was attributed to the excessive distance of the source, which would have made these signals too weak to trigger the GW detector searches \citep{Baibhav2026arXiv}. Therefore, while none of the three cases presents GW emission that fully adheres to the predictions of the scenario, this does not suffice to falsify the model.}

\revone{Because of the less definite predictions for the electromagnetic emission, it is more difficult to make strong statements in that regard. Still, it seems natural to expect a SN caused by a collapsar to present some distinguishing features with respect to ordinary neutrino-driven SNe, at least in terms of energetics and ejecta mass. Indeed, GRB-associated broad-lined type Ic SNe -- the prototypical collapsar-driven SNe -- show clear signs of high ejecta velocities (in excess of 0.1 c) and large ejecta masses (on the order of several solar masses), which translates to kinetic energies in excess of $10^{52}$ erg \citep{Cano2017}. Both \thisSN\ and SN2025adtq, on the other hand, have been successfully modelled as type IIb SNe with ejecta masses of a few $\mathrm{M_\odot}$ and kinetic energy of $10^{51}$ erg or less (see Sect.\ \ref{sec:lc_modeling} and \citep{Kasliwal2025,Hall2026arXiv}), and their spectra do not present broad lines indicative of high expansion velocities. As discussed in Sect.\ \ref{sec:radio_limits_on_jet}, the radio emission at the \thisSN\ position detected by \citet{ODwyer2026} is in line with that of other type IIb SNe. \citet{Kasliwal2025} claimed a peculiar color evolution for \thisSN, but the comparison with known type IIb SNe in Figure 5 of \citealt{Gillanders2025} does not seem to show any unusual behavior instead. The lack of any clear peculiarity in these SNe must be interpreted as either evidence that most CCSNe are the result of a collapsar -- which we consider as very unlikely -- or that these are just ordinary neutrino-driven SNe, which again conflicts with the superkilonova interpretation. SN2019npv has not been studied in a similar detail, but we note that also in this case the available spectrum does not feature broad lines, and the identified P-Cygni feature at 1.08 $\mu$m is consistent with a modest expansion velocity of 7000 km/s \citep{Andreoni2020}.}


\subsection{Young Type IIb SNe as interlopers}

The difficulty in the search for a kilonova does not simply reside in the faintness or rapid fading of the source: one of the main challenges is in discriminating contaminants. In a typical search region of hundreds of square degrees, the contaminant sources are several orders of magnitude more common than kilonovae \citep{Oates2025, Fulton25}. In recent years, it has become clear that Type IIb SNe are particularly pernicious in this regard: observations have demonstrated that 30-50 percent of Type IIb SNe feature a shock cooling tail \citep{Ayala25}, where the lightcurve displays a rapid initial decline following shock breakout in a relatively extended progenitor, before rising in brightness again to the main $^{56}$Ni-powered peak. Unfortunately, during the first few days of this shock cooling tail, the SN has a relatively featureless spectrum \citep[e.g.][]{Wang2023,Subrayan2025,Xi2026,Farah2026}, similarly to KNe \citep{Pian2017}.

Focusing on SNe IIb only, assuming these to comprise 10\% of the CCSN population (see sect. \ref{sec:rate}), and further assuming only 30\% of these to feature a shock cooling tail, we conservatively find that the volumetric rate of Type IIb SN shock cooling tails is about $R_\mathrm{SNIIb-SC}\sim0.3\times10^{-5}$~yr$^{-1}$~Mpc$^{-3}$.
 
The 90\% localisation area of S250828k was $\Omega_{90\%}=949$  deg$^2$ on the sky. Assuming that anything within the 90\% region and within the 2 sigma GW distance uncertainty will appear as a plausible candidate, our potential search volume is therefore $V_\mathrm{search}\sim 6.5\times 10^6$ Mpc$^{3}$ (same calculation as in section \ref{sec:rate}, but with a different sky area). Consequently, within this region there will be around $V_\mathrm{search}\times R_\mathrm{SNIIb-SC}\sim19\,\mathrm{yr^{-1}}$ Type IIb shock cooling tails.

The final assumption we must make is for how long a shock-cooling tail can be mistaken for a KN. \cite{Crawford25} find that the mean duration for a cooling tail is 6 days, and so we assume that any such SN that explodes up to a week before the KN is discovered will appear as a plausible counterpart.

Putting all of this together, and modeling the occurrence of SNe as a Poissonian process, we find that there is a roughly 30\% chance of one or more Type IIb cooling tail interlopers in the case of S250818k, and indeed any comparable event. In fact, the probability is likely even higher: in the updated GW skymap from the offline parameter estimation \citep[][available on the second day after the GW trigger]{2025GCN41440}, the \thisSN\,distance was at the 98.8$^\mathrm{th}$ percentile of the sky-marginalised distance posterior, evidence that the community will follow up targets where the spatial association with a GW is quite marginal.

\subsubsection{Consequences for future KN searches}

It is interesting to consider the implications of the above discussion in the context of GW counterpart searches.
While a 30\% chance of a IIb cooling tail in a GW localisation region does not appear as a strong limiting factor, we already noted that this is likely an underestimate (and of course, does not consider other contaminants such as luminous blue variable outbursts, cataclysmic variable stars in chance alignment with galaxies and shock cooling in peculiar thermonuclear SNe). In fact, the experiences of O3 and O4 suggest that every GW will have at least one initially plausible candidate for the counterpart that ultimately turns out to be unrelated, unless the localisation of the event is particularly good (as it was the case for GW190814 
-- \citealt{Ackley2020,GW190814bvLVK}). Clearly, as large amounts of telescope time are invested in the followup of such events, it behooves us to try and reduce this.

A possible strategy is to take advantage of the fact that most of the localisation probability is concentrated in the central region of the skymap. If we can significantly reduce the counterpart search area, then rather than expecting 0.3 - 1 plausible appearing interlopers, we would expect zero. In other words, if we want to find one KN, then rather than exhaustively searching the error box of one GW trigger, we should search the most probable 10-30\% region of 3-10 BNS mergers. Of course, such a strategy relies on GW detectors providing a steady stream of BNS alerts, \revone{as it would systematically miss 70-90\% of the counterparts by definition. As provocative as it may sound, this might ultimately prove to be the most effective strategy to limit the observing time spent in characterizing contaminants, in a future where BNS mergers are routinely detected by the GW network, but still with typical localizations of hundreds of square degrees}. 

The implication of the IIb shock-cooling tail rate is more problematic for untriggered KN searches \citep[e.g.][]{Andreoni2022,Zhu2023}, where an optical survey is used to look for transients that appear plausibly consistent with KNe without any high energy or GW detection. If we take a volumetric BNS rate of $\sim10^{-8}-10^{-7}$~yr$^{-1}$~Mpc$^{-3}$ \citep{Abac2025_LVKpop4}, the shock cooling tails are 30-300 times more common. In fact, the only winning move in such a scenario is to target passive galaxies. Around 70\% of the mass in the local Universe lies in non-star-forming galaxies \citep{Kelvin14}, opening the prospect of only following up candidates found in such hosts (pre-selecting galaxies either through colour cuts or spectroscopy). The IIb shock-cooling tails are inevitably followed by luminous $^{56}$Ni powered rising lightcurves which peak on the timescales of 20 days. These easily distinguish SNe from KNe, but if we wait for the re-brightening signature, we miss the opportunity to observe over the first 4-7 days, at which point a KN at 400 Mpc would likely be fainter than $r\sim24$ when we realise it has not re-brightened. The era of Rubin and LSST promises many candidates, but challenges are significant, and the role of spectroscopy in rejecting contaminants remains fundamental.


\begin{acknowledgements}
Based on observations collected at the European Organisation for Astronomical Research in the Southern Hemisphere under ESO programme 108.22JF.
G.B. acknowledges support from the European Union’s Horizon 2020 Programme under the AHEAD2020 project (grant agreement n. 871158). 
T.-W.C. acknowledges financial support from the Yushan Fellow Program of the Ministry of Education, Taiwan (MOE-111-YSFMS-0008-001-P1), and from the National Science and Technology Council, Taiwan (NSTC 114-2112-M-008-021-MY3).
Dimple acknowledges support from STFC grant No. ST/Y002253/1.
This publication has emanated from research conducted with the financial support of Taighde Éireann – Research Ireland under Grant number 24/FFP-P/12959.
G.G. acknowledges support by ASI (Italian Space Agency) through the Contract no. 2019-27-HH.0. 
C.P.G. acknowledges financial support from the Secretary of Universities and Research (Government of Catalonia) and by the Horizon 2020 Research and Innovation Programme of the European Union under the Marie Sk\l{}odowska-Curie and the Beatriu de Pin\'os 2021 BP 00168 programme, from the Spanish Ministerio de Ciencia e Innovaci\'on (MCIN) and the Agencia Estatal de Investigaci\'on (AEI) 10.13039/501100011033 under the PID2023-151307NB-I00 SNNEXT project, from Centro Superior de Investigaciones Cient\'ificas (CSIC) under the PIE project 20215AT016 and the program Unidad de Excelencia Mar\'ia de Maeztu CEX2020-001058-M, and from the Departament de Recerca i Universitats de la Generalitat de Catalunya through the 2021-SGR-01270 grant.
L.I. acknowledges financial support from the INAF Data Grant Program 'YES' (PI: Izzo) {\it Multi-wavelength and multi messenger analysis of relativistic supernovae}.
P.G.J. is supported by the European Union (ERC, Starstruck, 101095973, PI Jonker). Views and opinions expressed are, however, those of the author(s) only and do not necessarily reflect those of the European Union or the European Research Council Executive Agency. Neither the European Union nor the granting authority can be held responsible for them.
G.P.L. is supported by a Royal Society Dorothy Hodgkin Fellowship, grant No. DHF-R1-221175 and DHF-ERE-221005.
G.L. was supported by a research grant (VIL60862) from VILLUM FONDEN.
J.D.L. acknowledges support from a UK Research and Innovation Future Leaders Fellowship (grant references MR/T020784/1 and UKRI1062).
K.M. acknowledges funding from Horizon Europe ERC grant no. 101125877.
This research was funded in part by National Science Centre, Poland (grant number  2023/49/B/ST9/00066). For the purpose of Open Access, the author has applied a CC-BY public copyright licence to any Author Accepted Manuscript (AAM) version arising from this submission.
F.O. acknowledges support from the INAF-Large Grant 2024:"Envisioning Tomorrow: prospects and challenges for multimessenger astronomy in the era of Rubin and Einstein Telescope"; the INAF-GO Large Grant: "Exploitation of optical and near-infrared followup data of Gamma-ray Bursts" and the  INAF-MINIGRANT (2023): "SeaTiDE - Searching for Tidal Disruption Events with ZTF: the Tidal Disruption Event population in the era of wide field surveys".
S.P. acknowledges funding from the Large Grant INAF 2024.
L.P. acknowledges support by ASI (Italian Space Agency) through the Contract no. 2019-27-HH.0. 
A.S. acknowledges financial support from the Centre national d’études spatiales (CNES), France (ROR: \url{https://ror.org/04h1h0y33}) within the framework of the SVOM mission.
This work has been funded by the European Union-Next Generation EU, PRIN 2022 RFF M4C21.1 (202298J7KT - PEACE). O.S.S.\ acknowledges funding from INAF through grant 1.05.23.04.04.
S.J.S. acknowledges funding from STFC Grants ST/Y001605/1, ST/X001253/1, a Royal Society Research Professorship and the Hintze Family Charitable Foundation. 
D.S. acknowledges support from The Science and Technology Facilities Council (STFC) via grants ST/T007184/1, ST/T003103/1, ST/T000406/1, ST/X001121/1.
N.R.T. acknowledges funding from the UK STFC grant UKRI1200.
A.L.T. acknowledges support by ASI (Italian Space Agency) through the Contract no. 2019-27-HH.0 and from the European Union’s Horizon 2020 Programme under the AHEAD2020 project (grant agreement n. 871158).
e-MERLIN is a National Facility operated by the University of Manchester at Jodrell Bank Observatory on behalf of STFC, part of UK Research and Innovation. The MeerKAT telescope is operated by the South African Radio Astronomy Observatory, which is a facility of the National Research Foundation, an agency of the Department of Science, Technology and Innovation. This work has made use of the “MPIfR S-band receiver system” designed, constructed and maintained by funding of the MPI für Radioastronomy and the Max-Planck-Society.  We thank the staff of the GMRT that made these observations possible. GMRT is run by the National Centre for Radio Astrophysics of the Tata Institute of Fundamental Research.
\end{acknowledgements}

\bibliographystyle{Bibtex/aa}
\footnotesize
\bibliography{References}

@ARTICLE{OConnor2025,
       author = {{O'Connor}, Brendan and {Ricci}, Roberto and {Troja}, Eleonora and {Palmese}, Antonella and {Yang}, Yu-Han and {Ryan}, Geoffrey and {van Eerten}, Hendrik and {Yadav}, Muskan and {Hall}, Xander J. and {Amsellem}, Ariel and {Becerra}, Rosa L. and {Busmann}, Malte and {Cabrera}, Tomas and {Dichiara}, Simone and {Hu}, Lei and {Kaur}, Ravjit and {Kunnumkai}, Keerthi and {Magana Hernandez}, Ignacio},
        title = "{AT2025ulz and S250818k: Deep X-ray and radio limits on off-axis afterglow emission and prospects for future discovery}",
      journal = {arXiv e-prints},
         year = 2025,
        month = oct,
          eid = {arXiv:2510.23728},
        pages = {arXiv:2510.23728},
          doi = {10.48550/arXiv.2510.23728},
archivePrefix = {arXiv},
       eprint = {2510.23728},
 primaryClass = {astro-ph.HE},
       adsurl = {https://ui.adsabs.harvard.edu/abs/2025arXiv251023728O}
}

@ARTICLE{Hall2025b,
       author = {{Hall}, Xander J. and {Busmann}, Malte and {Koehn}, Hauke and {Kunnumkai}, Keerthi and {Palmese}, Antonella and {O'Connor}, Brendan and {Freeburn}, James and {Hu}, Lei and {Gruen}, Daniel and {Dietrich}, Tim and {Bulla}, Mattia and {Coughlin}, Michael W. and {Antier}, Sarah and {Pillas}, Marion and {Price}, Paul A. and {Ahumada}, Tom{\'a}s and {Amsellem}, Ariel and {Andreoni}, Igor and {Augustin}, Jule and {Cabrera}, Tom'as and {Deshpande}, Rasika and {Fab{\`a}-Moreno}, Jennifer and {Gassert}, Julius and {Karpov}, Sergey and {Kasliwal}, Mansi and {Maga{\~n}a Hernandez}, Ignacio and {Mandelbaum}, Rachel and {Fontinele Nunes}, Felipe and {Pang}, Peter T.~H. and {Sommer}, Julian and {Stein}, Robert and {Tabor}, Constantin and {Vega}, Pablo and {Wouters}, Thibeau and {Zuo}, Xiaoxiong},
        title = "{AT2025ulz and S250818k: Investigating early time observations of a subsolar mass gravitational-wave binary neutron star merger candidate}",
      journal = {arXiv e-prints},
         year = 2025,
        month = oct,
          eid = {arXiv:2510.24620},
        pages = {arXiv:2510.24620},
          doi = {10.48550/arXiv.2510.24620},
archivePrefix = {arXiv},
       eprint = {2510.24620},
 primaryClass = {astro-ph.HE},
       adsurl = {https://ui.adsabs.harvard.edu/abs/2025arXiv251024620H}
}

@ARTICLE{Lamb2017,
       author = {{Lamb}, Gavin P. and {Kobayashi}, Shiho},
        title = "{Electromagnetic counterparts to structured jets from gravitational wave detected mergers}",
      journal = {\mnras},
         year = 2017,
        month = dec,
       volume = {472},
       number = {4},
        pages = {4953-4964},
          doi = {10.1093/mnras/stx2345},
archivePrefix = {arXiv},
       eprint = {1706.03000},
 primaryClass = {astro-ph.HE},
       adsurl = {https://ui.adsabs.harvard.edu/abs/2017MNRAS.472.4953L}
}

@ARTICLE{Hall2025a,
       author = {{Hall}, Xander J. and {Palmese}, Antonella and {O'Connor}, Brendan and {Gruen}, Daniel and {Busmann}, Malte and {Gassert}, Julius and {Hu}, Lei and {Magana Hernandez}, Ignacio and {Aguilar}, Jessica Nicole and {Amsellem}, Ariel and {Ahlen}, Steven and {Banovetz}, John and {BenZvi}, Segev and {Bianchi}, Davide and {Brooks}, David and {Castander}, Francisco Javier and {Claybaugh}, Todd and {Cuceu}, Andrei and {Dey}, Arjun and {Doel}, Peter and {Faba-Moreno}, Jennifer and {Ferraro}, Simone and {Font-Ribera}, Andreu and {Forero-Romero}, Jaime E. and {Gutierrez}, Gaston and {Le Guillou}, Laurent and {Joyce}, Dick and {Kisner}, Theodore and {Kremin}, Anthony and {Lahav}, Ofer and {Lamman}, Claire and {Landriau}, Martin and {Levi}, Michael and {de la Macorra}, Axel and {Manera}, Marc and {Meisner}, Aaron and {Miquel}, Ramon and {Moustakas}, John and {Nadathur}, Seshadri and {Prada}, Francisco and {Perez-Rafols}, Ignasi and {Rossi}, Graziano and {Sanchez}, Eusebio and {Schlegel}, David and {Schubnell}, Michael and {Sprayberry}, David and {Tarle}, Gregory and {Weaver}, Benjamin Alan and {Zhou}, Rongpu and {Zou}, Hu},
        title = "{AT2025ulz and S250818k: Leveraging DESI spectroscopy in the hunt for a kilonova associated with a sub-solar mass gravitational wave candidate}",
      journal = {arXiv e-prints},
         year = 2025,
        month = oct,
          eid = {arXiv:2510.23723},
        pages = {arXiv:2510.23723},
          doi = {10.48550/arXiv.2510.23723},
archivePrefix = {arXiv},
       eprint = {2510.23723},
 primaryClass = {astro-ph.HE},
       adsurl = {https://ui.adsabs.harvard.edu/abs/2025arXiv251023723H}
}

@ARTICLE{Yang2025,
       author = {{Yang}, Yu-Han and {Troja}, Eleonora and {Risti{\'c}}, Marko and {Yadav}, Muskan and {El Kabir}, Massine and {S{\'a}nchez-Ram{\'\i}rez}, Rub{\'e}n and {Becerra}, Rosa L. and {Fryer}, Chris L. and {O'Connor}, Brendan and {Dichiara}, Simone and {Castro-Tirado}, Alberto J. and {Angulo-Valdez}, Camila and {Becerra Gonz{\'a}lez}, Josefa and {Font}, Jos{\'e} A. and {Fox}, Ori and {Hu}, Lei and {Hu}, Youdong and {Lee}, William H. and {Pereyra}, Margarita and {Sintes}, Alicia M. and {Watson}, Alan M. and {Oc{\'e}lotl C}, L{\'o}pez Mendoza K.},
        title = "{AT2025ulz and S250818k: zooming in with the Hubble Space Telescope}",
      journal = {arXiv e-prints},
         year = 2025,
        month = oct,
          eid = {arXiv:2510.18854},
        pages = {arXiv:2510.18854},
          doi = {10.48550/arXiv.2510.18854},
archivePrefix = {arXiv},
       eprint = {2510.18854},
 primaryClass = {astro-ph.HE},
       adsurl = {https://ui.adsabs.harvard.edu/abs/2025arXiv251018854Y}
}

@ARTICLE{Yang2024,
       author = {{Yang}, Yu-Han and {Troja}, Eleonora and {O'Connor}, Brendan and {Fryer}, Chris L. and {Im}, Myungshin and {Durbak}, Joe and {Paek}, Gregory S.~H. and {Ricci}, Roberto and {Bom}, Cl{\'e}cio R. and {Gillanders}, James H. and {Castro-Tirado}, Alberto J. and {Peng}, Zong-Kai and {Dichiara}, Simone and {Ryan}, Geoffrey and {van Eerten}, Hendrik and {Dai}, Zi-Gao and {Chang}, Seo-Won and {Choi}, Hyeonho and {De}, Kishalay and {Hu}, Youdong and {Kilpatrick}, Charles D. and {Kutyrev}, Alexander and {Jeong}, Mankeun and {Lee}, Chung-Uk and {Makler}, Martin and {Navarete}, Felipe and {P{\'e}rez-Garc{\'\i}a}, Ignacio},
        title = "{A lanthanide-rich kilonova in the aftermath of a long gamma-ray burst}",
      journal = {\nat},
         year = 2024,
        month = feb,
       volume = {626},
       number = {8000},
        pages = {742-745},
          doi = {10.1038/s41586-023-06979-5},
archivePrefix = {arXiv},
       eprint = {2308.00638},
 primaryClass = {astro-ph.HE},
       adsurl = {https://ui.adsabs.harvard.edu/abs/2024Natur.626..742Y}
}

@ARTICLE{Franz2025,
       author = {{Franz}, Noah and {Subrayan}, Bhagya and {Kilpatrick}, Charles D. and {Hosseinzadeh}, Griffin and {Sand}, David J. and {Alexander}, Kate D. and {Fong}, Wen-fai and {Christy}, Collin T. and {Pearson}, Jeniveve and {Laskar}, Tanmoy and {Hsu}, Brian and {Rastinejad}, Jillian and {Lundquist}, Michael J. and {Berger}, Edo and {Bostroem}, K. Azalee and {Bom}, Clecio R. and {Darc}, Phelipe and {Gurwell}, Mark and {Hostler Schimpf}, Shelbi and {Keating}, Garrett K. and {Noel}, Phillip and {Ransome}, Conor and {Rao}, Ramprasad and {Santana-Silva}, Luidhy and {Souza Santos}, A. and {Shrestha}, Manisha and {Anche}, Ramya and {Andrews}, Jennifer E. and {Borthakur}, Sanchayeeta and {Butler}, Nathaniel R. and {Coppejans}, Deanne L. and {Daly}, Philip N and {Daniel}, Kathryne J. and {Duffell}, Paul C. and {Eftekhari}, Tarraneh and {Fields}, Carl E. and {Gagliano}, Alexander T. and {Golay}, Walter W. and {Grichener}, Aldana and {Hamden}, Erika T. and {Hiramatsu}, Daichi and {Kumar}, Harsh and {Manikantan}, Vikram and {Margutti}, Raffaella and {Paschalidis}, Vasileios and {Paterson}, Kerry and {Reichart}, Daniel E. and {Renzo}, Mathieu and {Salmas}, Kali and {Schroeder}, Genevieve and {Smith}, Nathan and {Spekkens}, Kristine and {Strader}, Jay and {Trilling}, David E. and {Vieira}, Nicholas and {Weiner}, Benjamin and {Williams}, Peter K.~G.},
        title = "{Optimizing Kilonova Searches: A Case Study of the Type IIb SN 2025ulz in the Localization Volume of the Low-Significance Gravitational Wave Event S250818k}",
      journal = {arXiv e-prints},
         year = 2025,
        month = oct,
          eid = {arXiv:2510.17104},
        pages = {arXiv:2510.17104},
          doi = {10.48550/arXiv.2510.17104},
archivePrefix = {arXiv},
       eprint = {2510.17104},
 primaryClass = {astro-ph.HE},
       adsurl = {https://ui.adsabs.harvard.edu/abs/2025arXiv251017104F}
}

@ARTICLE{Kasliwal2025,
       author = {{Kasliwal}, Mansi M. and {Ahumada}, Tomas and {Stein}, Robert and {Karambelkar}, Viraj and {Hall}, Xander J. and {Singh}, Avinash and {Fremling}, Christoffer and {Metzger}, Brian D. and {Bulla}, Mattia and {Swain}, Vishwajeet and {Antier}, Sarah and {Pillas}, Marion and {Busmann}, Malte and {Freeburn}, James and {Karpov}, Sergey and {Bochenek}, Aleksandra and {O'Connor}, Brendan and {Perley}, Daniel A. and {Akl}, Dalya and {Anand}, Shreya and {Toivonen}, Andrew and {Rose}, Sam and {Jegou du Laz}, Theophile and {Liu}, Chang and {Das}, Kaustav and {Sharma Chaudhary}, Sushant and {Barna}, Tyler and {Pawan Saikia}, Aditya and {Andreoni}, Igor and {Bellm}, Eric C. and {Bhalerao}, Varun and {Cenko}, S. Bradley and {Coughlin}, Michael W. and {Gruen}, Daniel and {Kasen}, Daniel and {Miller}, Adam A. and {Nissanke}, Samaya and {Palmese}, Antonella and {Sollerman}, Jesper and {Sravan}, Niharika and {Anupama}, G.~C. and {Banerjee}, Smaranika and {Barway}, Sudhanshu and {Bloom}, Joshua S. and {Cabrera}, Tomas and {Chen}, Tracy and {Copperwheat}, Chris and {Corsi}, Alessandra and {Dekany}, Richard and {Earley}, Nicholas and {Graham}, Matthew and {Hello}, Patrice and {Helou}, George and {Hu}, Lei and {Kini}, Yves and {Mahabal}, Ashish and {Masci}, Frank and {Mohan}, Tanishk and {Pletskova}, Natalya and {Purdum}, Josiah and {Qin}, Yu-Jing and {Rehemtulla}, Nabeel and {Salgundi}, Anirudh and {Wang}, Yuankun},
        title = "{ZTF25abjmnps (AT2025ulz) and S250818k: A Candidate Superkilonova from a Sub-threshold Sub-Solar Gravitational Wave Trigger}",
      journal = {arXiv e-prints},
         year = 2025,
        month = oct,
          eid = {arXiv:2510.23732},
        pages = {arXiv:2510.23732},
          doi = {10.48550/arXiv.2510.23732},
archivePrefix = {arXiv},
       eprint = {2510.23732},
 primaryClass = {astro-ph.HE},
       adsurl = {https://ui.adsabs.harvard.edu/abs/2025arXiv251023732K}
}

@ARTICLE{Gillanders2023,
       author = {{Gillanders}, James H. and {Troja}, Eleonora and {Fryer}, Chris L. and {Ristic}, Marko and {O'Connor}, Brendan and {Fontes}, Christopher J. and {Yang}, Yu-Han and {Domoto}, Nanae and {Rahmouni}, Salma and {Tanaka}, Masaomi and {Fox}, Ori D. and {Dichiara}, Simone},
        title = "{Heavy element nucleosynthesis associated with a gamma-ray burst}",
      journal = {arXiv e-prints},
         year = 2023,
        month = aug,
          eid = {arXiv:2308.00633},
        pages = {arXiv:2308.00633},
          doi = {10.48550/arXiv.2308.00633},
archivePrefix = {arXiv},
       eprint = {2308.00633},
 primaryClass = {astro-ph.HE},
       adsurl = {https://ui.adsabs.harvard.edu/abs/2023arXiv230800633G}
}

@ARTICLE{Gillanders2025,
       author = {{Gillanders}, J.~H. and {Huber}, M.~E. and {Nicholl}, M. and {Smartt}, S.~J. and {Smith}, K.~W. and {Chambers}, K.~C. and {Young}, D.~R. and {Tweddle}, J.~W. and {Srivastav}, S. and {Fulton}, M.~D. and {Stoppa}, F. and {Paek}, G.~S.~H. and {Aamer}, A. and {Alarcon}, M.~R. and {Andersson}, A. and {Aryan}, A. and {Auchettl}, K. and {Chen}, T.-W. and {de Boer}, T. and {Kong}, A.~K.~H. and {Licandro}, J. and {Lowe}, T. and {Magill}, D. and {Magnier}, E.~A. and {Minguez}, P. and {Moore}, T. and {Pignata}, G. and {Rest}, A. and {Serra-Ricart}, M. and {Shappee}, B.~J. and {Smith}, I.~A. and {Tucker}, M.~A. and {Wainscoat}, R.},
        title = "{Pan-STARRS Follow-up of the Gravitational-wave Event S250818k and the Light Curve of SN2025ulz}",
      journal = {\apjl},
         year = 2025,
        month = dec,
       volume = {995},
       number = {1},
          eid = {L27},
        pages = {L27},
          doi = {10.3847/2041-8213/ae2125},
archivePrefix = {arXiv},
       eprint = {2510.01142},
 primaryClass = {astro-ph.HE},
       adsurl = {https://ui.adsabs.harvard.edu/abs/2025ApJ...995L..27G}
}

@ARTICLE{Gillanders2025_AT2023vfi,
       author = {{Gillanders}, J.~H. and {Smartt}, S.~J.},
        title = "{Analysis of the JWST spectra of the kilonova AT 2023vfi accompanying GRB 230307A}",
      journal = {\mnras},
         year = 2025,
        month = apr,
       volume = {538},
       number = {3},
        pages = {1663-1689},
          doi = {10.1093/mnras/staf287},
archivePrefix = {arXiv},
       eprint = {2408.11093},
 primaryClass = {astro-ph.HE},
       adsurl = {https://ui.adsabs.harvard.edu/abs/2025MNRAS.538.1663G}
}

@ARTICLE{2025GCN41666,
       author = {{Rhodes}, Lauren and {Smirnov}, Oleg and {Mooley}, Kunal and {Woudt}, Patrick},
        title = "{LIGO/VIRGO/KAGRA S250818k: No evidence for radio variability of AT2025ulz in MeerKAT 3 GHz data}",
      journal = {GRB Coordinates Network},
         year = 2025,
        month = sep,
       volume = {41666},
        pages = {1},
       adsurl = {https://ui.adsabs.harvard.edu/abs/2025GCN.41666....1R}
}

@ARTICLE{2025GCN41594,
       author = {{Bruni}, G. and {Piro}, L. and {Gianfagna}, G. and {Thakur}, A.~L.},
        title = "{LIGO/Virgo/KAGRA S250818k: MeerKAT detection of an increase in radio flux from AT2025ulz}",
      journal = {GRB Coordinates Network},
         year = 2025,
        month = aug,
       volume = {41594},
        pages = {1},
       adsurl = {https://ui.adsabs.harvard.edu/abs/2025GCN.41594....1B}
}

@ARTICLE{2025GCN41540,
       author = {{Gillanders}, J.~H. and {Huber}, M.~E. and {Chambers}, K.~C. and {Smartt}, S.~J. and {Smith}, K.~W. and {Srivastav}, S. and {Stoppa}, F. and {Stevance}, H. and {Tweddle}, J. and {Nicholl}, M. and {Young}, D.~R. and {Aamer}, A. and {Angus}, C.~R. and {Fulton}, M.~D. and {Magill}, D. and {McCollum}, M. and {Moore}, T. and {Sim}, S. and {Weston}, J. and {Sheng}, X. and {Chen}, T. -W. and {Shingles}, L. and {Ramsden}, P. and {Schultz}, A.~S.~B. and {de Boer}, T. and {Fairlamb}, J. and {Lin}, C.~C. and {Lowe}, T. and {Magnier}, E. and {Minguez}, P. and {Paek}, G. and {Smith}, I.~A. and {Wainscoat}, R.~J. and {Rest}, A. and {Stubbs}, C.},
        title = "{LIGO/Virgo/KAGRA S250818k: Pan-STARRS imaging confirms re-brightening of SN2025ulz}",
      journal = {GRB Coordinates Network},
         year = 2025,
        month = aug,
       volume = {41540},
        pages = {1},
       adsurl = {https://ui.adsabs.harvard.edu/abs/2025GCN.41540....1G}
}

@ARTICLE{2025GCN41538,
       author = {{Kasliwal}, Mansi M. and {Karambelkar}, Viraj and {Fremling}, Christoffer and {Ahumada}, Tomas and {Hall}, Xander J. and {Perley}, Daniel A. and {Anand}, Shreya and {Liu}, Chang and {Das}, Kaustav and {Bhalerao}, Varun and {Swain}, Vishwajeet and {Saikia}, Aditya and {Ztf Collaboration} and {Growth Collaboration}},
        title = "{LIGO/Virgo/KAGRA S250818k: Continued Keck I LRIS spectroscopy of ZTF25abjmnps (AT2025ulz)}",
      journal = {GRB Coordinates Network},
         year = 2025,
        month = aug,
       volume = {41538},
        pages = {1},
       adsurl = {https://ui.adsabs.harvard.edu/abs/2025GCN.41538....1K}
}

@ARTICLE{2025GCN41532,
       author = {{Banerjee}, Smaranika and {Botticella}, Maria-Teresa and {Brennan}, Se{\'a}n J. and {Cappellaro}, Enrico and {Chen}, Ting-Wan and {D'Avanzo}, Paolo and {D'Elia}, Valerio and {de Pasquale}, Massimiliano and {Eyles-Ferris}, Rob A.~J. and {Fraser}, Morgan and {Gillanders}, James H. and {Gompertz}, Ben and {Habeeb}, Nusrin and {Izzo}, Luca and {Jonker}, Peter G. and {Levan}, Andrew J. and {Bj{\o}rn Malesani}, Daniele and {Martin-Carrillo}, Antonio and {Nicholl}, Matt and {Oates}, Sam and {Piranomonte}, Silvia and {Piro}, Luigi and {Rossi}, Andrea and {Sharan Salafia}, Om and {Sarin}, Nikhil and {Schulze}, Steve and {Singh}, Avinash and {Smartt}, Stephen J. and {Sneppen}, Albert and {Sollerman}, Jesper and {Steeghs}, Danny and {Tanvir}, Nial R. and {Thakur}, Aishwarya L. and {Engrave Collaboration}},
        title = "{LIGO/Virgo/KAGRA S250818k: ENGRAVE observations of SN 2025ulz as a type II supernova}",
      journal = {GRB Coordinates Network},
         year = 2025,
        month = aug,
       volume = {41532},
        pages = {1},
       adsurl = {https://ui.adsabs.harvard.edu/abs/2025GCN.41532....1B}
}

@ARTICLE{2025GCN41528,
       author = {{Becerra}, Rosa L. and {Troja}, Eleonora and {Dichiara}, Simone},
        title = "{LIGO/Virgo/KAGRA S250818k: Swift Observations of AT 2025ulz - Second Epoch}",
      journal = {GRB Coordinates Network},
         year = 2025,
        month = aug,
       volume = {41528},
        pages = {1},
       adsurl = {https://ui.adsabs.harvard.edu/abs/2025GCN.41528....1B}
}

@ARTICLE{2025GCN41518,
       author = {{Angulo}, Camila and {Watson}, Alan M. and {Dornic}, Damien and {Basa}, St{\'e}phane and {Lee}, William H. and {S{\'a}nchez {\'A}lvarez}, Fredd and {Akl}, Dalya and {Antier}, Sarah and {Atteia}, Jean-Luc and {Becerra}, Rosa L. and {Butler}, Nathaniel R. and {Ducoin}, Jean-Gr{\'e}goire and {Fortin}, Francis and {Garc{\'\i}a Garc{\'\i}a}, Leonardo and {Gill}, Ramandeep and {Globus}, No{\'e}mie and {Ocelotl L{\'o}pez}, Kin and {L{\'o}pez-C{\'a}mara}, Diego and {Magnani}, Francesco and {Moreno M{\'e}ndez}, Enrique and {Pereyra}, Margarita and {Avo Rakotondrainibe}, Ny and {Schneider}, Benjamin and {de Ugarte Postigo}, Antonio},
        title = "{LIGO/Virgo/KAGRA S250818k: COLIBR{\'I} confirmation of rebrightening of AT2025ulz}",
      journal = {GRB Coordinates Network},
         year = 2025,
        month = aug,
       volume = {41518},
        pages = {1},
       adsurl = {https://ui.adsabs.harvard.edu/abs/2025GCN.41518....1A}
}

@ARTICLE{2025GCN41507,
       author = {{Freeburn}, J. and {O'Connor}, B. and {Hall}, X.~J. and {Busmann}, M. and {Andreoni}, I. and {Palmese}, A. and {Gruen}, D. and {Hu}, L. and {Cabrera}, T. and {Kunnumkai}, K. and {Amsellem}, A.},
        title = "{LIGO/Virgo/KAGRA S250818k: Rebrightening detected with Gemini/GMOS}",
      journal = {GRB Coordinates Network},
         year = 2025,
        month = aug,
       volume = {41507},
        pages = {1},
       adsurl = {https://ui.adsabs.harvard.edu/abs/2025GCN.41507....1F}
}

@ARTICLE{2025GCN41506,
       author = {{Troja}, Eleonora and {O'Connor}, Brendan and {Becerra}, Rosa L.},
        title = "{LIGO/Virgo/KAGRA S250818k: HST nIR detection of AT 2025ulz}",
      journal = {GRB Coordinates Network},
         year = 2025,
        month = aug,
       volume = {41506},
        pages = {1},
       adsurl = {https://ui.adsabs.harvard.edu/abs/2025GCN.41506....1T}
}

@ARTICLE{2025GCN41500,
       author = {{Bruni}, G. and {Piro}, L. and {Gianfagna}, G. and {Thakur}, A.~L.},
        title = "{LIGO/Virgo/KAGRA S250818k: 3 GHz MeerKAT observations of AT2025ulz}",
      journal = {GRB Coordinates Network},
         year = 2025,
        month = aug,
       volume = {41500},
        pages = {1},
       adsurl = {https://ui.adsabs.harvard.edu/abs/2025GCN.41500....1B}
}

@ARTICLE{2025GCN41492,
       author = {{Malesani}, D.~B. and {Boye}, A. and {Izzo}, L. and {Leloudas}, G. and {An}, J. and {Liu}, X. and {Xu}, D. and {Fraser}, M. and {Brennan}, S.~J. and {Broe Bendsten}, J. and {Koch}, I.~A. and {Wagner}, S. Lund and {Magaard Knudsen}, J. and {Hein Pedersen}, J. and {Fynbo}, J.~P.~U. and {Holmberg Rasmussen}, R. and {Valeckas}, K. and {De Pasquale}, M.},
        title = "{LIGO/Virgo/KAGRA S250818k: NOT optical observations of AT2025ulz}",
      journal = {GRB Coordinates Network},
         year = 2025,
        month = aug,
       volume = {41492},
        pages = {1},
       adsurl = {https://ui.adsabs.harvard.edu/abs/2025GCN.41492....1M}
}

@ARTICLE{2025GCN41480,
       author = {{Perley}, Daniel A. and {Kasliwal}, Mansi M. and {Karambelkar}, Viraj and {Lundquist}, Michael and {Hall}, Xander and {Ahumada}, Tomas and {Sharma}, Kritti and {Rose}, Sam and {Ztf Collaboration} and {Growth Collaboration}},
        title = "{LIGO/Virgo/KAGRA S250818k: Continued fading and reddening of AT2025ulz from Keck/LRIS imaging observations}",
      journal = {GRB Coordinates Network},
         year = 2025,
        month = aug,
       volume = {41480},
        pages = {1},
       adsurl = {https://ui.adsabs.harvard.edu/abs/2025GCN.41480....1P}
}

@ARTICLE{2025GCN41476,
       author = {{Banerjee}, Smaranika and {Botticella}, Maria-Teresa and {Brennan}, Se{\'a}n J. and {Cappellaro}, Enrico and {Chen}, Ting-Wan and {D'Avanzo}, Paolo and {de Pasquale}, Max and {Eyles-Ferris}, Rob A.~J. and {Fraser}, Morgan and {Gillanders}, James H. and {Gompertz}, Ben and {Habeeb}, Nusrin and {Izzo}, Luca and {Jonker}, Peter G. and {Levan}, Andrew J. and {Bj{\o}rn Malesani}, Daniele and {Martin-Carrillo}, Antonio and {Oates}, Sam and {Nicholl}, Matt and {Rossi}, Andrea and {Sharan Salafia}, Om and {Sarin}, Nikhil and {Schulze}, Steve and {Smartt}, Stephen J. and {Steeghs}, Danny and {Tanvir}, Nial R. and {Thakur}, Aishwarya L. and {Engrave Collaboration}},
        title = "{LIGO/Virgo/KAGRA S250818k: ENGRAVE observations of AT 2025ulz}",
      journal = {GRB Coordinates Network},
         year = 2025,
        month = aug,
       volume = {41476},
        pages = {1},
       adsurl = {https://ui.adsabs.harvard.edu/abs/2025GCN.41476....1B}
}

@ARTICLE{2025GCN41454,
       author = {{Gillanders}, J.~H. and {Huber}, M.~E. and {Chambers}, K.~C. and {Smartt}, S.~J. and {Smith}, K.~W. and {Srivastav}, S. and {Stoppa}, F. and {Stevance}, H. and {Tweddle}, J. and {Nicholl}, M. and {Young}, D.~R. and {Aamer}, A. and {Angus}, C.~R. and {Fulton}, M.~D. and {Magill}, D. and {McCollum}, M. and {Moore}, T. and {Sim}, S. and {Weston}, J. and {Sheng}, X. and {Chen}, T. -W. and {Shingles}, L. and {Ramsden}, P. and {Schultz}, A.~S.~B. and {de Boer}, T. and {Fairlamb}, J. and {Lin}, C.~C. and {Lowe}, T. and {Magnier}, E. and {Minguez}, P. and {Paek}, G. and {Smith}, I.~A. and {Wainscoat}, R.~J. and {Rest}, A. and {Stubbs}, C.},
        title = "{LIGO/Virgo/KAGRA S250818k: Pan-STARRS grizy-band imaging and photometry of AT2025ulz}",
      journal = {GRB Coordinates Network},
         year = 2025,
        month = aug,
       volume = {41454},
        pages = {1},
       adsurl = {https://ui.adsabs.harvard.edu/abs/2025GCN.41454....1G}
}

@ARTICLE{2025GCN41453,
       author = {{Hall}, Xander J. and {Stein}, Robert and {O'Connor}, Brendan and {Palmese}, Antonella},
        title = "{LIGO/Virgo/KAGRA S250818k: Swift observations of AT 2025ulz}",
      journal = {GRB Coordinates Network},
         year = 2025,
        month = aug,
       volume = {41453},
        pages = {1},
       adsurl = {https://ui.adsabs.harvard.edu/abs/2025GCN.41453....1H}
}

@ARTICLE{2025GCN41452,
       author = {{O'Connor}, B. and {Freeburn}, J. and {Hall}, X.~J. and {Busmann}, Malte and {Andreoni}, I. and {Palmese}, A. and {Gruen}, D. and {Hu}, L. and {Cabrera}, T. and {Kunnumkai}, K. and {Amsellem}, A.},
        title = "{LIGO/Virgo/KAGRA S250818k: Multi-band Gemini GMOS Detections}",
      journal = {GRB Coordinates Network},
         year = 2025,
        month = aug,
       volume = {41452},
        pages = {1},
       adsurl = {https://ui.adsabs.harvard.edu/abs/2025GCN.41452....1O}
}

@ARTICLE{2026arXiv260328741S,
       author = {{Stoppa}, Fiorenzo and {Smartt}, Stephen J.},
        title = "{SNID-SAGE: A Modern Framework for Interactive Supernova Classification and Spectral Analysis}",
      journal = {arXiv e-prints},
         year = 2026,
        month = mar,
          eid = {arXiv:2603.28741},
        pages = {arXiv:2603.28741},
archivePrefix = {arXiv},
       eprint = {2603.28741},
 primaryClass = {astro-ph.IM},
       adsurl = {https://ui.adsabs.harvard.edu/abs/2026arXiv260328741S}
}

@ARTICLE{2025GCN41440,
       author = {{LVK Collaboration}},
        title = "{LIGO/Virgo/KAGRA S250818k: Updated Sky localization and EM Bright Classification}",
      journal = {GRB Coordinates Network},
         year = 2025,
        month = aug,
       volume = {41440},
        pages = {1},
       adsurl = {https://ui.adsabs.harvard.edu/abs/2025GCN.41440....1L}
}

@ARTICLE{2025GCN41437,
       author = {{LVK Collaboration}},
        title = "{LIGO/Virgo/KAGRA S250818k: Properties of the low-significance GW compact binary merger candidate potentially associated with AT 2025ulz}",
      journal = {GRB Coordinates Network},
         year = 2025,
        month = aug,
       volume = {41437},
        pages = {1},
       adsurl = {https://ui.adsabs.harvard.edu/abs/2025GCN.41437....1L}
}

@ARTICLE{2025GCN41436,
       author = {{Karambelkar}, Viraj and {Kasliwal}, Mansi M. and {Hall}, Xander J. and {Ztf Collaboration} and {Growth Collaboration}},
        title = "{LIGO/Virgo/KAGRA S250818k: Keck I LRIS spectroscopy of ZTF25abjmnps (AT2025ulz)}",
      journal = {GRB Coordinates Network},
         year = 2025,
        month = aug,
       volume = {41436},
        pages = {1},
       adsurl = {https://ui.adsabs.harvard.edu/abs/2025GCN.41436....1K}
}

@ARTICLE{2025GCN41414,
       author = {{Stein}, Robert and {Ahumada}, Tom{\'a}s and {Kasliwal}, Mansi and {Du Laz}, Theophile and {Pathak}, Utkarsh and {Swain}, Vishwajeet and {Salgundi}, Anirudh and {Bhalerao}, Varun and {Hall}, Xander J. and {Ztf Collaboration} and {Growth Collaboration}},
        title = "{LIGO/Virgo/KAGRA S250818k: Candidates from the Zwicky Transient Facility}",
      journal = {GRB Coordinates Network},
         year = 2025,
        month = aug,
       volume = {41414},
        pages = {1},
       adsurl = {https://ui.adsabs.harvard.edu/abs/2025GCN.41414....1S}
}

@article{Nayana2022,
       author = {{Nayana}, A.~J. and {Chandra}, Poonam and {Krishna}, Anoop and {Anupama}, G.~C.},
        title = "{Radio Evolution of a Type IIb Supernova SN 2016gkg}",
      journal = {\apj},
         year = 2022,
        month = aug,
       volume = {934},
       number = {2},
          eid = {186},
        pages = {186},
          doi = {10.3847/1538-4357/ac7c1e},
archivePrefix = {arXiv},
       eprint = {2206.12103},
 primaryClass = {astro-ph.HE},
       adsurl = {https://ui.adsabs.harvard.edu/abs/2022ApJ...934..186N}
}

@ARTICLE{Weiler2007,
       author = {{Weiler}, Kurt W. and {Williams}, Christopher L. and {Panagia}, Nino and {Stockdale}, Christopher J. and {Kelley}, Matthew T. and {Sramek}, Richard A. and {Van Dyk}, Schuyler D. and {Marcaide}, J.~M.},
        title = "{Long-Term Radio Monitoring of SN 1993J}",
      journal = {\apj},
         year = 2007,
        month = dec,
       volume = {671},
       number = {2},
        pages = {1959-1980},
          doi = {10.1086/523258},
archivePrefix = {arXiv},
       eprint = {0709.1136},
 primaryClass = {astro-ph},
       adsurl = {https://ui.adsabs.harvard.edu/abs/2007ApJ...671.1959W}
}

@article{Soderberg2006,
doi = {10.1086/507571},
url = {https://doi.org/10.1086/507571},
year = {2006},
month = {nov},
publisher = {},
volume = {651},
number = {2},
pages = {1005},
author = {Soderberg, A. M. and Chevalier, R. A. and Kulkarni, S. R. and Frail, D. A.},
title = {The Radio and X-Ray Luminous SN 2003bg and the Circumstellar Density Variations around Radio Supernovae},
journal = {The Astrophysical Journal}
}

@article{Ryder2004,
    author = {Ryder, Stuart D. and Sadler, Elaine M. and Subrahmanyan, Ravi and Weiler, Kurt W. and Panagia, Nino and Stockdale, Christopher},
    title = {Modulations in the radio light curve of the Type IIb supernova 2001ig: evidence for a Wolf–Rayet binary progenitor?},
    journal = {Monthly Notices of the Royal Astronomical Society},
    volume = {349},
    number = {3},
    pages = {1093-1100},
    year = {2004},
    month = {04},
    issn = {0035-8711},
    doi = {10.1111/j.1365-2966.2004.07589.x},
    url = {https://doi.org/10.1111/j.1365-2966.2004.07589.x},
    eprint = {https://academic.oup.com/mnras/article-pdf/349/3/1093/18647605/349-3-1093.pdf},
}

@article{Bietenholz2021,
doi = {10.3847/1538-4357/abccd9},
url = {https://doi.org/10.3847/1538-4357/abccd9},
year = {2021},
month = {feb},
publisher = {The American Astronomical Society},
volume = {908},
number = {1},
pages = {75},
author = {Bietenholz, M. F. and Bartel, N. and Argo, M. and Dua, R. and Ryder, S. and Soderberg, A.},
title = {The Radio Luminosity-risetime Function of Core-collapse Supernovae},
journal = {The Astrophysical Journal}
}

@article{Kamble2016,
doi = {10.3847/0004-637X/818/2/111},
url = {https://doi.org/10.3847/0004-637X/818/2/111},
year = {2016},
month = {feb},
publisher = {The American Astronomical Society},
volume = {818},
number = {2},
pages = {111},
author = {Kamble, Atish and Margutti, Raffaella and Soderberg, Alicia M. and Chakraborti, Sayan and Fransson, Claes and Chevalier, Roger and Powell, Diana and Milisavljevic, Dan and Parrent, Jerod and Bietenholz, Michael},
title = {PROGENITORS OF TYPE IIb SUPERNOVAE IN THE LIGHT OF RADIO AND X-RAYS FROM SN 2013df},
journal = {The Astrophysical Journal}
}

@article{Krauss2012,
doi = {10.1088/2041-8205/750/2/L40},
url = {https://doi.org/10.1088/2041-8205/750/2/L40},
year = {2012},
month = {apr},
publisher = {The American Astronomical Society},
volume = {750},
number = {2},
pages = {L40},
author = {Krauss, M. I. and Soderberg, A. M. and Chomiuk, L. and Zauderer, B. A. and Brunthaler, A. and Bietenholz, M. F. and Chevalier, R. A. and Fransson, C. and Rupen, M.},
title = {EXPANDED VERY LARGE ARRAY OBSERVATIONS OF THE RADIO EVOLUTION OF SN 2011dh},
journal = {The Astrophysical Journal Letters}
}

@article{Bufano2011,
    author = {Bufano, F. and Pignata, G. and Bersten, M. and Mazzali, P. A. and Ryder, S. D. and Margutti, R. and Milisavljevic, D. and Morelli, L. and Benetti, S. and Cappellaro, E. and Gonzalez-Gaitan, S. and Romero-Cañizales, C. and Stritzinger, M. and Walker, E. S. and Anderson, J. P. and Contreras, C. and de Jaeger, T. and Förster, F. and Gutierrez, C. and Hamuy, M. and Hsiao, E. and Morrell, N. and Olivares E., F. and Paillas, E. and Parker, S. and Pian, E. and Pickering, T. E. and Sanders, N. and Stockdale, C. and Turatto, M. and Valenti, S. and Fesen, R. A. and Maza, J. and Nomoto, K. and Phillips, M. M. and Soderberg, A.},
    title = {SN 2011hs: a fast and faint Type IIb supernova from a supergiant progenitor},
    journal = {Monthly Notices of the Royal Astronomical Society},
    volume = {439},
    number = {2},
    pages = {1807-1828},
    year = {2014},
    month = {02},
    issn = {0035-8711},
    doi = {10.1093/mnras/stu065},
    url = {https://doi.org/10.1093/mnras/stu065},
    eprint = {https://academic.oup.com/mnras/article-pdf/439/2/1807/18471549/stu065.pdf},
}

@article{Stockdale2007,
doi = {10.1086/522584},
url = {https://doi.org/10.1086/522584},
year = {2007},
month = {dec},
publisher = {},
volume = {671},
number = {1},
pages = {689},
author = {Stockdale, Christopher J. and Williams, Christopher L. and Weiler, Kurt W. and Panagia, Nino and Sramek, Richard A. and Van Dyk, Schuyler D. and Kelley, Matthew T.},
title = {The Radio Evolution of SN 2001gd},
journal = {The Astrophysical Journal}
}

@article{Romero-Canizales2014,
    author = {Romero-Cañizales, C. and Herrero-Illana, R. and Pérez-Torres, M. A. and Alberdi, A. and Kankare, E. and Bauer, F. E. and Ryder, S. D. and Mattila, S. and Conway, J. E. and Beswick, R. J. and Muxlow, T. W. B.},
    title = {The nature of supernovae 2010O and 2010P in Arp 299 – II. Radio emission},
    journal = {Monthly Notices of the Royal Astronomical Society},
    volume = {440},
    number = {2},
    pages = {1067-1079},
    year = {2014},
    month = {03},
    issn = {0035-8711},
    doi = {10.1093/mnras/stu430},
    url = {https://doi.org/10.1093/mnras/stu430},
    eprint = {https://academic.oup.com/mnras/article-pdf/440/2/1067/18503040/stu430.pdf},
}

@software{2020ascl.soft09003H,
       author = {{Heywood}, Ian},
        title = "{oxkat: Semi-automated imaging of MeerKAT observations}",
 howpublished = {Astrophysics Source Code Library, record ascl:2009.003},
         year = 2020,
        month = sep,
          eid = {ascl:2009.003},
archivePrefix = {ascl},
       eprint = {2009.003},
       adsurl = {https://ui.adsabs.harvard.edu/abs/2020ascl.soft09003H}
}

@ARTICLE{wsclean1,
       author = {{Offringa}, A.~R. and {McKinley}, B. and {Hurley-Walker}, N. and {Briggs}, F.~H. and {Wayth}, R.~B. and {Kaplan}, D.~L. and {Bell}, M.~E. and {Feng}, L. and {Neben}, A.~R. and {Hughes}, J.~D. and {Rhee}, J. and {Murphy}, T. and {Bhat}, N.~D.~R. and {Bernardi}, G. and {Bowman}, J.~D. and {Cappallo}, R.~J. and {Corey}, B.~E. and {Deshpande}, A.~A. and {Emrich}, D. and {Ewall-Wice}, A. and {Gaensler}, B.~M. and {Goeke}, R. and {Greenhill}, L.~J. and {Hazelton}, B.~J. and {Hindson}, L. and {Johnston-Hollitt}, M. and {Jacobs}, D.~C. and {Kasper}, J.~C. and {Kratzenberg}, E. and {Lenc}, E. and {Lonsdale}, C.~J. and {Lynch}, M.~J. and {McWhirter}, S.~R. and {Mitchell}, D.~A. and {Morales}, M.~F. and {Morgan}, E. and {Kudryavtseva}, N. and {Oberoi}, D. and {Ord}, S.~M. and {Pindor}, B. and {Procopio}, P. and {Prabu}, T. and {Riding}, J. and {Roshi}, D.~A. and {Shankar}, N. Udaya and {Srivani}, K.~S. and {Subrahmanyan}, R. and {Tingay}, S.~J. and {Waterson}, M. and {Webster}, R.~L. and {Whitney}, A.~R. and {Williams}, A. and {Williams}, C.~L.},
        title = "{WSCLEAN: an implementation of a fast, generic wide-field imager for radio astronomy}",
      journal = {\mnras},
         year = 2014,
        month = oct,
       volume = {444},
       number = {1},
        pages = {606-619},
          doi = {10.1093/mnras/stu1368},
archivePrefix = {arXiv},
       eprint = {1407.1943},
 primaryClass = {astro-ph.IM},
       adsurl = {https://ui.adsabs.harvard.edu/abs/2014MNRAS.444..606O}
}

@ARTICLE{wsclean2,
       author = {{Offringa}, A.~R. and {Smirnov}, O.},
        title = "{An optimized algorithm for multiscale wideband deconvolution of radio astronomical images}",
      journal = {\mnras},
         year = 2017,
        month = oct,
       volume = {471},
       number = {1},
        pages = {301-316},
          doi = {10.1093/mnras/stx1547},
archivePrefix = {arXiv},
       eprint = {1706.06786},
 primaryClass = {astro-ph.IM},
       adsurl = {https://ui.adsabs.harvard.edu/abs/2017MNRAS.471..301O}
}

@INPROCEEDINGS{casa,
       author = {{McMullin}, J.~P. and {Waters}, B. and {Schiebel}, D. and {Young}, W. and {Golap}, K.},
        title = "{CASA Architecture and Applications}",
    booktitle = {Astronomical Data Analysis Software and Systems XVI},
         year = 2007,
       editor = {{Shaw}, R.~A. and {Hill}, F. and {Bell}, D.~J.},
       series = {Astronomical Society of the Pacific Conference Series},
       volume = {376},
        month = oct,
        pages = {127},
       adsurl = {https://ui.adsabs.harvard.edu/abs/2007ASPC..376..127M}
}

@article{Gianfagna2024,
    author = {Gianfagna, Giulia and Piro, Luigi and Pannarale, Francesco and Van Eerten, Hendrik and Ricci, Fulvio and Ryan, Geoffrey},
    title = {Potential biases and prospects for the Hubble constant estimation via electromagnetic and gravitational-wave joint analyses},
    journal = {Monthly Notices of the Royal Astronomical Society},
    volume = {528},
    number = {2},
    pages = {2600-2613},
    year = {2024},
    month = {01},
    issn = {0035-8711},
    doi = {10.1093/mnras/stae198},
    url = {https://doi.org/10.1093/mnras/stae198},
    eprint = {https://academic.oup.com/mnras/article-pdf/528/2/2600/56485474/stae198.pdf},
}

@ARTICLE{Ryan2024,
       author = {{Ryan}, Geoffrey and {van Eerten}, Hendrik and {Troja}, Eleonora and {Piro}, Luigi and {O'Connor}, Brendan and {Ricci}, Roberto},
        title = "{Modeling of Long-term Afterglow Counterparts to Gravitational Wave Events: The Full View of GRB 170817A}",
      journal = {\apj},
         year = 2024,
        month = nov,
       volume = {975},
       number = {1},
          eid = {131},
        pages = {131},
          doi = {10.3847/1538-4357/ad6a14},
archivePrefix = {arXiv},
       eprint = {2310.02328},
 primaryClass = {astro-ph.HE},
       adsurl = {https://ui.adsabs.harvard.edu/abs/2024ApJ...975..131R}
}

@article{Troja2019,
    author = {Troja, E and van Eerten, H and Ryan, G and Ricci, R and Burgess, J M and Wieringa, M H and Piro, L and Cenko, S B and Sakamoto, T},
    title = {A year in the life of GW 170817: the rise and fall of a structured jet from a binary neutron star merger},
    journal = {Monthly Notices of the Royal Astronomical Society},
    volume = {489},
    number = {2},
    pages = {1919-1926},
    year = {2019},
    month = {08},
    issn = {0035-8711},
    doi = {10.1093/mnras/stz2248},
    url = {https://doi.org/10.1093/mnras/stz2248},
    eprint = {https://academic.oup.com/mnras/article-pdf/489/2/1919/30033101/stz2248.pdf},
}

@article{Hajela2022,
doi = {10.3847/2041-8213/ac504a},
url = {https://doi.org/10.3847/2041-8213/ac504a},
year = {2022},
month = {mar},
publisher = {The American Astronomical Society},
volume = {927},
number = {1},
pages = {L17},
author = {Hajela, A. and Margutti, R. and Bright, J. S. and Alexander, K. D. and Metzger, B. D. and Nedora, V. and Kathirgamaraju, A. and Margalit, B. and Radice, D. and Guidorzi, C. and Berger, E. and MacFadyen, A. and Giannios, D. and Chornock, R. and Heywood, I. and Sironi, L. and Gottlieb, O. and Coppejans, D. and Laskar, T. and Cendes, Y. and Duran, R. Barniol and Eftekhari, T. and Fong, W. and McDowell, A. and Nicholl, M. and Xie, X. and Zrake, J. and Bernuzzi, S. and Broekgaarden, F. S. and Kilpatrick, C. D. and Terreran, G. and Villar, V. A. and Blanchard, P. K. and Gomez, S. and Hosseinzadeh, G. and Matthews, D. J. and Rastinejad, J. C.},
title = {Evidence for X-Ray Emission in Excess to the Jet-afterglow Decay 3.5 yr after the Binary Neutron Star Merger GW 170817: A New Emission Component},
journal = {The Astrophysical Journal Letters}
}

@article{Oates2025,
    author = {Oates, Samantha},
    title = {False positives in gravitational wave campaigns: the electromagnetic perspective},
    journal = {Phil. Trans. R. Soc. A.},
    year = {2025},
    volume = {383},
    pages = {20240120}
}

@ARTICLE{Fulton25,
       author = {{Fulton}, M.~D. and {Smartt}, S.~J. and {Huber}, M.~E. and {Smith}, K.~W. and {Chambers}, K.~C. and {Nicholl}, M. and {Srivastav}, S. and {Young}, D.~R. and {Magnier}, E.~A. and {Lin}, C.-C. and {Minguez}, P. and {de Boer}, T. and {Lowe}, T. and {Wainscoat}, R.},
        title = "{Results from the Pan-STARRS search for kilonovae: contamination by massive stellar outbursts}",
      journal = {\mnras},
         year = 2025,
        month = sep,
       volume = {542},
       number = {2},
        pages = {541-559},
          doi = {10.1093/mnras/staf1165},
archivePrefix = {arXiv},
       eprint = {2506.07082},
 primaryClass = {astro-ph.HE},
       adsurl = {https://ui.adsabs.harvard.edu/abs/2025MNRAS.542..541F}
}

@ARTICLE{NakarPiran2011,
       author = {{Nakar}, Ehud and {Piran}, Tsvi},
        title = "{Detectable radio flares following gravitational waves from mergers of binary neutron stars}",
      journal = {\nat},
         year = 2011,
        month = oct,
       volume = {478},
       number = {7367},
        pages = {82-84},
          doi = {10.1038/nature10365},
archivePrefix = {arXiv},
       eprint = {1102.1020},
 primaryClass = {astro-ph.HE},
       adsurl = {https://ui.adsabs.harvard.edu/abs/2011Natur.478...82N}
}

@ARTICLE{Bertin1996,
       author = {{Bertin}, E. and {Arnouts}, S.},
        title = "{SExtractor: Software for source extraction.}",
      journal = {\aaps},
         year = 1996,
        month = jun,
       volume = {117},
        pages = {393-404},
          doi = {10.1051/aas:1996164},
       adsurl = {https://ui.adsabs.harvard.edu/abs/1996A&AS..117..393B}
}

@INPROCEEDINGS{2006ASPC..351..112B,
       author = {{Bertin}, E.},
        title = "{Automatic Astrometric and Photometric Calibration with SCAMP}",
    booktitle = {Astronomical Data Analysis Software and Systems XV},
         year = 2006,
       editor = {{Gabriel}, C. and {Arviset}, C. and {Ponz}, D. and {Enrique}, S.},
       series = {Astronomical Society of the Pacific Conference Series},
       volume = {351},
        month = jul,
        pages = {112},
       adsurl = {https://ui.adsabs.harvard.edu/abs/2006ASPC..351..112B}
}

@software{Becker2015,
       author = {{Becker}, Andrew},
        title = "{HOTPANTS: High Order Transform of PSF ANd Template Subtraction}",
 howpublished = {Astrophysics Source Code Library, record ascl:1504.004},
         year = 2015,
        month = apr,
          eid = {ascl:1504.004},
archivePrefix = {ascl},
       eprint = {1504.004},
       adsurl = {https://ui.adsabs.harvard.edu/abs/2015ascl.soft04004B}
}

@ARTICLE{Dey2019,
       author = {{Dey}, Arjun and {Schlegel}, David J. and {Lang}, Dustin and {Blum}, Robert and {Burleigh}, Kaylan and {Fan}, Xiaohui and {Findlay}, Joseph R. and {Finkbeiner}, Doug and {Herrera}, David and {Juneau}, St{\'e}phanie and {Landriau}, Martin and {Levi}, Michael and {McGreer}, Ian and {Meisner}, Aaron and {Myers}, Adam D. and {Moustakas}, John and {Nugent}, Peter and {Patej}, Anna and {Schlafly}, Edward F. and {Walker}, Alistair R. and {Valdes}, Francisco and {Weaver}, Benjamin A. and {Y{\`e}che}, Christophe and {Zou}, Hu and {Zhou}, Xu and {Abareshi}, Behzad and {Abbott}, T.~M.~C. and {Abolfathi}, Bela and {Aguilera}, C. and {Alam}, Shadab and {Allen}, Lori and {Alvarez}, A. and {Annis}, James and {Ansarinejad}, Behzad and {Aubert}, Marie and {Beechert}, Jacqueline and {Bell}, Eric F. and {BenZvi}, Segev Y. and {Beutler}, Florian and {Bielby}, Richard M. and {Bolton}, Adam S. and {Brice{\~n}o}, C{\'e}sar and {Buckley-Geer}, Elizabeth J. and {Butler}, Karen and {Calamida}, Annalisa and {Carlberg}, Raymond G. and {Carter}, Paul and {Casas}, Ricard and {Castander}, Francisco J. and {Choi}, Yumi and {Comparat}, Johan and {Cukanovaite}, Elena and {Delubac}, Timoth{\'e}e and {DeVries}, Kaitlin and {Dey}, Sharmila and {Dhungana}, Govinda and {Dickinson}, Mark and {Ding}, Zhejie and {Donaldson}, John B. and {Duan}, Yutong and {Duckworth}, Christopher J. and {Eftekharzadeh}, Sarah and {Eisenstein}, Daniel J. and {Etourneau}, Thomas and {Fagrelius}, Parker A. and {Farihi}, Jay and {Fitzpatrick}, Mike and {Font-Ribera}, Andreu and {Fulmer}, Leah and {G{\"a}nsicke}, Boris T. and {Gaztanaga}, Enrique and {George}, Koshy and {Gerdes}, David W. and {Gontcho}, Satya Gontcho A. and {Gorgoni}, Claudio and {Green}, Gregory and {Guy}, Julien and {Harmer}, Diane and {Hernandez}, M. and {Honscheid}, Klaus and {Huang}, Lijuan Wendy and {James}, David J. and {Jannuzi}, Buell T. and {Jiang}, Linhua and {Joyce}, Richard and {Karcher}, Armin and {Karkar}, Sonia and {Kehoe}, Robert and {Kneib}, Jean-Paul and {Kueter-Young}, Andrea and {Lan}, Ting-Wen and {Lauer}, Tod R. and {Le Guillou}, Laurent and {Le Van Suu}, Auguste and {Lee}, Jae Hyeon and {Lesser}, Michael and {Perreault Levasseur}, Laurence and {Li}, Ting S. and {Mann}, Justin L. and {Marshall}, Robert and {Mart{\'\i}nez-V{\'a}zquez}, C.~E. and {Martini}, Paul and {du Mas des Bourboux}, H{\'e}lion and {McManus}, Sean and {Meier}, Tobias Gabriel and {M{\'e}nard}, Brice and {Metcalfe}, Nigel and {Mu{\~n}oz-Guti{\'e}rrez}, Andrea and {Najita}, Joan and {Napier}, Kevin and {Narayan}, Gautham and {Newman}, Jeffrey A. and {Nie}, Jundan and {Nord}, Brian and {Norman}, Dara J. and {Olsen}, Knut A.~G. and {Paat}, Anthony and {Palanque-Delabrouille}, Nathalie and {Peng}, Xiyan and {Poppett}, Claire L. and {Poremba}, Megan R. and {Prakash}, Abhishek and {Rabinowitz}, David and {Raichoor}, Anand and {Rezaie}, Mehdi and {Robertson}, A.~N. and {Roe}, Natalie A. and {Ross}, Ashley J. and {Ross}, Nicholas P. and {Rudnick}, Gregory and {Safonova}, Sasha and {Saha}, Abhijit and {S{\'a}nchez}, F. Javier and {Savary}, Elodie and {Schweiker}, Heidi and {Scott}, Adam and {Seo}, Hee-Jong and {Shan}, Huanyuan and {Silva}, David R. and {Slepian}, Zachary and {Soto}, Christian and {Sprayberry}, David and {Staten}, Ryan and {Stillman}, Coley M. and {Stupak}, Robert J. and {Summers}, David L. and {Sien Tie}, Suk and {Tirado}, H. and {Vargas-Maga{\~n}a}, Mariana and {Vivas}, A. Katherina and {Wechsler}, Risa H. and {Williams}, Doug and {Yang}, Jinyi and {Yang}, Qian and {Yapici}, Tolga and {Zaritsky}, Dennis and {Zenteno}, A. and {Zhang}, Kai and {Zhang}, Tianmeng and {Zhou}, Rongpu and {Zhou}, Zhimin},
        title = "{Overview of the DESI Legacy Imaging Surveys}",
      journal = {\aj},
         year = 2019,
        month = may,
       volume = {157},
       number = {5},
          eid = {168},
        pages = {168},
          doi = {10.3847/1538-3881/ab089d},
archivePrefix = {arXiv},
       eprint = {1804.08657},
 primaryClass = {astro-ph.IM},
       adsurl = {https://ui.adsabs.harvard.edu/abs/2019AJ....157..168D}
}

@software{Bradley2024,
       author = {{Bradley}, Larry and {Sip{\H{o}}cz}, Brigitta and {Robitaille}, Thomas and {Tollerud}, Erik and {Vin{\'\i}cius}, Z{\'e} and {Deil}, Christoph and {Barbary}, Kyle and {Wilson}, Tom J and {Busko}, Ivo and {Donath}, Axel and {G{\"u}nther}, Hans Moritz and {Cara}, Mihai and {Lim}, P.~L. and {Me{\ss}linger}, Sebastian and {Conseil}, Simon and {Burnett}, Zach and {Bostroem}, Azalee and {Droettboom}, Michael and {Bray}, E.~M. and {Andersen Bratholm}, Lars and {Ginsburg}, Adam and {Jamieson}, William and {Barentsen}, Geert and {Craig}, Matt and {Morris}, Brett M. and {Perrin}, Marshall and {Rathi}, Shivangee and {Pascual}, Sergio and {Georgiev}, Iskren Y.},
        title = "{astropy/photutils: 2.0.2}",
         year = 2024,
        month = oct,
          eid = {10.5281/zenodo.13989456},
          doi = {10.5281/zenodo.13989456},
      version = {2.0.2},
    publisher = {Zenodo},
       adsurl = {https://ui.adsabs.harvard.edu/abs/2024zndo..13989456B}
}

@ARTICLE{Planck2020,
       author = {{Planck Collaboration} and {Aghanim}, N. and {Akrami}, Y. and {Ashdown}, M. and {Aumont}, J. and {Baccigalupi}, C. and {Ballardini}, M. and {Banday}, A.~J. and {Barreiro}, R.~B. and {Bartolo}, N. and {Basak}, S. and others},
        title = "{Planck 2018 results. VI. Cosmological parameters}",
      journal = {\aap},
         year = 2020,
        month = sep,
       volume = {641},
          eid = {A6},
        pages = {A6},
          doi = {10.1051/0004-6361/201833910},
archivePrefix = {arXiv},
       eprint = {1807.06209},
 primaryClass = {astro-ph.CO},
       adsurl = {https://ui.adsabs.harvard.edu/abs/2020A&A...641A...6P}
}

@ARTICLE{Greco2022,
       author = {{Greco}, G. and {Punturo}, M. and {Allen}, M. and {Nebot}, A. and {Fernique}, P. and {Baumann}, M. and {Pineau}, F.-X. and {Boch}, T. and {Derriere}, S. and {Branchesi}, M. and {Bawaj}, M. and {Vocca}, H.},
        title = "{Multi Order Coverage data structure to plan multi-messenger observations}",
      journal = {Astronomy and Computing},
         year = 2022,
        month = apr,
       volume = {39},
          eid = {100547},
        pages = {100547},
          doi = {10.1016/j.ascom.2022.100547},
archivePrefix = {arXiv},
       eprint = {2201.05191},
 primaryClass = {astro-ph.IM},
       adsurl = {https://ui.adsabs.harvard.edu/abs/2022A&C....3900547G}
}

@ARTICLE{Singer2016_bayestar,
       author = {{Singer}, Leo P. and {Price}, Larry R.},
        title = "{Rapid Bayesian position reconstruction for gravitational-wave transients}",
      journal = {\prd},
         year = 2016,
        month = jan,
       volume = {93},
       number = {2},
          eid = {024013},
        pages = {024013},
          doi = {10.1103/PhysRevD.93.024013},
archivePrefix = {arXiv},
       eprint = {1508.03634},
 primaryClass = {gr-qc},
       adsurl = {https://ui.adsabs.harvard.edu/abs/2016PhRvD..93b4013S}
}

@ARTICLE{Singer2016_3D,
       author = {{Singer}, Leo P. and {Chen}, Hsin-Yu and {Holz}, Daniel E. and {Farr}, Will M. and {Price}, Larry R. and {Raymond}, Vivien and {Cenko}, S. Bradley and {Gehrels}, Neil and {Cannizzo}, John and {Kasliwal}, Mansi M. and {Nissanke}, Samaya and {Coughlin}, Michael and {Farr}, Ben and {Urban}, Alex L. and {Vitale}, Salvatore and {Veitch}, John and {Graff}, Philip and {Berry}, Christopher P.~L. and {Mohapatra}, Satya and {Mandel}, Ilya},
        title = "{Supplement: {\textquotedblleft}Going the Distance: Mapping Host Galaxies of LIGO and Virgo Sources in Three Dimensions Using Local Cosmography and Targeted Follow-up{\textquotedblright} (2016, ApJL, 829, L15)}",
      journal = {\apjs},
         year = 2016,
        month = sep,
       volume = {226},
       number = {1},
          eid = {10},
        pages = {10},
          doi = {10.3847/0067-0049/226/1/10},
archivePrefix = {arXiv},
       eprint = {1605.04242},
 primaryClass = {astro-ph.IM},
       adsurl = {https://ui.adsabs.harvard.edu/abs/2016ApJS..226...10S}
}

@ARTICLE{Schlegel1998,
       author = {{Schlegel}, David J. and {Finkbeiner}, Douglas P. and {Davis}, Marc},
        title = "{Maps of Dust Infrared Emission for Use in Estimation of Reddening and Cosmic Microwave Background Radiation Foregrounds}",
      journal = {\apj},
         year = 1998,
        month = jun,
       volume = {500},
       number = {2},
        pages = {525-553},
          doi = {10.1086/305772},
archivePrefix = {arXiv},
       eprint = {astro-ph/9710327},
 primaryClass = {astro-ph},
       adsurl = {https://ui.adsabs.harvard.edu/abs/1998ApJ...500..525S}
}

@ARTICLE{Ashton2019,
       author = {{Ashton}, Gregory and {H{\"u}bner}, Moritz and {Lasky}, Paul D. and {Talbot}, Colm and et al.},
        title = "{BILBY: A User-friendly Bayesian Inference Library for Gravitational-wave Astronomy}",
      journal = {\apjs},
         year = 2019,
        month = apr,
       volume = {241},
       number = {2},
          eid = {27},
        pages = {27},
          doi = {10.3847/1538-4365/ab06fc},
archivePrefix = {arXiv},
       eprint = {1811.02042},
 primaryClass = {astro-ph.IM},
       adsurl = {https://ui.adsabs.harvard.edu/abs/2019ApJS..241...27A}
}

@misc{Buchner2016,
       author = {{Buchner}, Johannes},
        title = "{PyMultiNest: Python interface for MultiNest}",
 howpublished = {Astrophysics Source Code Library, record ascl:1606.005},
         year = 2016,
        month = jun,
          eid = {ascl:1606.005},
       adsurl = {https://ui.adsabs.harvard.edu/abs/2016ascl.soft06005B}
}

@ARTICLE{Feroz2009,
       author = {{Feroz}, F. and {Hobson}, M.~P. and {Bridges}, M.},
        title = "{MULTINEST: an efficient and robust Bayesian inference tool for cosmology and particle physics}",
      journal = {\mnras},
         year = 2009,
        month = oct,
       volume = {398},
       number = {4},
        pages = {1601-1614},
          doi = {10.1111/j.1365-2966.2009.14548.x},
archivePrefix = {arXiv},
       eprint = {0809.3437},
 primaryClass = {astro-ph},
       adsurl = {https://ui.adsabs.harvard.edu/abs/2009MNRAS.398.1601F}
}

@ARTICLE{Sarin2024,
       author = {{Sarin}, Nikhil and {H{\"u}bner}, Moritz and {Omand}, Conor M.~B. and {Setzer}, Christian N. and et al.},
        title = "{REDBACK: a Bayesian inference software package for electromagnetic transients}",
      journal = {\mnras},
         year = 2024,
        month = jun,
       volume = {531},
       number = {1},
        pages = {1203-1227},
          doi = {10.1093/mnras/stae1238},
archivePrefix = {arXiv},
       eprint = {2308.12806},
 primaryClass = {astro-ph.HE},
       adsurl = {https://ui.adsabs.harvard.edu/abs/2024MNRAS.531.1203S}
}

@ARTICLE{Pinto2000,
       author = {{Pinto}, Philip A. and {Eastman}, Ronald G.},
        title = "{The Physics of Type IA Supernova Light Curves. I. Analytic Results and Time Dependence}",
      journal = {\apj},
         year = 2000,
        month = feb,
       volume = {530},
       number = {2},
        pages = {744-756},
          doi = {10.1086/308376},
       adsurl = {https://ui.adsabs.harvard.edu/abs/2000ApJ...530..744P}
}

@ARTICLE{Arnett1982,
       author = {{Arnett}, W.~D.},
        title = "{Type I supernovae. I - Analytic solutions for the early part of the light curve}",
      journal = {\apj},
         year = 1982,
        month = feb,
       volume = {253},
        pages = {785-797},
          doi = {10.1086/159681},
       adsurl = {https://ui.adsabs.harvard.edu/abs/1982ApJ...253..785A}
}

@ARTICLE{Sarin2024_caution,
       author = {{Sarin}, Nikhil and {Rosswog}, Stephan},
        title = "{Cautionary Tales on Heating-rate Prescriptions in Kilonovae}",
      journal = {\apjl},
         year = 2024,
        month = sep,
       volume = {973},
       number = {1},
          eid = {L24},
        pages = {L24},
          doi = {10.3847/2041-8213/ad739d},
archivePrefix = {arXiv},
       eprint = {2404.07271},
 primaryClass = {astro-ph.HE},
       adsurl = {https://ui.adsabs.harvard.edu/abs/2024ApJ...973L..24S}
}

@ARTICLE{Ayala25,
       author = {{Ayala}, Bastian and {Anderson}, Joseph P. and {Pignata}, G. and {F{\"o}rster}, Francisco and {Smartt}, S.~J. and {Rest}, A. and {Solar}, Mart{\'\i}n and {Erasmus}, Nicolas and {Dastidar}, Raya and {Ramirez}, Mauricio and {Pineda-Garc{\'\i}a}, Jonathan},
        title = "{Early light curve excess in Type IIb supernovae observed with ATLAS: Qualitative constraints on progenitor systems}",
      journal = {\aap},
         year = 2025,
        month = sep,
       volume = {701},
          eid = {A128},
        pages = {A128},
          doi = {10.1051/0004-6361/202554370},
archivePrefix = {arXiv},
       eprint = {2503.05909},
 primaryClass = {astro-ph.HE},
       adsurl = {https://ui.adsabs.harvard.edu/abs/2025A&A...701A.128A}
}

@ARTICLE{Pessi25,
       author = {{Pessi}, T. and {Desai}, D.~D. and {Prieto}, J.~L. and {Kochanek}, C.~S. and {Shappee}, B.~J. and {Anderson}, J.~P. and {Beacom}, J.~F. and {Dong}, S. and {Stanek}, K.~Z. and {Thompson}, T.~A.},
        title = "{Supernova rates and luminosity functions from ASAS-SN: II. 2014─2017 core-collapse supernovae and their subtypes}",
      journal = {\aap},
         year = 2025,
        month = nov,
       volume = {703},
          eid = {A34},
        pages = {A34},
          doi = {10.1051/0004-6361/202556799},
archivePrefix = {arXiv},
       eprint = {2508.10985},
 primaryClass = {astro-ph.HE},
       adsurl = {https://ui.adsabs.harvard.edu/abs/2025A&A...703A..34P}
}

@ARTICLE{Frohmaier21,
       author = {{Frohmaier}, C. and {Angus}, C.~R. and {Vincenzi}, M. and {Sullivan}, M. and {Smith}, M. and {Nugent}, P.~E. and {Cenko}, S.~B. and {Gal-Yam}, A. and {Kulkarni}, S.~R. and {Law}, N.~M. and {Quimby}, R.~M.},
        title = "{From core collapse to superluminous: the rates of massive stellar explosions from the Palomar Transient Factory}",
      journal = {\mnras},
         year = 2021,
        month = jan,
       volume = {500},
       number = {4},
        pages = {5142-5158},
          doi = {10.1093/mnras/staa3607},
archivePrefix = {arXiv},
       eprint = {2010.15270},
 primaryClass = {astro-ph.HE},
       adsurl = {https://ui.adsabs.harvard.edu/abs/2021MNRAS.500.5142F}
}

@ARTICLE{Ma25,
       author = {{Ma}, Xiaoran and {Wang}, Xiaofeng and {Mo}, Jun and {Andrew Howell}, D. and {Pellegrino}, Craig and {Zhang}, Jujia and {Wu}, Chengyuan and {Yan}, Shengyu and {Liu}, Dongdong and {Arcavi}, Iair and {Chen}, Zhihao and {Farah}, Joseph and {Padilla Gonzalez}, Estefania and {Guo}, Fangzhou and {Hiramatsu}, Daichi and {Li}, Gaici and {Lin}, Han and {Liu}, Jialian and {McCully}, Curtis and {Newsome}, Megan and {Sai}, Hanna and {Terreran}, Giacomo and {Xiang}, Danfeng and {Zhang}, Xinhan},
        title = "{Supernovae at distances <40 Mpc: II. Supernova rate in the local Universe}",
      journal = {\aap},
         year = 2025,
        month = jun,
       volume = {698},
          eid = {A306},
        pages = {A306},
          doi = {10.1051/0004-6361/202452685},
archivePrefix = {arXiv},
       eprint = {2504.04507},
 primaryClass = {astro-ph.HE},
       adsurl = {https://ui.adsabs.harvard.edu/abs/2025A&A...698A.306M}
}

@ARTICLE{Crawford25,
       author = {{Crawford}, Adrian and {Pritchard}, Tyler A. and {Modjaz}, Maryam and {Pellegrino}, Craig and {Kumar}, Sahana and {Baer-Way}, Raphael},
        title = "{Peaky Finders: Characterizing Double-peaked Type IIb Supernovae in Large-scale Live-stream Photometric Surveys}",
      journal = {\apj},
         year = 2025,
        month = aug,
       volume = {989},
       number = {2},
          eid = {192},
        pages = {192},
          doi = {10.3847/1538-4357/adea3a},
archivePrefix = {arXiv},
       eprint = {2503.03735},
 primaryClass = {astro-ph.HE},
       adsurl = {https://ui.adsabs.harvard.edu/abs/2025ApJ...989..192C}
}

@ARTICLE{Kelvin14,
       author = {{Kelvin}, Lee S. and {Driver}, Simon P. and {Robotham}, Aaron S.~G. and {Taylor}, Edward N. and {Graham}, Alister W. and {Alpaslan}, Mehmet and {Baldry}, Ivan and {Bamford}, Steven P. and {Bauer}, Amanda E. and {Bland-Hawthorn}, Joss and {Brown}, Michael J.~I. and {Colless}, Matthew and {Conselice}, Christopher J. and {Holwerda}, Benne W. and {Hopkins}, Andrew M. and {Lara-L{\'o}pez}, Maritza A. and {Liske}, Jochen and {L{\'o}pez-S{\'a}nchez}, {\'A}ngel R. and {Loveday}, Jon and {Norberg}, Peder and {Phillipps}, Steven and {Popescu}, Cristina C. and {Prescott}, Matthew and {Sansom}, Anne E. and {Tuffs}, Richard J.},
        title = "{Galaxy And Mass Assembly (GAMA): stellar mass functions by Hubble type}",
      journal = {\mnras},
         year = 2014,
        month = oct,
       volume = {444},
       number = {2},
        pages = {1647-1659},
          doi = {10.1093/mnras/stu1507},
archivePrefix = {arXiv},
       eprint = {1407.7555},
 primaryClass = {astro-ph.GA},
       adsurl = {https://ui.adsabs.harvard.edu/abs/2014MNRAS.444.1647K}
}

@ARTICLE{SNID,
       author = {{Blondin}, St{\'e}phane and {Tonry}, John L.},
        title = "{Determining the Type, Redshift, and Age of a Supernova Spectrum}",
      journal = {\apj},
         year = 2007,
        month = sep,
       volume = {666},
       number = {2},
        pages = {1024-1047},
          doi = {10.1086/520494},
archivePrefix = {arXiv},
       eprint = {0709.4488},
 primaryClass = {astro-ph},
       adsurl = {https://ui.adsabs.harvard.edu/abs/2007ApJ...666.1024B}
}

@ARTICLE{Tartaglia2017,
       author = {{Tartaglia}, L. and {Fraser}, M. and {Sand}, D.~J. and {Valenti}, S. and {Smartt}, S.~J. and {McCully}, C. and {Anderson}, J.~P. and {Arcavi}, I. and {Elias-Rosa}, N. and {Galbany}, L. and {Gal-Yam}, A. and {Haislip}, J.~B. and {Hosseinzadeh}, G. and {Howell}, D.~A. and {Inserra}, C. and {Jha}, S.~W. and {Kankare}, E. and {Lundqvist}, P. and {Maguire}, K. and {Mattila}, S. and {Reichart}, D. and {Smith}, K.~W. and {Smith}, M. and {Stritzinger}, M. and {Sullivan}, M. and {Taddia}, F. and {Tomasella}, L.},
        title = "{The Progenitor and Early Evolution of the Type IIb SN 2016gkg}",
      journal = {\apjl},
         year = 2017,
        month = feb,
       volume = {836},
       number = {1},
          eid = {L12},
        pages = {L12},
          doi = {10.3847/2041-8213/aa5c7f},
archivePrefix = {arXiv},
       eprint = {1611.00419},
 primaryClass = {astro-ph.HE},
       adsurl = {https://ui.adsabs.harvard.edu/abs/2017ApJ...836L..12T}
}

@ARTICLE{Arcavi2017,
       author = {{Arcavi}, Iair and {Hosseinzadeh}, Griffin and {Brown}, Peter J. and {Smartt}, Stephen J. and {Valenti}, Stefano and {Tartaglia}, Leonardo and {Piro}, Anthony L. and {Sanchez}, Jos{\'e} L. and {Nicholls}, Brent and {Monard}, Berto L.~A.~G. and {Howell}, D. Andrew and {McCully}, Curtis and {Sand}, David J. and {Tonry}, John and {Denneau}, Larry and {Stalder}, Brian and {Heinze}, Ari and {Rest}, Armin and {Smith}, Ken W. and {Bishop}, David},
        title = "{Constraints on the Progenitor of SN 2016gkg from Its Shock-cooling Light Curve}",
      journal = {\apjl},
         year = 2017,
        month = mar,
       volume = {837},
       number = {1},
          eid = {L2},
        pages = {L2},
          doi = {10.3847/2041-8213/aa5be1},
archivePrefix = {arXiv},
       eprint = {1611.06451},
 primaryClass = {astro-ph.HE},
       adsurl = {https://ui.adsabs.harvard.edu/abs/2017ApJ...837L...2A}
}

@ARTICLE{Arcavi2011,
       author = {{Arcavi}, Iair and {Gal-Yam}, Avishay and {Yaron}, Ofer and {Sternberg}, Assaf and {Rabinak}, Itay and {Waxman}, Eli and {Kasliwal}, Mansi M. and {Quimby}, Robert M. and {Ofek}, Eran O. and {Horesh}, Assaf and {Kulkarni}, Shrinivas R. and {Filippenko}, Alexei V. and {Silverman}, Jeffrey M. and {Cenko}, S. Bradley and {Li}, Weidong and {Bloom}, Joshua S. and {Sullivan}, Mark and {Nugent}, Peter E. and {Poznanski}, Dovi and {Gorbikov}, Evgeny and {Fulton}, Benjamin J. and {Howell}, D. Andrew and {Bersier}, David and {Riou}, Amedee and {Lamotte-Bailey}, Stephane and {Griga}, Thomas and {Cohen}, Judith G. and {Hachinger}, Stephan and {Polishook}, David and {Xu}, Dong and {Ben-Ami}, Sagi and {Manulis}, Ilan and {Walker}, Emma S. and {Maguire}, Kate and {Pan}, Yen-Chen and {Matheson}, Thomas and {Mazzali}, Paolo A. and {Pian}, Elena and {Fox}, Derek B. and {Gehrels}, Neil and {Law}, Nicholas and {James}, Philip and {Marchant}, Jonathan M. and {Smith}, Robert J. and {Mottram}, Chris J. and {Barnsley}, Robert M. and {Kandrashoff}, Michael T. and {Clubb}, Kelsey I.},
        title = "{SN 2011dh: Discovery of a Type IIb Supernova from a Compact Progenitor in the Nearby Galaxy M51}",
      journal = {\apjl},
         year = 2011,
        month = dec,
       volume = {742},
       number = {2},
          eid = {L18},
        pages = {L18},
          doi = {10.1088/2041-8205/742/2/L18},
archivePrefix = {arXiv},
       eprint = {1106.3551},
 primaryClass = {astro-ph.CO},
       adsurl = {https://ui.adsabs.harvard.edu/abs/2011ApJ...742L..18A}
}

@ARTICLE{Bersten2012,
       author = {{Bersten}, Melina C. and {Benvenuto}, Omar G. and {Nomoto}, Ken'ichi and {Ergon}, Mattias and {Folatelli}, Gast{\'o}n and {Sollerman}, Jesper and {Benetti}, Stefano and {Botticella}, Maria Teresa and {Fraser}, Morgan and {Kotak}, Rubina and {Maeda}, Keiichi and {Ochner}, Paolo and {Tomasella}, Lina},
        title = "{The Type IIb Supernova 2011dh from a Supergiant Progenitor}",
      journal = {\apj},
         year = 2012,
        month = sep,
       volume = {757},
       number = {1},
          eid = {31},
        pages = {31},
          doi = {10.1088/0004-637X/757/1/31},
archivePrefix = {arXiv},
       eprint = {1207.5975},
 primaryClass = {astro-ph.HE},
       adsurl = {https://ui.adsabs.harvard.edu/abs/2012ApJ...757...31B}
}

@ARTICLE{Ergon2014,
       author = {{Ergon}, M. and {Sollerman}, J. and {Fraser}, M. and {Pastorello}, A. and {Taubenberger}, S. and {Elias-Rosa}, N. and {Bersten}, M. and {Jerkstrand}, A. and {Benetti}, S. and {Botticella}, M.~T. and {Fransson}, C. and {Harutyunyan}, A. and {Kotak}, R. and {Smartt}, S. and {Valenti}, S. and {Bufano}, F. and {Cappellaro}, E. and {Fiaschi}, M. and {Howell}, A. and {Kankare}, E. and {Magill}, L. and {Mattila}, S. and {Maund}, J. and {Naves}, R. and {Ochner}, P. and {Ruiz}, J. and {Smith}, K. and {Tomasella}, L. and {Turatto}, M.},
        title = "{Optical and near-infrared observations of SN 2011dh - The first 100 days}",
      journal = {\aap},
         year = 2014,
        month = feb,
       volume = {562},
          eid = {A17},
        pages = {A17},
          doi = {10.1051/0004-6361/201321850},
archivePrefix = {arXiv},
       eprint = {1305.1851},
 primaryClass = {astro-ph.SR},
       adsurl = {https://ui.adsabs.harvard.edu/abs/2014A&A...562A..17E}
}

@INPROCEEDINGS{Bacon2010,
       author = {{Bacon}, R. and {Accardo}, M. and {Adjali}, L. and {Anwand}, H. and {Bauer}, S. and {Biswas}, I. and {Blaizot}, J. and {Boudon}, D. and {Brau-Nogue}, S. and {Brinchmann}, J. and {Caillier}, P. and {Capoani}, L. and {Carollo}, C.~M. and {Contini}, T. and {Couderc}, P. and {Daguis{\'e}}, E. and {Deiries}, S. and {Delabre}, B. and {Dreizler}, S. and {Dubois}, J. and {Dupieux}, M. and {Dupuy}, C. and {Emsellem}, E. and {Fechner}, T. and {Fleischmann}, A. and {Fran{\c{c}}ois}, M. and {Gallou}, G. and {Gharsa}, T. and {Glindemann}, A. and {Gojak}, D. and {Guiderdoni}, B. and {Hansali}, G. and {Hahn}, T. and {Jarno}, A. and {Kelz}, A. and {Koehler}, C. and {Kosmalski}, J. and {Laurent}, F. and {Le Floch}, M. and {Lilly}, S.~J. and {Lizon}, J.-L. and {Loupias}, M. and {Manescau}, A. and {Monstein}, C. and {Nicklas}, H. and {Olaya}, J.-C. and {Pares}, L. and {Pasquini}, L. and {P{\'e}contal-Rousset}, A. and {Pell{\'o}}, R. and {Petit}, C. and {Popow}, E. and {Reiss}, R. and {Remillieux}, A. and {Renault}, E. and {Roth}, M. and {Rupprecht}, G. and {Serre}, D. and {Schaye}, J. and {Soucail}, G. and {Steinmetz}, M. and {Streicher}, O. and {Stuik}, R. and {Valentin}, H. and {Vernet}, J. and {Weilbacher}, P. and {Wisotzki}, L. and {Yerle}, N.},
        title = "{The MUSE second-generation VLT instrument}",
    booktitle = {Ground-based and Airborne Instrumentation for Astronomy III},
         year = 2010,
       editor = {{McLean}, Ian S. and {Ramsay}, Suzanne K. and {Takami}, Hideki},
       series = {Society of Photo-Optical Instrumentation Engineers (SPIE) Conference Series},
       volume = {7735},
        month = jul,
          eid = {773508},
        pages = {773508},
          doi = {10.1117/12.856027},
archivePrefix = {arXiv},
       eprint = {2211.16795},
 primaryClass = {astro-ph.IM},
       adsurl = {https://ui.adsabs.harvard.edu/abs/2010SPIE.7735E..08B}
}

@ARTICLE{Lyman2018,
       author = {{Lyman}, J.~D. and {Taddia}, F. and {Stritzinger}, M.~D. and {Galbany}, L. and {Leloudas}, G. and {Anderson}, J.~P. and {Eldridge}, J.~J. and {James}, P.~A. and {Kr{\"u}hler}, T. and {Levan}, A.~J. and {Pignata}, G. and {Stanway}, E.~R.},
        title = "{Investigating the diversity of supernovae type Iax: a MUSE and NOT spectroscopic study of their environments}",
      journal = {\mnras},
         year = 2018,
        month = jan,
       volume = {473},
       number = {1},
        pages = {1359-1387},
          doi = {10.1093/mnras/stx2414},
archivePrefix = {arXiv},
       eprint = {1707.04270},
 primaryClass = {astro-ph.HE},
       adsurl = {https://ui.adsabs.harvard.edu/abs/2018MNRAS.473.1359L}
}

@ARTICLE{2023GCN33889,
       author = {{LVK Collaboration}},
        title = "{LIGO/Virgo/KAGRA S230529ay: Identification of a GW compact binary merger candidate}",
      journal = {GRB Coordinates Network},
         year = 2023,
        month = may,
       volume = {33889},
        pages = {1},
       adsurl = {https://ui.adsabs.harvard.edu/abs/2023GCN.33889....1L}
}

@ARTICLE{2025GCN39175,
       author = {{LVK Collaboration}},
        title = "{LIGO/Virgo/KAGRA S250206dm: Identification of a GW compact binary merger candidate}",
      journal = {GRB Coordinates Network},
         year = 2025,
        month = feb,
       volume = {39175},
        pages = {1},
       adsurl = {https://ui.adsabs.harvard.edu/abs/2025GCN.39175....1L}
}

@ARTICLE{astropy2013,
       author = {{Astropy Collaboration} and {Robitaille}, Thomas P. and {Tollerud}, Erik J. and {Greenfield}, Perry and {Droettboom}, Michael and {Bray}, Erik and {Aldcroft}, Tom and {Davis}, Matt and {Ginsburg}, Adam and {Price-Whelan}, Adrian M. and {Kerzendorf}, Wolfgang E. and {Conley}, Alexander and {Crighton}, Neil and {Barbary}, Kyle and {Muna}, Demitri and {Ferguson}, Henry and {Grollier}, Fr{\'e}d{\'e}ric and {Parikh}, Madhura M. and {Nair}, Prasanth H. and {Unther}, Hans M. and {Deil}, Christoph and {Woillez}, Julien and {Conseil}, Simon and {Kramer}, Roban and {Turner}, James E.~H. and {Singer}, Leo and {Fox}, Ryan and {Weaver}, Benjamin A. and {Zabalza}, Victor and {Edwards}, Zachary I. and {Azalee Bostroem}, K. and {Burke}, D.~J. and {Casey}, Andrew R. and {Crawford}, Steven M. and {Dencheva}, Nadia and {Ely}, Justin and {Jenness}, Tim and {Labrie}, Kathleen and {Lim}, Pey Lian and {Pierfederici}, Francesco and {Pontzen}, Andrew and {Ptak}, Andy and {Refsdal}, Brian and {Servillat}, Mathieu and {Streicher}, Ole},
        title = "{Astropy: A community Python package for astronomy}",
      journal = {\aap},
         year = 2013,
        month = oct,
       volume = {558},
          eid = {A33},
        pages = {A33},
          doi = {10.1051/0004-6361/201322068},
archivePrefix = {arXiv},
       eprint = {1307.6212},
 primaryClass = {astro-ph.IM},
       adsurl = {https://ui.adsabs.harvard.edu/abs/2013A&A...558A..33A}
}

@ARTICLE{Kilpatrick2022,
       author = {{Kilpatrick}, Charles D. and {Fong}, Wen-fai and {Blanchard}, Peter K. and {Leja}, Joel and {Nugent}, Anya E. and {Palmese}, Antonella and {Paterson}, Kerry and {Starkenburg}, Tjitske and {Alexander}, Kate D. and {Berger}, Edo and {Chornock}, Ryan and {Hajela}, Aprajita and {Margutti}, Raffaella},
        title = "{Hubble Space Telescope Observations of GW170817: Complete Light Curves and the Properties of the Galaxy Merger of NGC 4993}",
      journal = {\apj},
         year = 2022,
        month = feb,
       volume = {926},
       number = {1},
          eid = {49},
        pages = {49},
          doi = {10.3847/1538-4357/ac3e59},
archivePrefix = {arXiv},
       eprint = {2109.06211},
 primaryClass = {astro-ph.HE},
       adsurl = {https://ui.adsabs.harvard.edu/abs/2022ApJ...926...49K}
}

@software{STSCI2012,
       author = {{STSCI Development Team}},
        title = "{DrizzlePac: HST image software}",
 howpublished = {Astrophysics Source Code Library, record ascl:1212.011},
         year = 2012,
        month = dec,
          eid = {ascl:1212.011},
archivePrefix = {ascl},
       eprint = {1212.011},
       adsurl = {https://ui.adsabs.harvard.edu/abs/2012ascl.soft12011S}
}

@software{Dolphin2016,
       author = {{Dolphin}, Andrew},
        title = "{DOLPHOT: Stellar photometry}",
 howpublished = {Astrophysics Source Code Library, record ascl:1608.013},
         year = 2016,
        month = aug,
          eid = {ascl:1608.013},
archivePrefix = {ascl},
       eprint = {1608.013},
       adsurl = {https://ui.adsabs.harvard.edu/abs/2016ascl.soft08013D}
}

@ARTICLE{Skrutskie2006,
       author = {{Skrutskie}, M.~F. and {Cutri}, R.~M. and {Stiening}, R. and {Weinberg}, M.~D. and {Schneider}, S. and {Carpenter}, J.~M. and {Beichman}, C. and {Capps}, R. and {Chester}, T. and {Elias}, J. and {Huchra}, J. and {Liebert}, J. and {Lonsdale}, C. and {Monet}, D.~G. and {Price}, S. and {Seitzer}, P. and {Jarrett}, T. and {Kirkpatrick}, J.~D. and {Gizis}, J.~E. and {Howard}, E. and {Evans}, T. and {Fowler}, J. and {Fullmer}, L. and {Hurt}, R. and {Light}, R. and {Kopan}, E.~L. and {Marsh}, K.~A. and {McCallon}, H.~L. and {Tam}, R. and {Van Dyk}, S. and {Wheelock}, S.},
        title = "{The Two Micron All Sky Survey (2MASS)}",
      journal = {\aj},
         year = 2006,
        month = feb,
       volume = {131},
       number = {2},
        pages = {1163-1183},
          doi = {10.1086/498708},
       adsurl = {https://ui.adsabs.harvard.edu/abs/2006AJ....131.1163S}
}

@software{Moldon2021,
       author = {{Moldon}, Javier},
        title = "{eMCP: e-MERLIN CASA pipeline}",
 howpublished = {Astrophysics Source Code Library, record ascl:2109.006},
         year = 2021,
        month = sep,
          eid = {ascl:2109.006},
archivePrefix = {ascl},
       eprint = {2109.006},
       adsurl = {https://ui.adsabs.harvard.edu/abs/2021ascl.soft09006M}
}

@INPROCEEDINGS{Intema2014,
       author = {{Intema}, H.~T.},
        title = "{SPAM: A data reduction recipe for high-resolution,low-frequency radio-interferometric observations}",
    booktitle = {Astronomical Society of India Conference Series},
         year = 2014,
       series = {Astronomical Society of India Conference Series},
       volume = {13},
        month = jan,
        pages = {469},
          doi = {10.48550/arXiv.1402.4889},
archivePrefix = {arXiv},
       eprint = {1402.4889},
 primaryClass = {astro-ph.IM},
       adsurl = {https://ui.adsabs.harvard.edu/abs/2014ASInC..13..469I}
}

@ARTICLE{Kawaguchi2020,
       author = {{Kawaguchi}, Kyohei and {Shibata}, Masaru and {Tanaka}, Masaomi},
        title = "{Diversity of Kilonova Light Curves}",
      journal = {\apj},
         year = 2020,
        month = feb,
       volume = {889},
       number = {2},
          eid = {171},
        pages = {171},
          doi = {10.3847/1538-4357/ab61f6},
archivePrefix = {arXiv},
       eprint = {1908.05815},
 primaryClass = {astro-ph.HE},
       adsurl = {https://ui.adsabs.harvard.edu/abs/2020ApJ...889..171K}
}

@ARTICLE{Levan2024,
       author = {{Levan}, Andrew J. and {Gompertz}, Benjamin P. and {Salafia}, Om Sharan and {Bulla}, Mattia and {Burns}, Eric and {Hotokezaka}, Kenta and {Izzo}, Luca and {Lamb}, Gavin P. and {Malesani}, Daniele B. and {Oates}, Samantha R. and {Ravasio}, Maria Edvige and {Rouco Escorial}, Alicia and {Schneider}, Benjamin and {Sarin}, Nikhil and {Schulze}, Steve and {Tanvir}, Nial R. and {Ackley}, Kendall and {Anderson}, Gemma and {Brammer}, Gabriel B. and {Christensen}, Lise and {Dhillon}, Vikram S. and {Evans}, Phil A. and {Fausnaugh}, Michael and {Fong}, Wen-fai and {Fruchter}, Andrew S. and {Fryer}, Chris and {Fynbo}, Johan P.~U. and {Gaspari}, Nicola and {Heintz}, Kasper E. and {Hjorth}, Jens and {Kennea}, Jamie A. and {Kennedy}, Mark R. and {Laskar}, Tanmoy and {Leloudas}, Giorgos and {Mandel}, Ilya and {Martin-Carrillo}, Antonio and {Metzger}, Brian D. and {Nicholl}, Matt and {Nugent}, Anya and {Palmerio}, Jesse T. and {Pugliese}, Giovanna and {Rastinejad}, Jillian and {Rhodes}, Lauren and {Rossi}, Andrea and {Saccardi}, Andrea and {Smartt}, Stephen J. and {Stevance}, Heloise F. and {Tohuvavohu}, Aaron and {van der Horst}, Alexander and {Vergani}, Susanna D. and {Watson}, Darach and {Barclay}, Thomas and {Bhirombhakdi}, Kornpob and {Breedt}, Elm{\'e} and {Breeveld}, Alice A. and {Brown}, Alexander J. and {Campana}, Sergio and {Chrimes}, Ashley A. and {D'Avanzo}, Paolo and {D'Elia}, Valerio and {De Pasquale}, Massimiliano and {Dyer}, Martin J. and {Galloway}, Duncan K. and {Garbutt}, James A. and {Green}, Matthew J. and {Hartmann}, Dieter H. and {Jakobsson}, P{\'a}ll and {Kerry}, Paul and {Kouveliotou}, Chryssa and {Langeroodi}, Danial and {Le Floc'h}, Emeric and {Leung}, James K. and {Littlefair}, Stuart P. and {Munday}, James and {O'Brien}, Paul and {Parsons}, Steven G. and {Pelisoli}, Ingrid and {Sahman}, David I. and {Salvaterra}, Ruben and {Sbarufatti}, Boris and {Steeghs}, Danny and {Tagliaferri}, Gianpiero and {Th{\"o}ne}, Christina C. and {de Ugarte Postigo}, Antonio and {Kann}, David Alexander},
        title = "{Heavy-element production in a compact object merger observed by JWST}",
      journal = {\nat},
         year = 2024,
        month = feb,
       volume = {626},
       number = {8000},
        pages = {737-741},
          doi = {10.1038/s41586-023-06759-1},
archivePrefix = {arXiv},
       eprint = {2307.02098},
 primaryClass = {astro-ph.HE},
       adsurl = {https://ui.adsabs.harvard.edu/abs/2024Natur.626..737L}
}

@ARTICLE{Pognan2026,
       author = {{Pognan}, Quentin and {Kawaguchi}, Kyohei and {Wanajo}, Shinya and {Fujibayashi}, Sho and {Jerkstrand}, Anders and {Grumer}, Jon},
        title = "{Lanthanide impact on the infrared spectra of nebular phase kilonovae}",
      journal = {\mnras},
         year = 2026,
        month = apr,
       volume = {547},
       number = {3},
          eid = {stag441},
        pages = {stag441},
          doi = {10.1093/mnras/stag441},
       adsurl = {https://ui.adsabs.harvard.edu/abs/2026MNRAS.547ag441P}
}

@ARTICLE{Chevalier1982,
       author = {{Chevalier}, R.~A.},
        title = "{The radio and X-ray emission from type II supernovae.}",
      journal = {\apj},
         year = 1982,
        month = aug,
       volume = {259},
        pages = {302-310},
          doi = {10.1086/160167},
       adsurl = {https://ui.adsabs.harvard.edu/abs/1982ApJ...259..302C}
}

@ARTICLE{Taddia2018,
       author = {{Taddia}, F. and {Stritzinger}, M.~D. and {Bersten}, M. and {Baron}, E. and {Burns}, C. and {Contreras}, C. and {Holmbo}, S. and {Hsiao}, E.~Y. and {Morrell}, N. and {Phillips}, M.~M. and {Sollerman}, J. and {Suntzeff}, N.~B.},
        title = "{The Carnegie Supernova Project I. Analysis of stripped-envelope supernova light curves}",
      journal = {\aap},
         year = 2018,
        month = feb,
       volume = {609},
          eid = {A136},
        pages = {A136},
          doi = {10.1051/0004-6361/201730844},
archivePrefix = {arXiv},
       eprint = {1707.07614},
 primaryClass = {astro-ph.HE},
       adsurl = {https://ui.adsabs.harvard.edu/abs/2018A&A...609A.136T}
}

@ARTICLE{Chevalier10,
       author = {{Chevalier}, Roger A. and {Soderberg}, Alicia M.},
        title = "{Type IIb Supernovae with Compact and Extended Progenitors}",
      journal = {\apjl},
         year = 2010,
        month = mar,
       volume = {711},
       number = {1},
        pages = {L40-L43},
          doi = {10.1088/2041-8205/711/1/L40},
archivePrefix = {arXiv},
       eprint = {0911.3408},
 primaryClass = {astro-ph.HE},
       adsurl = {https://ui.adsabs.harvard.edu/abs/2010ApJ...711L..40C}
}

@ARTICLE{Farah2026,
       author = {{Farah}, Joseph R. and {Howell}, D. Andrew and {Hiramatsu}, Daichi and {McCully}, Curtis and {Andrews}, Moira and {Newsome}, Megan and {Padilla Gonzalez}, Estefania and {Pellegrino}, Craig and {Berger}, Edo and {Blanchard}, Peter and {Gomez}, Sebastian and {Kumar}, Harsh and {Bostroem}, K. Azalee and {Ni}, Yuan Qi and {Gagliano}, A. and {Ravi}, Aravind P.},
        title = "{When IIb Ceases To Be: Bridging the Gap between IIb and Short-plateau Supernovae}",
      journal = {\apj},
         year = 2026,
        month = feb,
       volume = {998},
       number = {2},
          eid = {321},
        pages = {321},
          doi = {10.3847/1538-4357/ae3a71},
archivePrefix = {arXiv},
       eprint = {2509.12470},
 primaryClass = {astro-ph.HE},
       adsurl = {https://ui.adsabs.harvard.edu/abs/2026ApJ...998..321F}
}

@ARTICLE{Xi2026,
       author = {{Xi}, Qiang and {Sun}, Ning-Chen and {Aguado}, David and {P{\'e}rez-Fournon}, Ismael and {Poidevin}, Fr{\'e}d{\'e}rick and {Jin}, Junjie and {Mao}, Yiming and {Niu}, Zexi and {Wang}, Beichuan and {Zhang}, Yu and {Misra}, Kuntal and {Janghel}, Divyanshu and {Maund}, Justyn R. and {Kumar}, Amit and {Tinyanont}, Samaporn and {Liu}, Liang-Duan and {Zhang}, Yu-Hao and {Ailawadhi}, Bhavya and {Dubey}, Monalisa and {Guo}, Zhen and {Gupta}, Anshika and {He}, Min and {Jain}, Dhruv and {Kar}, Debalina and {Li}, Wenxiong and {Lyman}, Joe D. and {Mu}, Haiyang and {Pranshu}, Kumar and {Sun}, Xinxiang and {Wang}, Lingzhi and {Yadav}, Sarvesh Kumar and {Zhao}, Yi-Han and {Zheng}, Jie and {Zhu}, Yinan and {L{\'o}pez Fern{\'a}ndez-Nespral}, David and {L{\'o}pez Oramas}, Alicia and {Wang}, Yanan and {Wiersema}, Klaas and {Liu}, Jifeng},
        title = "{SN 2024aecx: A Fast-evolving Type IIb Supernova with a Prominent Shock-cooling Peak}",
      journal = {\apj},
         year = 2026,
        month = feb,
       volume = {998},
       number = {1},
          eid = {98},
        pages = {98},
          doi = {10.3847/1538-4357/ae2d06},
archivePrefix = {arXiv},
       eprint = {2509.12343},
 primaryClass = {astro-ph.SR},
       adsurl = {https://ui.adsabs.harvard.edu/abs/2026ApJ...998...98X}
}

@ARTICLE{Subrayan2025,
       author = {{Subrayan}, Bhagya M. and {Sand}, David J. and {Bostroem}, K. Azalee and {Jha}, Saurabh W. and {Ravi}, Aravind P. and {Schwab}, Michaela and {Andrews}, Jennifer E. and {Hosseinzadeh}, Griffin and {Valenti}, Stefano and {Dong}, Yize and {Pearson}, Jeniveve and {Shrestha}, Manisha and {Kwok}, Lindsey A. and {Hoang}, Emily and {Rho}, Jeonghee and {Park}, Seong Hyun and {Yoon}, Sung-Chul and {Geballe}, T.~R. and {Haislip}, Joshua and {Janzen}, Daryl and {Kouprianov}, Vladimir and {Mehta}, Darshana and {Meza Retamal}, Nicol{\'a}s and {Reichart}, Daniel E. and {Andrews}, Moira and {Farah}, Joseph and {Newsome}, Megan and {Howell}, D. Andrew and {McCully}, Curtis},
        title = "{Early Shock Cooling Observations and Progenitor Constraints of Type IIb Supernova SN 2024uwq}",
      journal = {\apjl},
         year = 2025,
        month = sep,
       volume = {990},
       number = {2},
          eid = {L68},
        pages = {L68},
          doi = {10.3847/2041-8213/adfe52},
archivePrefix = {arXiv},
       eprint = {2505.02908},
 primaryClass = {astro-ph.HE},
       adsurl = {https://ui.adsabs.harvard.edu/abs/2025ApJ...990L..68S}
}

@ARTICLE{Wang2023,
       author = {{Wang}, Qinan and {Armstrong}, Patrick and {Zenati}, Yossef and {Ridden-Harper}, Ryan and {Rest}, Armin and {Arcavi}, Iair and {Kilpatrick}, Charles D. and {Foley}, Ryan J. and {Tucker}, Brad E. and {Lidman}, Chris and {Killestein}, Thomas L. and {Shahbandeh}, Melissa and {Anderson}, Joseph P. and {Angulo}, Rodrigo and {Ashall}, Chris and {Burke}, Jamison and {Chen}, Ting-Wan and {von Coelln}, Sophie and {Dalrymple}, Kyle A. and {Davis}, Kyle W. and {Fulton}, Michael D. and {Galbany}, Llu{\'\i}s and {Padilla Gonzalez}, Estefania and {Gao}, Bore and {Gromadzki}, Mariusz and {Howell}, D. Andrew and {Ihanec}, Nada and {Jencson}, Jacob E. and {Jones}, David O. and {Lyman}, Joseph D. and {McCully}, Curtis and {M{\"u}ller-Bravo}, Tom{\'a}s E. and {Newsome}, Megan and {Nicholl}, Matt and {O'Neill}, David and {Pellegrino}, Craig and {Rest}, Sofia and {Smartt}, Stephen J. and {Smith}, Ken and {Srivastav}, Shubham and {Terreran}, Giacomo and {Tinyanont}, Samaporn and {Young}, David R. and {Zenteno}, Alfredo},
        title = "{Revealing the Progenitor of SN 2021zby through Analysis of the TESS Shock-cooling Light Curve}",
      journal = {\apjl},
         year = 2023,
        month = feb,
       volume = {943},
       number = {2},
          eid = {L15},
        pages = {L15},
          doi = {10.3847/2041-8213/acb0d0},
archivePrefix = {arXiv},
       eprint = {2211.03811},
 primaryClass = {astro-ph.HE},
       adsurl = {https://ui.adsabs.harvard.edu/abs/2023ApJ...943L..15W}
}

@ARTICLE{Pian2017,
       author = {{Pian}, E. and {D'Avanzo}, P. and {Benetti}, S. and {Branchesi}, M. and {Brocato}, E. and {Campana}, S. and {Cappellaro}, E. and {Covino}, S. and {D'Elia}, V. and {Fynbo}, J.~P.~U. and {Getman}, F. and {Ghirlanda}, G. and {Ghisellini}, G. and {Grado}, A. and {Greco}, G. and {Hjorth}, J. and {Kouveliotou}, C. and {Levan}, A. and {Limatola}, L. and {Malesani}, D. and {Mazzali}, P.~A. and {Melandri}, A. and {M{\o}ller}, P. and {Nicastro}, L. and {Palazzi}, E. and {Piranomonte}, S. and {Rossi}, A. and {Salafia}, O.~S. and {Selsing}, J. and {Stratta}, G. and {Tanaka}, M. and {Tanvir}, N.~R. and {Tomasella}, L. and {Watson}, D. and {Yang}, S. and {Amati}, L. and {Antonelli}, L.~A. and {Ascenzi}, S. and {Bernardini}, M.~G. and {Bo{\"e}r}, M. and {Bufano}, F. and {Bulgarelli}, A. and {Capaccioli}, M. and {Casella}, P. and {Castro-Tirado}, A.~J. and {Chassande-Mottin}, E. and {Ciolfi}, R. and {Copperwheat}, C.~M. and {Dadina}, M. and {De Cesare}, G. and {di Paola}, A. and {Fan}, Y.~Z. and {Gendre}, B. and {Giuffrida}, G. and {Giunta}, A. and {Hunt}, L.~K. and {Israel}, G.~L. and {Jin}, Z.-P. and {Kasliwal}, M.~M. and {Klose}, S. and {Lisi}, M. and {Longo}, F. and {Maiorano}, E. and {Mapelli}, M. and {Masetti}, N. and {Nava}, L. and {Patricelli}, B. and {Perley}, D. and {Pescalli}, A. and {Piran}, T. and {Possenti}, A. and {Pulone}, L. and {Razzano}, M. and {Salvaterra}, R. and {Schipani}, P. and {Spera}, M. and {Stamerra}, A. and {Stella}, L. and {Tagliaferri}, G. and {Testa}, V. and {Troja}, E. and {Turatto}, M. and {Vergani}, S.~D. and {Vergani}, D.},
        title = "{Spectroscopic identification of r-process nucleosynthesis in a double neutron-star merger}",
      journal = {\nat},
         year = 2017,
        month = nov,
       volume = {551},
       number = {7678},
        pages = {67-70},
          doi = {10.1038/nature24298},
archivePrefix = {arXiv},
       eprint = {1710.05858},
 primaryClass = {astro-ph.HE},
       adsurl = {https://ui.adsabs.harvard.edu/abs/2017Natur.551...67P}
}

@ARTICLE{Aasi2015,
       author = {{Aasi}, J. and {Abbott}, B.~P. and {Abbott}, R. and {Abbott}, T. and {Abernathy}, M.~R. and {Ackley}, K. and {Adams}, C. and {Adams}, T. and {Addesso}, P. and others},
        title = "{Advanced LIGO}",
      journal = {Classical and Quantum Gravity},
         year = 2015,
        month = apr,
       volume = {32},
       number = {7},
          eid = {074001},
        pages = {074001},
          doi = {10.1088/0264-9381/32/7/074001},
archivePrefix = {arXiv},
       eprint = {1411.4547},
 primaryClass = {gr-qc},
       adsurl = {https://ui.adsabs.harvard.edu/abs/2015CQGra..32g4001L}
}

@ARTICLE{Acernese2015,
       author = {{Acernese}, F. and {Agathos}, M. and {Agatsuma}, K. and {Aisa}, D. and {Allemandou}, N. and {Allocca}, A. and {Amarni}, J. and {Astone}, P. and {Balestri}, G. and others},
        title = "{Advanced Virgo: a second-generation interferometric gravitational wave detector}",
      journal = {Classical and Quantum Gravity},
         year = 2015,
        month = jan,
       volume = {32},
       number = {2},
          eid = {024001},
        pages = {024001},
          doi = {10.1088/0264-9381/32/2/024001},
archivePrefix = {arXiv},
       eprint = {1408.3978},
 primaryClass = {gr-qc},
       adsurl = {https://ui.adsabs.harvard.edu/abs/2015CQGra..32b4001A}
}

@ARTICLE{Somiya2012,
       author = {{Somiya}, Kentaro},
        title = "{Detector configuration of KAGRA-the Japanese cryogenic gravitational-wave detector}",
      journal = {Classical and Quantum Gravity},
         year = 2012,
        month = jun,
       volume = {29},
       number = {12},
          eid = {124007},
        pages = {124007},
          doi = {10.1088/0264-9381/29/12/124007},
archivePrefix = {arXiv},
       eprint = {1111.7185},
 primaryClass = {gr-qc},
       adsurl = {https://ui.adsabs.harvard.edu/abs/2012CQGra..29l4007S}
}

@ARTICLE{Chen2025,
       author = {{Chen}, Yi-Xian and {Metzger}, Brian D.},
        title = "{Gravitational Instability and Fragmentation in Collapsar Disks Supports the Formation of Subsolar Neutron Stars}",
      journal = {\apjl},
         year = 2025,
        month = sep,
       volume = {991},
       number = {1},
          eid = {L22},
        pages = {L22},
          doi = {10.3847/2041-8213/ae045d},
archivePrefix = {arXiv},
       eprint = {2508.17183},
 primaryClass = {astro-ph.HE},
       adsurl = {https://ui.adsabs.harvard.edu/abs/2025ApJ...991L..22C}
}

@ARTICLE{Abac2025_LVKpop4,
       author = {{Abac}, A.~G. and {Abouelfettouh}, I. and {Acernese}, F. and {Ackley}, K. and {Adamcewicz}, C. and {Adhicary}, S. and {Adhikari}, D. and {Adhikari}, N. and {Adhikari}, R.~X. and {Adkins}, V.~K. and {Afroz}, S. and {Agarwal}, D. and {Agathos}, M. and {Aghaei Abchouyeh}, M. and others},
        title = "{GWTC-4.0: Population Properties of Merging Compact Binaries}",
      journal = {arXiv e-prints},
         year = 2025,
        month = aug,
          eid = {arXiv:2508.18083},
        pages = {arXiv:2508.18083},
          doi = {10.48550/arXiv.2508.18083},
archivePrefix = {arXiv},
       eprint = {2508.18083},
 primaryClass = {astro-ph.HE},
       adsurl = {https://ui.adsabs.harvard.edu/abs/2025arXiv250818083T}
}

@ARTICLE{Andreoni2022,
       author = {{Andreoni}, Igor and {Coughlin}, Michael W. and {Almualla}, Mouza and {Bellm}, Eric C. and {Bianco}, Federica B. and {Bulla}, Mattia and {Cucchiara}, Antonino and {Dietrich}, Tim and {Goobar}, Ariel and {Kool}, Erik C. and {Li}, Xiaolong and {Ragosta}, Fabio and {Sagu{\'e}s-Carracedo}, Ana and {Singer}, Leo P.},
        title = "{Optimizing Cadences with Realistic Light-curve Filtering for Serendipitous Kilonova Discovery with Vera Rubin Observatory}",
      journal = {\apjs},
         year = 2022,
        month = jan,
       volume = {258},
       number = {1},
          eid = {5},
        pages = {5},
          doi = {10.3847/1538-4365/ac3bae},
archivePrefix = {arXiv},
       eprint = {2106.06820},
 primaryClass = {astro-ph.HE},
       adsurl = {https://ui.adsabs.harvard.edu/abs/2022ApJS..258....5A}
}

@ARTICLE{Zhu2023,
       author = {{Zhu}, Jin-Ping and {Wu}, Shichao and {Yang}, Yuan-Pei and {Liu}, Chang and {Zhang}, Bing and {Song}, Hao-Ran and {Gao}, He and {Cao}, Zhoujian and {Yu}, Yun-Wei and {Kang}, Yacheng and {Shao}, Lijing},
        title = "{Kilonovae and Optical Afterglows from Binary Neutron Star Mergers. II. Optimal Search Strategy for Serendipitous Observations and Target-of-opportunity Observations of Gravitational Wave Triggers}",
      journal = {\apj},
         year = 2023,
        month = jan,
       volume = {942},
       number = {2},
          eid = {88},
        pages = {88},
          doi = {10.3847/1538-4357/aca527},
archivePrefix = {arXiv},
       eprint = {2110.10469},
 primaryClass = {astro-ph.HE},
       adsurl = {https://ui.adsabs.harvard.edu/abs/2023ApJ...942...88Z}
}

@ARTICLE{Nicholl2021,
       author = {{Nicholl}, Matt and {Margalit}, Ben and {Schmidt}, Patricia and {Smith}, Graham P. and et al.},
        title = "{Tight multimessenger constraints on the neutron star equation of state from GW170817 and a forward model for kilonova light-curve synthesis}",
      journal = {\mnras},
         year = 2021,
        month = aug,
       volume = {505},
       number = {2},
        pages = {3016-3032},
          doi = {10.1093/mnras/stab1523},
archivePrefix = {arXiv},
       eprint = {2102.02229},
 primaryClass = {astro-ph.HE},
       adsurl = {https://ui.adsabs.harvard.edu/abs/2021MNRAS.505.3016N}
}

@ARTICLE{Villar2017,
       author = {{Villar}, V.~A. and {Guillochon}, J. and {Berger}, E. and {Metzger}, B.~D. and et al.},
        title = "{The Combined Ultraviolet, Optical, and Near-infrared Light Curves of the Kilonova Associated with the Binary Neutron Star Merger GW170817: Unified Data Set, Analytic Models, and Physical Implications}",
      journal = {\apjl},
         year = 2017,
        month = dec,
       volume = {851},
       number = {1},
          eid = {L21},
        pages = {L21},
          doi = {10.3847/2041-8213/aa9c84},
archivePrefix = {arXiv},
       eprint = {1710.11576},
 primaryClass = {astro-ph.HE},
       adsurl = {https://ui.adsabs.harvard.edu/abs/2017ApJ...851L..21V}
}

@ARTICLE{Sarin2022,
       author = {{Sarin}, Nikhil and {Omand}, Conor M.~B. and {Margalit}, Ben and {Jones}, David I.},
        title = "{On the diversity of magnetar-driven kilonovae}",
      journal = {\mnras},
         year = 2022,
        month = nov,
       volume = {516},
       number = {4},
        pages = {4949-4962},
          doi = {10.1093/mnras/stac2609},
archivePrefix = {arXiv},
       eprint = {2205.14159},
 primaryClass = {astro-ph.HE},
       adsurl = {https://ui.adsabs.harvard.edu/abs/2022MNRAS.516.4949S}
}

@ARTICLE{Metzger2020,
       author = {{Metzger}, Brian D.},
        title = "{Kilonovae}",
      journal = {Living Reviews in Relativity},
         year = 2020,
        month = dec,
       volume = {23},
       number = {1},
          eid = {1},
        pages = {1},
          doi = {10.1007/s41114-019-0024-0},
archivePrefix = {arXiv},
       eprint = {1910.01617},
 primaryClass = {astro-ph.HE},
       adsurl = {https://ui.adsabs.harvard.edu/abs/2020LRR....23....1M}
}

@ARTICLE{Siegel2022,
       author = {{Siegel}, Daniel M. and {Agarwal}, Aman and {Barnes}, Jennifer and {Metzger}, Brian D. and {Renzo}, Mathieu and {Villar}, V. Ashley},
        title = "{``Super-kilonovae'' from Massive Collapsars as Signatures of Black Hole Birth in the Pair-instability Mass Gap}",
      journal = {\apj},
         year = 2022,
        month = dec,
       volume = {941},
       number = {1},
          eid = {100},
        pages = {100},
          doi = {10.3847/1538-4357/ac8d04},
archivePrefix = {arXiv},
       eprint = {2111.03094},
 primaryClass = {astro-ph.HE},
       adsurl = {https://ui.adsabs.harvard.edu/abs/2022ApJ...941..100S}
}

@ARTICLE{Piro2021,
       author = {{Piro}, Anthony L. and {Haynie}, Annastasia and {Yao}, Yuhan},
        title = "{Shock Cooling Emission from Extended Material Revisited}",
      journal = {\apj},
         year = 2021,
        month = mar,
       volume = {909},
       number = {2},
          eid = {209},
        pages = {209},
          doi = {10.3847/1538-4357/abe2b1},
archivePrefix = {arXiv},
       eprint = {2007.08543},
 primaryClass = {astro-ph.HE},
       adsurl = {https://ui.adsabs.harvard.edu/abs/2021ApJ...909..209P}
}

@ARTICLE{Piro2018,
       author = {{Piro}, Anthony L. and {Kollmeier}, Juna A.},
        title = "{Evidence for Cocoon Emission from the Early Light Curve of SSS17a}",
      journal = {\apj},
         year = 2018,
        month = mar,
       volume = {855},
       number = {2},
          eid = {103},
        pages = {103},
          doi = {10.3847/1538-4357/aaaab3},
archivePrefix = {arXiv},
       eprint = {1710.05822},
 primaryClass = {astro-ph.HE},
       adsurl = {https://ui.adsabs.harvard.edu/abs/2018ApJ...855..103P}
}

@ARTICLE{ODwyer2026,
       author = {{O'Dwyer}, Tanner and {Corsi}, Alessandra and {Yadav}, Deepika and {Mooley}, Kunal P. and {Baer-Way}, Raphael and {Chandra}, Poonam and {Hallinan}, Gregg and {Kasliwal}, Mansi M. and {Rhodes}, Lauren and {Smirnov}, Oleg M. and {Lazzati}, Davide and {van Leeuwen}, Joeri and {Deller}, Adam and {Atri}, Pikky and {Khanam}, Tanazza},
        title = "{Identification of a Radio Counterpart to SN 2025ulz in the S250818k Localization Area}",
      journal = {arXiv e-prints},
         year = 2026,
        month = apr,
          eid = {arXiv:2604.05128},
        pages = {arXiv:2604.05128},
archivePrefix = {arXiv},
       eprint = {2604.05128},
 primaryClass = {astro-ph.HE},
       adsurl = {https://ui.adsabs.harvard.edu/abs/2026arXiv260405128O}
}
\normalsize

\appendix

\section{S250818k sky localisation}
\label{sec:localisation}

\begin{figure*}[t]
  \centering
  \includegraphics[width=\textwidth]{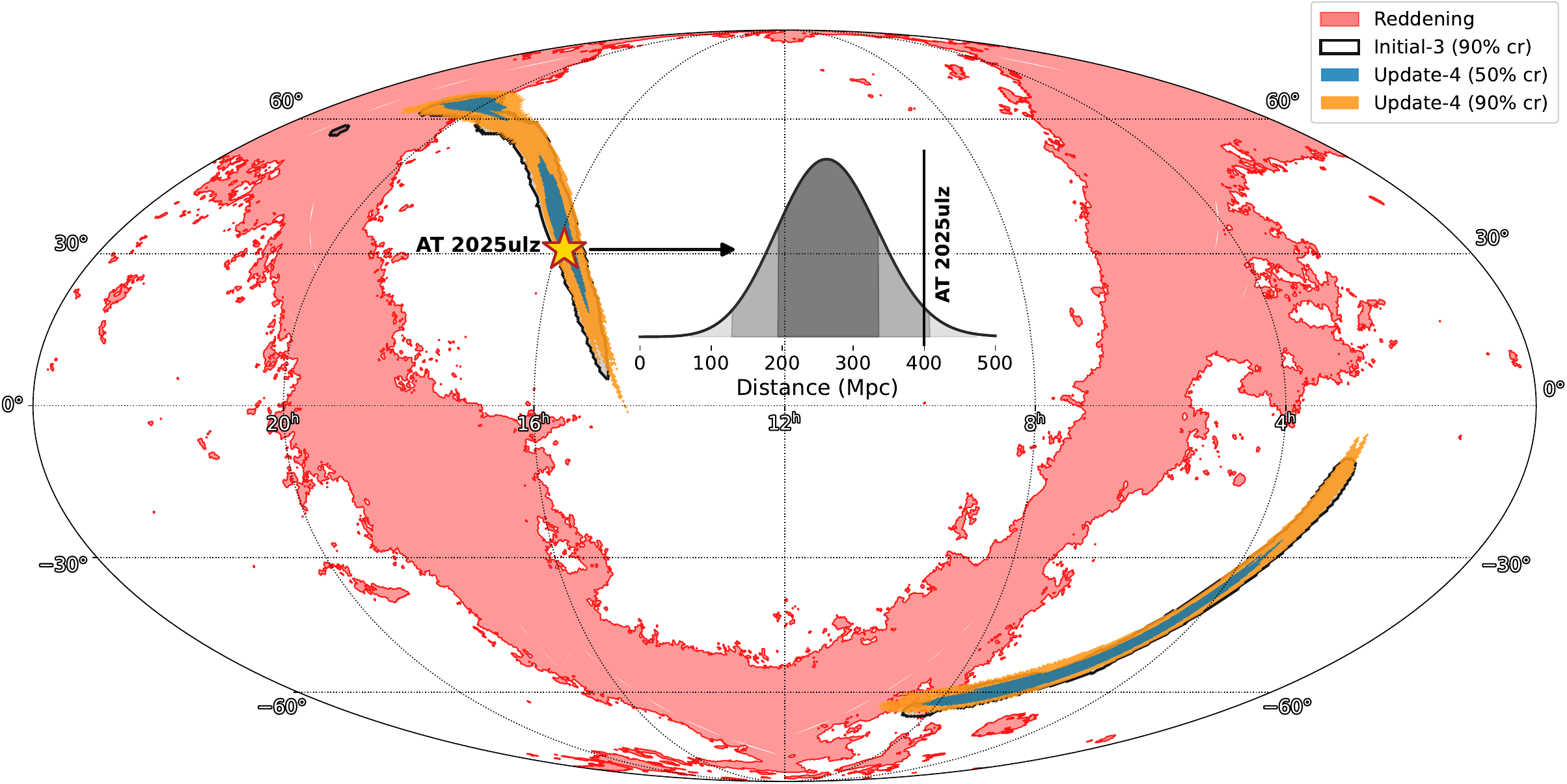}
  \caption{Localisation of the GW event candidate S250818k. The main plot shows a Mollweide projection of the 50\% and 90\% credible regions of the update skymap (light blue: 50\%, 276 deg$^2$; orange: 90\%, 949 deg$^{2}$) and the 90\% credible region of the initial skymap (black contour, 786 deg$^{2}$). The red shaded region marks the sky positions for which the \citet{Schlegel1998} Galactic reddening map predicts $A_{V}>1$. The sky position of the optical transient \thisSN\,is marked with a yellow star. The inset shows the distance posterior conditioned at the position of \thisSN\,(solid black line), with the distance of the source indicated by a vertical line. }
  \label{fig:skymap}
\end{figure*}

Figure \ref{fig:skymap} summarizes the sky localisation of S250818k, of which we show both the 90\% and 50\% credible regions of the sky-projected position (constructed following \citealt{Greco2022}), and the sky-position-conditional distance posterior (inset, constructed following \citealt{Singer2016_3D}). The sky position of \thisSN\,is marked by a star, and its distance is marked by a vertical black line in the inset, for comparison. The LVK Collaboration issued four notices about S250818k: two preliminary notices, followed by an initial and an update notice\footnote{\url{https://gracedb.ligo.org/superevents/S250818k/view/}}. The latter two were accompanied by GCN circulars \citep{2025GCN41437,2025GCN41440}. The plotted sky-localization areas refer to the initial circular\footnote{\url{bayestar.multiorder.fits,2}} (based on the rapid parameter estimation code \texttt{Bayestar}, \citealt{Singer2016_bayestar}) and the update  circular\footnote{\url{Bilby.offline0.multiorder.fits,0}} (based on offline full parameter estimation with \texttt{Bilby}, \citealt{Ashton2019}). The sky-position-conditional distance posterior refers to the latter. The black contour in the main panel (not readily discernible because it is largely coincident with the subsequent update) marks the 90\% credible region of the initial skymap, covering 786~deg$^{2}$. The 50\% and 90\% credible regions of the updated skymap are shown in light blue and orange, corresponding to sky areas of 276~deg$^{2}$ and 949~deg$^{2}$, respectively.
The position of \thisSN\,lies within the~35\% credible region of the update skymap, which covers an area of 160~deg$^{2}$. In distance, the object falls within the $2\sigma$ interval, toward the high-distance tail of the distribution. 
The main panel in the figure includes a Galactic dust reddening and extinction map (red) based on \cite{Schlegel1998}, restricted to the highest-extinction values ($A_{V}>1$). The sky localization of S250818k lies largely outside these high-reddening areas, indicating that dust-related absorption effects along the line of sight had a negligible impact on the search for candidate electromagnetic counterparts.

\section{Data tables}

Table \ref{tab:photometry} summarizes the results of our photometric measurements on difference images of \thisSN. Table \ref{tab:radio} shows a log of our radio observations at the source position, with a summary of the setup and resulting flux density measurements or 5-$\sigma$ upper limits.

\begin{table*}
\centering
\caption{Optical and near-infrared photometry of \thisSN.}
\label{tab:photometry}
\begin{tabular}{cccccc}
\hline\hline
Night  & Mid-time MJD & Instrument & Band & Mag \\
\hline
Aug 21 & 60909.00996 & VLT/FORS2  & $r$ & 22.70$\pm$0.15 \\
Aug 22 & 60910.00380 & VLT/FORS2  & $r$ & 22.95$\pm$0.17 \\
Aug 21 & 60909.01913 & VLT/FORS2  & $i$ & 22.78$\pm$0.15 \\
Aug 22 & 60910.01758 & VLT/FORS2  & $i$ & 22.57$\pm$0.15 \\
Aug 21 & 60908.98651 & VLT/FORS2  & $z$ & 22.21$\pm$0.10 \\
Aug 22 & 60909.98656 & VLT/FORS2  & $z$ & 22.47$\pm$0.15 \\
Aug 20 & 60907.96934 & VLT/HAWK-I & $K$ & 22.30$\pm$0.50 \\
Aug 22 & 60909.98755 & VLT/HAWK-I & $K$ & 22.50$\pm$0.50 \\
Aug 20 & 60907.87521 & NOT/ALFOSC & $r$ & 23.46$\pm$0.14 \\
Aug 21 & 60908.95761 & NOT/ALFOSC & $r$ & 23.13$\pm$0.11 \\
Aug 22 & 60909.91030 & NOT/ALFOSC & $r$ & 22.37$\pm$0.11 \\
Aug 26 & 60913.91633 & NOT/ALFOSC & $r$ & 21.80$\pm$0.11 \\
Sep 22 & 60940.84640 & NOT/ALFOSC & $r$ & 21.83$\pm$0.06 \\
Oct 11 & 60959.82483 & NOT/ALFOSC & $r$ & 22.88$\pm$0.15 \\
\hline
\end{tabular}
\tablefoot{Magnitudes are in the AB system and not corrected for Galactic extinction.}
\end{table*}

\begin{table*}
\caption{Radio observations log and results.}                 
\label{tab:radio}    
\centering                        
\begin{tabular}{c c c c c c c}      
\hline\hline               
Telescope   & Date & Frequency &FWHM & P.A. & RMS & Flux density \\      
            & (dd-mm-yyyy) & (GHz) & (arcsec$\times$arcsec) & (deg) & ($\mu$Jy/beam) & ($\mu$Jy) \\
\hline  
VLA & 29-08-2025 & 3 & 1.97$\times$1.37 & 74 & 7 & $< 21$\\
(22A-414)& & 6 & 1.06$\times$0.74 & 77 & 8 & $< 24$\\
& & 10 & 0.58$\times$0.46 & 56 & 9 & $< 27$\\
& & 15 & 0.35$\times$0.31 & 68 & 7 & $< 21$\\
\hline                      
   MeerKAT  & 21-08-2025    & 3.06  & 4.04$\times$4.04  & 0 & 6&70$\pm$9  \\
            & 28-08-2025    & 3.06  & 3.47$\times$3.47  & 0 & 5&76$\pm$9  \\
           
            & 14-09-2025    & 1.28  & 6.42$\times$6.42  & 0 & 8&116$\pm$14 \\
            & 13-09-2025    & 3.06  & 3.31$\times$3.31  & 0 & 4&69$\pm$8  \\
\hline
  uGMRT     & 26-08-2025    & 1.23  & 7.24$\times$2.19  & -25  & 23 & $<$115  \\
\hline
  e-MERLIN  & 19-09-2025    & 5     & 0.049$\times$0.030 & 32  & 26&$<$130  \\   
\hline                                  
\end{tabular}
\end{table*}

\section{Additional information}

\begin{figure*}
\centering
\includegraphics[width=0.8\textwidth]{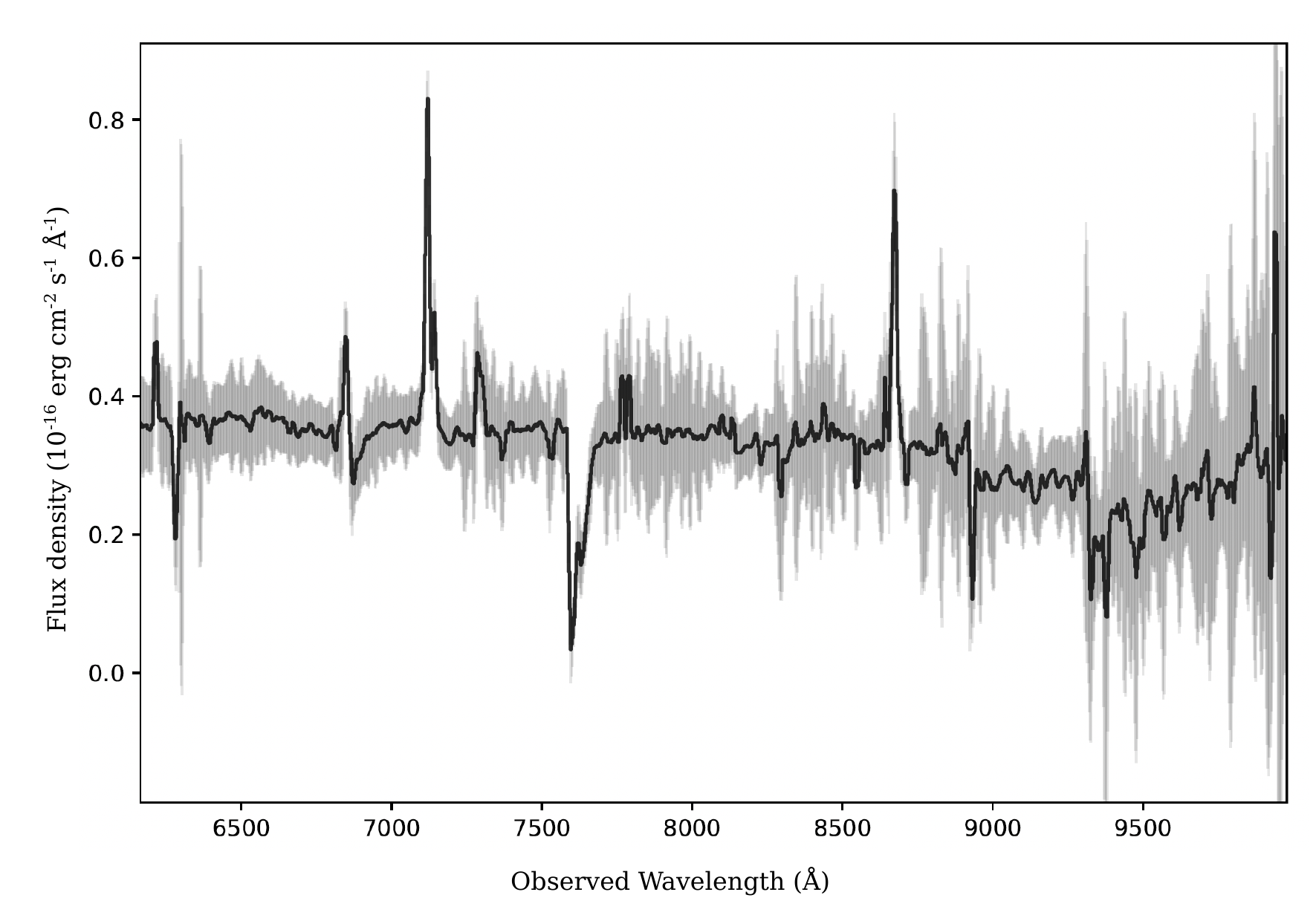}
\caption{The FORS2 spectrum of \thisSN\,taken on the night of Aug 20. We show in grey the spectrum at the native pixel sampling. The black line shows a slightly smoothed spectrum to mitigate contamination from sky line residuals.}
\vspace{-10pt}
\label{fig:FORS_spec}
\end{figure*}

Figure \ref{fig:FORS_spec} shows our FORS2 spectrum of \thisSN\,from the night of Aug 20. Unfortunately, this spectrum did not provide useful information for the classification of the transient. 

\begin{figure}
\centering
\includegraphics[width=\columnwidth]{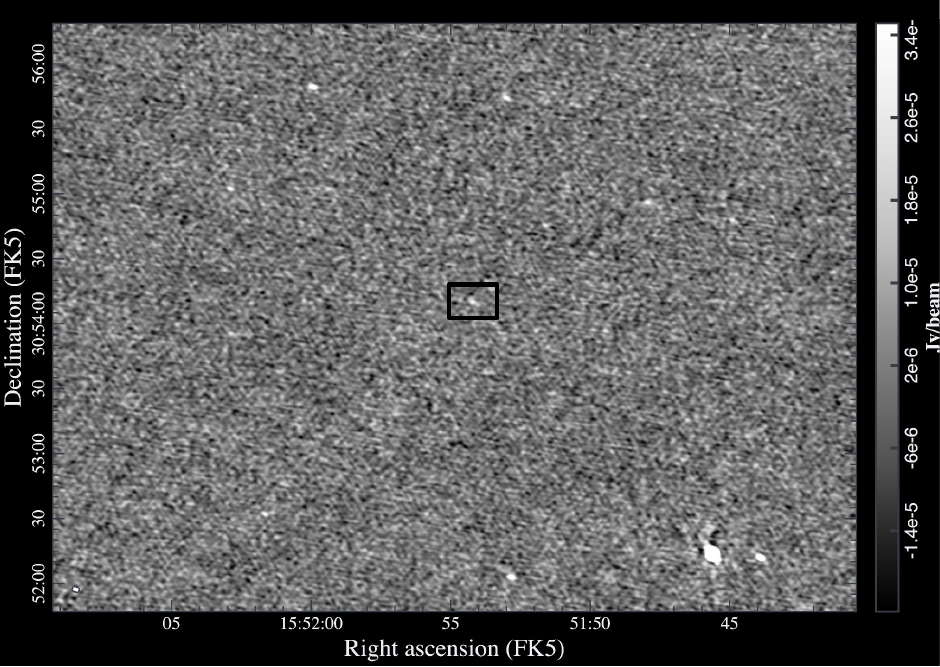}
\caption{Wider-field VLA S-band image. The black rectangle in the centre shows the region covered by the image in Fig.\ \ref{fig:VLA_Sband_image}. The wider field in this image shows the bright close-by source (visible in the lower-right corner) that made our analysis more complex.}
\vspace{-10pt}
\label{fig:VLA_Sband_image_wide}
\end{figure}

Figure \ref{fig:VLA_Sband_image_wide} shows a zoom-out with respect to the VLA S-band image in Fig.\ \ref{fig:VLA_Sband_image}. This figure was included to demonstrate the presence of a bright nearby source which made the reduction slightly more difficult (see \ref{subsec:vla}).

\begin{figure*}
\centering
\includegraphics[width=\textwidth]{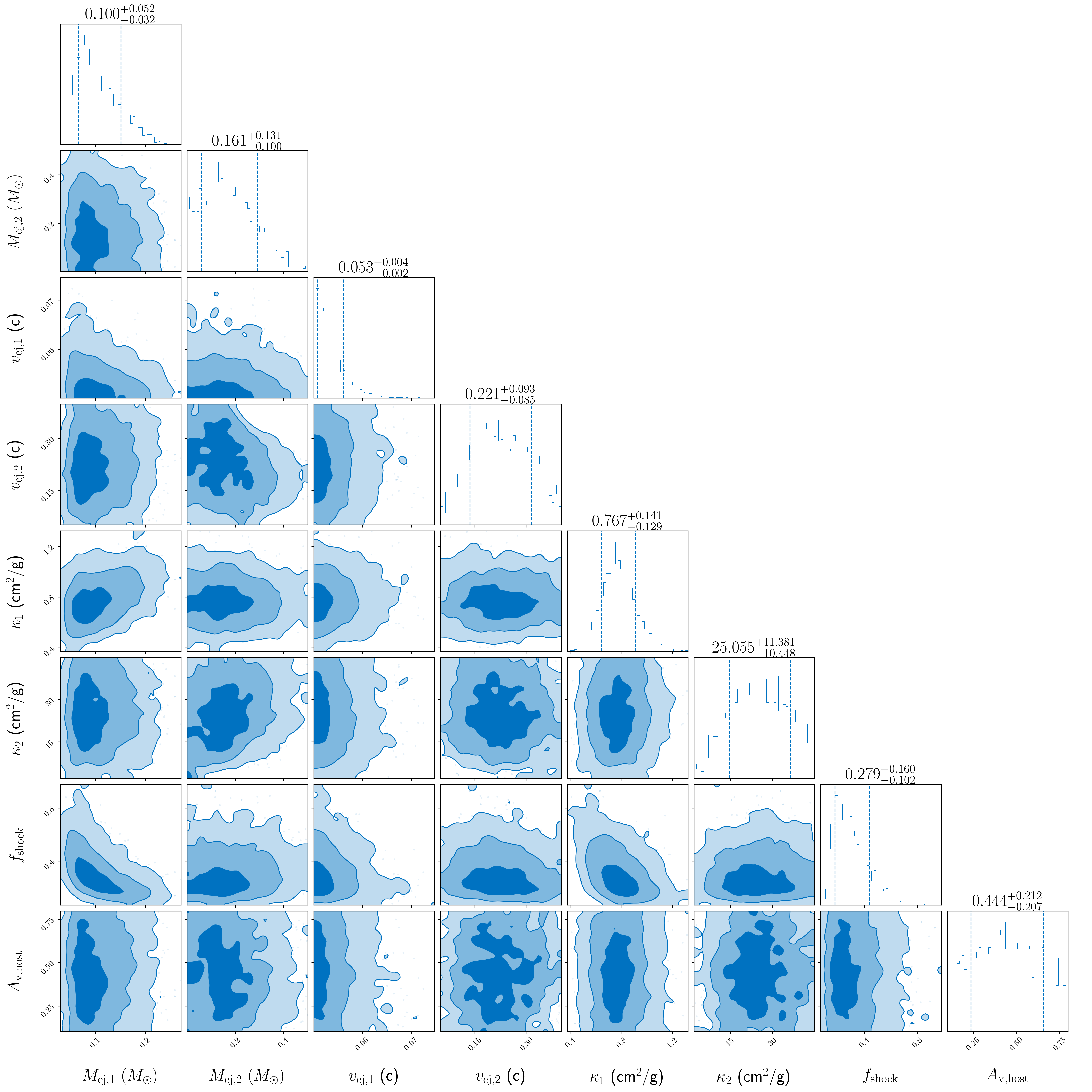}
\caption{Corner plot from our fit to the photometry (shown in the left-hand panel of Fig.~\ref{fig:fittedlc}) with a two-component kilonova model.}
\vspace{-10pt}
\label{fig:kn_model_corner}
\end{figure*}

\begin{figure*}
\centering
\includegraphics[width=0.8\textwidth]{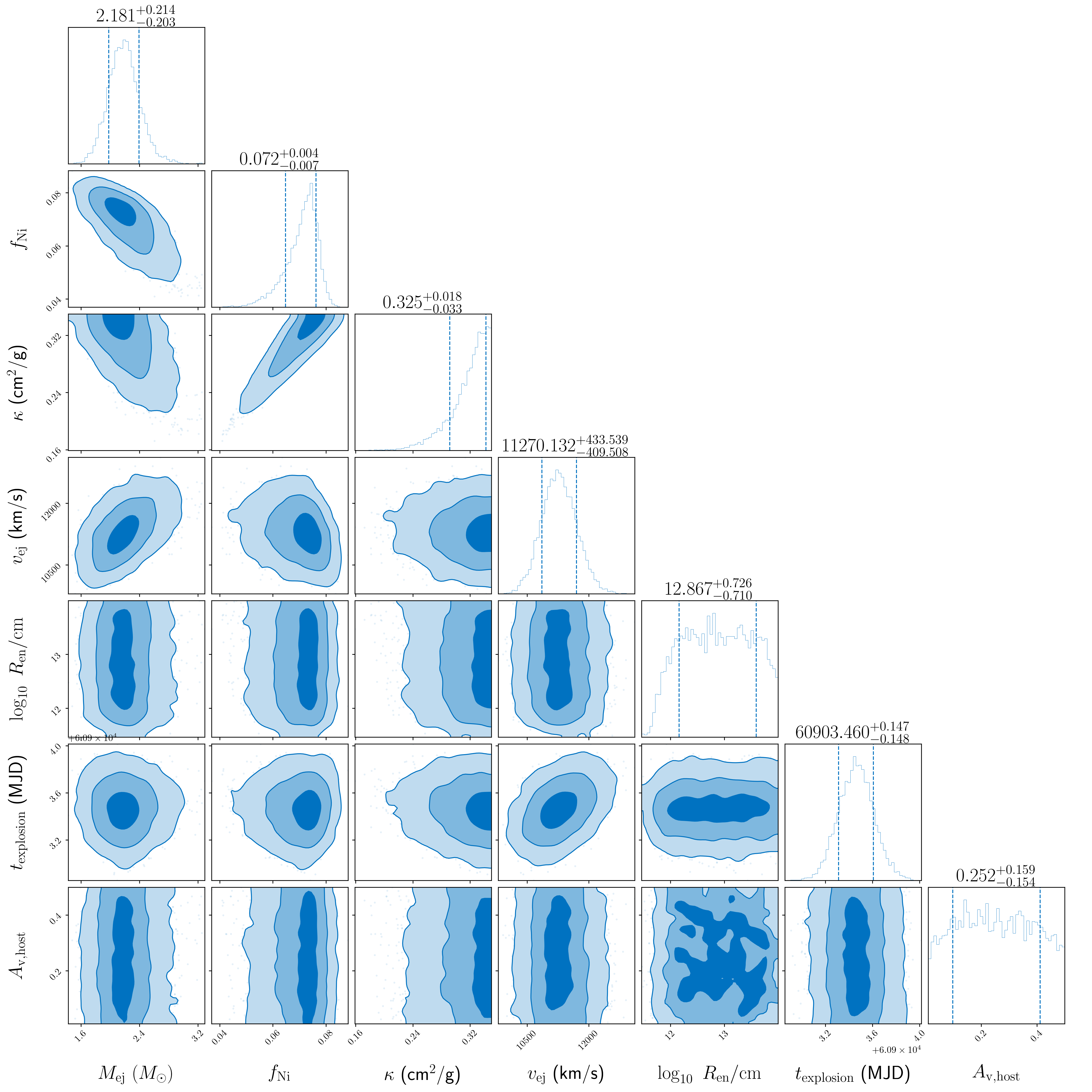}
\caption{Corner plot showing salient parameters from our fit to the photometry (shown in the right-hand panel of Fig.~\ref{fig:fittedlc}) with a shock cooling and radioactive decay model.}
\vspace{-10pt}
\label{fig:model_corner}
\end{figure*}

Figure \ref{fig:kn_model_corner} shows the corner plot demonstrating the posterior probability density of the two-component KN model parameters from the fit to photometric data up to 5 days after the GW trigger (see sect.\ \ref{sec:lc_modeling}. Figure \ref{fig:model_corner} shows a similar plot for the shock cooling plus $^{56}$Ni-decay powered SN model, fitted to the complete photometric dataset.

\end{document}